\documentclass{aa}  
\usepackage{amsmath}
\usepackage{graphicx}
\usepackage{natbib}
\usepackage{multicol}
\usepackage{longtable}
\usepackage{graphicx}
\usepackage{natbib,twoopt}
\usepackage{supertabular,booktabs}
\usepackage{sidecap} 
\usepackage{enumitem}
\usepackage{txfonts}
\usepackage{epstopdf}
\usepackage{tabularx}
\usepackage{longtable}
\usepackage{ltablex}
\usepackage[normalem]{ulem}
\usepackage[pdftex]{color}
\newcommand{\Msun}{M$_{\odot}$}

\begin{document}

   \title{When binaries keep track of recent nucleosynthesis}

   \subtitle{The Zr - Nb pair in extrinsic stars as an $s$-process diagnostic}

   \author{D. Karinkuzhi 
          \inst{1}
          \and
          S. Van Eck\inst{1}
          \and
          A. Jorissen\inst{1}
          \and
          S. Goriely\inst{1}
          \and
          L. Siess\inst{1}
          \and
          T. Merle\inst{1}
          \and
          A. Escorza\inst{1,2}
          \and
          M. Van der Swaelmen\inst{1}
          \and
          H.M.J. Boffin\inst{3}
          \and
          T. Masseron\inst{4,5}
          \and
          S. Shetye\inst{1,2}
          \and
          B. Plez\inst{6}
          }

 \institute{Institut d'Astronomie et d'Astrophysique, Universit\'e Libre de Bruxelles, ULB, Campus Plaine C.P. 226, Boulevard du Triomphe, B-1050 Bruxelles, Belgium
              \and
              Institute of Astronomy, KU Leuven, Celestijnenlaan  200D, 3001 Leuven, Belgium
              \and
             ESO, Karl Schwarzschild Stra\ss e 2, D-85748 Garching bei M\"unchen, Germany
              \and
              Instituto de Astrofísica de Canarias, E-38205 La Laguna, Tenerife, Spain
              \and
              Departamento de Astrofísica, Universidad de La Laguna, E-38206 La Laguna, Tenerife, Spain
              \and
              Laboratoire Univers et Particules de Montpellier, Université Montpellier, CNRS, 34095, Montpellier Cedex 05, France}

   \date{Received X, 2018; accepted Y, 2018}


  \abstract
   {Barium stars are $s$-process enriched giants. They owe their chemical peculiarities to a past mass transfer phase. During this phase they were polluted by their binary companion, which at the time was an asymptotic giant branch (AGB) star, but  is now an extinct white dwarf.  Barium stars are thus ideal targets for understanding and constraining the  $s$-process in low- and intermediate-mass AGB stars. }
   { We  derive the abundances of a large number of heavy elements in order to shed light on the conditions of operation of the neutron source responsible for the production of $s$-elements in the  former companions of the barium stars. 
   }
   {Adopting a recently used methodology \citep{Neyskens2015}, we analyse a sample of eighteen highly enriched barium stars observed with the high-resolution HERMES spectrograph  mounted on the MERCATOR telescope (La Palma). We determine the stellar parameters and abundances using MARCS model atmospheres. 
   In particular, we derive the Nb-to-Zr ratio which was previously shown to be a sensitive thermometer for the $s$-process nucleosynthesis.
   Indeed, in barium stars, $^{93}$Zr has fully decayed into mono-isotopic $^{93}$Nb, so  Nb/Zr  is a measure of the temperature-sensitive $^{93}$Zr/Zr isotopic ratio.
   }
  {HD~28159, previously classified as K5III and initially selected to serve as a reference cool K star for our abundance analysis, turns out to be enriched in $s$-process elements, and as such is a new barium star.
  Four stars, characterised by high nitrogen abundances, also tend to have high [Nb/Zr] and [hs/ls] ratios.
  The derived Zr and Nb abundances provide more accurate constraints on the $s$-process neutron source, identified to be $^{13}$C($\alpha$, n)$^{16}$O for barium stars. 
  The comparison with stellar evolution and nucleosynthesis models shows that the investigated barium stars were polluted by a low-mass (M$\sim$2--3~\Msun) AGB star. 
  HD 100503 is potentially identified as the highest metallicity CEMP-$rs$ star yet discovered. }
  {}


   \maketitle
%

\begin{table*}
\centering
\caption{Programme stars, and adopted atmospheric parameters ($\xi$ is the microturbulence velocity). 
The `Remark' column specifies the method used to derive the stellar parameters (see Sect.~\ref{Sect:parameters}).}
\label{Tab:program_stars}
\begin{tabular}{llccrrll}
\hline
\\
Name &  $T_{\rm eff}$&$\log g$           & $\xi$         &[Fe/H]& S/N  & Spectral & Remark \\
     &    (K)       &  (cm s$^{-2}$)    &  (km s$^{-1}$)& && type     &   \\
\hline\\
HD~5424      & $4728\pm 80  $& $2.53 \pm 0.02 $ & $1.30 \pm 0.05$& $-0.43 \pm 0.11  $  &75    &G8II     &{\sc bacchus}\\
HD~12392     & $4987\pm 17  $& $3.39 \pm 0.30 $ & $1.34 \pm 0.04$& $-0.38 \pm 0.09  $  &105   &G8II     &{\sc bacchus}\\
HD~16458     & $4550 \pm 25 $& $1.80 \pm 0.20 $ & $1.92 \pm 0.05$& $-0.64\pm 0.10   $  &139   &K1Ba5    &{$\chi^2$ method}\\
HD~27271     & $5022 \pm 40 $& $2.86 \pm 0.50 $ & $1.29 \pm 0.03$& $-0.07\pm 0.08   $  &151   &G8II     &{\sc bacchus}\\
HD~28159     & $3900 \pm 50 $& $2.00 \pm 0.50 $ & $2.00         $& $-0.50\pm 0.13   $  & 83   &K5III    &$\chi^2$ method\\
HD~31487     & $4960 \pm 50 $& $3.11 \pm 0.20 $ & $1.45 \pm 0.03$& $-0.04\pm 0.13   $  &120   &G8IIIBa3 &{\sc bacchus}\\
HD~43389     & $4000 \pm 50 $& $2.00 \pm 0.50 $ & $2.00         $& $-0.35\pm 0.13   $  &131   &K0/1IIICNIIp &$\chi^2$ method\\
HD~46407     & $4854 \pm 100$& $2.24 \pm 0.40 $ & $1.34 \pm 0.05$& $-0.36\pm 0.11   $  &127   & G9III   &{\sc bacchus}\\
HD~50082     & $4789 \pm 100$& $2.44 \pm 0.50 $ & $1.36 \pm 0.05$& $-0.32\pm 0.11   $  & 86   &K0Ba3    &{\sc bacchus}\\
HD~60197     & $3800 \pm 50 $& $2.00 \pm 0.50 $ & $2.00         $& $-0.60\pm 0.12   $  &105   &K3.5III  &$\chi^2$ method\\
HD~88562     & $4000 \pm 50 $& $2.00 \pm 0.50 $ & $2.00         $& $-0.53\pm 0.12   $  & 83   &K1III    &$\chi^2$ method\\
HD~100503    & $4000 \pm 50 $& $2.00 \pm 0.50 $ & $2.00         $& $-0.72\pm 0.13   $  & 79   &G/KpBa   &$\chi^2$ method \\
HD~116869    & $4892 \pm 30 $& $2.59 \pm 0.07 $ & $1.39 \pm 0.04$& $-0.44\pm 0.09   $  & 64   &G8III    &{\sc bacchus}\\
HD~120620    & $4831 \pm 13 $& $3.03 \pm 0.30 $ & $1.11 \pm 0.05$& $-0.30\pm 0.10   $  & 69   &K0III    &{\sc bacchus}\\
HD~121447    & $4000 \pm 50 $& $1.00 \pm 0.50 $ & $2.00         $& $-0.90\pm 0.13   $  & 73   &K4III    &$\chi^2$ method\\
HD~123949    & $4378 \pm 80 $& $1.78 \pm 0.53 $ & $1.37 \pm 0.07$& $-0.31\pm 0.13   $  & 88   &K1pBa    &{\sc bacchus}\\
HD~178717    & $3800 \pm 50 $& $1.00 \pm 0.50 $ & $2.00         $& $-0.52\pm 0.11   $  & 94   &K3.5III  &$\chi^2$ method\\
HD~199939    & $4710 \pm 9  $& $2.35 \pm 0.40 $ & $1.49 \pm 0.05$& $-0.22\pm 0.11   $  & 163  &G9III    &{\sc bacchus}\\
Arcturus$^*$ & $4250 \pm 35 $& $1.50 \pm 0.06 $ & $1.58 \pm 0.12$& $-0.62\pm 0.08   $  & 1000 & K1.5III&{\sc bacchus}   \\
V762 Cas$^*$ & $3800 \pm 50 $& $0.00 \pm 0.50 $ & $2.00         $& $-0.08\pm 0.15   $  &110   & M3       &$\chi^2$ method\\
\hline
\end{tabular}

$^*$ Reference objects\\
\end{table*}

\section{Introduction}
Low- and intermediate-mass asymptotic giant branch (AGB) stars are major contributors to the chemical evolution of the
Galaxy, especially in the case of carbon and elements heavier than iron produced by the $s$-process \citep{Kappeler1999,Kappeler2011} and perhaps for nitrogen as well \citep[e.g.][]{Merle2016}.  
There are still many open questions about the exact physical conditions required for the occurrence of these nucleosynthesis processes. 
On the AGB, the first stars to exhibit surface signatures of $s$-process nucleosynthesis are S-type stars, which have prominent ZrO bands
and $s$-process overabundances.
However, their low photospheric temperatures cause a strong molecular blending throughout their spectrum, making it very difficult to study their abundances in detail \citep{Smith-Lambert-1990,VanEck2017}. 
The problem is even more accute for carbon stars.
In this respect, extrinsic stars can be an invaluable help.

The concept of extrinsic stars was first introduced by \citet{Iben1983} to distinguish the S stars, which owe their $s$-process enrichment to mass transfer from an AGB companion, from the intrinsic S stars, which are genuine AGB stars whose atmospheres are enriched by $s$-process material produced in their interiors. Extrinsic S stars, which do not exhibit Tc lines (an element with no stable isotopes), populate the tip of the RGB or the early AGB; as expected, they are all binaries and do not show infrared excesses, unlike their more evolved intrinsic counterparts \citep{Jorissen1993, VanEck1999I, VanEck2000II, VanEck2000III}. The subclass of K giants known as barium stars \citep{Bidelman1951} are the warmer counterparts of extrinsic S stars \citep{Smith1988,Jorissen1988}, whereas CH giants \citep{Keenan1942} and CEMP stars \citep[at least those enriched with $s$-process elements, the so-called CEMP-s;][]{Masseron2010} are the low-metallicity counterparts of the barium stars.
A more detailed discussion about the properties of the various families of extrinsic stars may be found in \citet{Jorissen2004} for example. 

Extrinsic stars being less evolved and warmer than AGB stars, their abundance analysis is much easier and they are ideal targets for the study of AGB nucleosynthesis.
A recent study by \citet{Neyskens2015} is a good illustration  of this; these authors propose abundance diagnostics which act as $s$-process  chronometers and thermometers (respectively Tc/Zr and Nb/Zr). These ratios may be used in both intrinsic and extrinsic stars, and they may also serve to distinguish extrinsic from intrinsic stars.

In the present study, we derive Nb-to-Zr ratios in a representative sample of extrinsic stars to investigate the detailed operation of the $s$-process. Following the method described by \citet{Neyskens2015}, our aim is to confirm that the neutron source responsible for the production of $s$-process elements is the $^{13}$C($\alpha$,n)$^{16}$O reaction, which operates in radiative layers during interpulse periods at temperatures of the order of $10^8$~K  \citep[e.g.][]{Kappeler2011,Bisterzo2015}, rather than the $^{22}$Ne($\alpha$, n)$^{25}$Mg reaction activated at temperatures in excess of $3.2\times10^8$~K in the convective thermal pulses of massive AGB stars ($M \gtrsim 4-5$~M$_\odot$).


The list of target stars and the observations are described in Sect.~\ref{Sect:observations}. The derivation of the stellar atmospheric parameters is presented in Sect.~\ref{Sect:parameters}. All the derived abundances 
are presented in Sect.~\ref{Sect:abundances}. 
Light- and heavy-element abundance profiles, as well as comparison with predictions from stellar evolution and nucleosynthesis models,
are presented in Sects.~\ref{Sect:STAREVOL} and \ref{Sect:heavy}.
The $s$-process diagnostic based on the ([Zr/Fe], [Nb/Fe]) abundance pair is discussed in Sect.~\ref{Sect:thermometer}.

\begin{table*}
\caption{Abundances in the reference stars: $\sigma_s$ is the line-to-line scatter; $N$ the number of lines used; $\log \epsilon$ is the logarithm of the elemental abundance by number, in the scale where $\log {\epsilon_H} = 12$. For $^{12}$C/$^{13}$C the column lists the number ratio.} \label{Tab:benchmark_abundances}
\begin{tabular}{lrccrcrrrcrrrr}
\multicolumn{3}{c}{}& \multicolumn{6}{c}{Arcturus} && \multicolumn{4}{c}{V762 Cas} \\
\cline{4-9}\cline{11-14}\\
 &    $Z$  &    $\log {\epsilon_{\odot}}^a$ & $\log {\epsilon}$&$\sigma_{s} (N)$& [X/H]$^b$  & [X/Fe]$^b$ & [X/Fe]$^c$&[X/Fe]$^d$  && $\log {\epsilon}$&$\sigma_{s} (N)$& [X/H]  & [X/Fe]  \\
 &       &                         &                  &        &      \\  
\hline 
C     &  6  &    8.43  &   7.7 &0.15  & $-0.73$   &$-0.11$&-    &- &&  8.40&0.15 &   $-0.03$ &  0.05\\ 
$^{12}$C/$^{13}$C&  &  &    12 &      &         &     &      &  &&   19  &     &        &       \\
N     &  7  &    7.83  &   7.80    &0.06(25)  & $-0.03$  & 0.59 &-    &- &&  7.78&0.10 &   $-0.05$ &   0.03\\
O     &  8  &    8.69  &   8.40&0.15  & $-0.29$   & 0.33& - &-&&  8.69:&- &   0.00 &   0.08\\
Na    &  11 &    6.24  &   5.64&0.03(2)  & $-0.60$   & 0.02& $0.09$&0.15&&  -&- &  - &   -\\
Mg    &  12 &    7.60  &   7.46&0.06(2)  & $-0.14$   & 0.48& $0.33$ &0.37&&  -&- &   - &   -\\
Fe    & 26  &    7.50  &   6.88&0.11(45)  & $-0.62$ & -  & $-0.69$   &  -0.60    &&  7.42&0.10(40) &   $-0.08$ &       \\
Rb I  & 37  &    2.52  &    -  &    -  &  -     &   -     &    -    &&&    2.23&0.22(2)& $-0.29$&$-0.21$\\
Sr II & 38  &    2.87  &   2.20&0.00(2)   & $-0.67$   & $-0.05$& - &  -&&  -   &     &      -  &     -\\
Y II  & 39  &    2.21  &   1.48&0.10(4)&$-0.73$  &  $-0.11$& - &  -&&  2.15&0.03(6) & $-0.06$ &  0.02\\
Zr I  & 40  &    2.58  &   1.84&0.19(2)& $-0.74$  & $-0.12$& $-0.09$&0.01  &&  2.40&0.14(2) & $-0.18$   & $-0.10$\\
Zr II & 40  &    2.58  &   1.93&0.07(3)& $-0.65$  & $-0.03$ & - & -&&  -        &        &   - &  -\\
Nb II & 41  &    1.46  &   -   &            & -      & -& - & -&&  1.36 &0.10(11)&   $-0.10$ & $-0.02$\\
Ba II & 56  &    2.18  &   -    &           & -      & - &$-0.18$&$-0.19$ &&  2.2:  &   -  &   0.02 &   0.10\\
La II & 57  &    1.10  &   0.39&0.04(9)&  $-0.71$  & $-0.09$&$-0.04$&0.04 &&  1.11&0.11(10) &   0.01 & 0.09\\
Ce II & 58  &    1.58  &   0.73&0.03(10)& $-0.85$ & $-0.23$& - &  -&&  1.59&0.14(7) &   0.01 &   0.09\\
Pr II & 59  &    0.72  &   0.05&0.10(4) & $-0.67$ & $-0.05$& -  & -&&  0.71&0.03(4) &   $-0.01$ &  0.08\\
Nd II & 60  &    1.42  &   0.71&0.09(14)& $-0.71$ & $-0.09$& - &  -&&  1.43&0.10(11) &   0.01 &  0.09\\
Sm II & 62  &    0.96  &   0.46&0.14(7) & $-0.50$ &  $0.12$& - &  -&&  0.80&0.07(2) &   $-0.16$ &  $-0.08$\\
Eu II & 63  &    0.52  &   0.15&0.07(2) & $-0.37$ &  $0.25$& $0.40$& 0.36&&  0.45&0.07(2) &   $-0.07$ &  0.01\\
\hline 
\end{tabular}
\mbox{}\medskip\\
$^{a}$ \citet{Asplund2009} \\
$^{b}$ This work\\
$^{c}$ \citet{vanderswaelman2013}\\
$^{d}$ \citet{Worley2009}\\
Note: For Arcturus, \cite{Jofre-2014} find [Fe/H]=-0.52.
\end{table*}

\section{Observational sample}
\label{Sect:observations}

In the present paper, we have selected a sample of barium  stars with photometric temperatures 
below 5500~K (Table~\ref{Tab:program_stars}) and known from former abundance analyses \citep[e.g.][]{Smith1984,Allen2006,Smiljanic2007}
to have large $s$-process enrichment levels. This ensures that the $s$-process signature will dominate over the pristine (presumably solar-scaled, before mass-transfer) abundance pattern.
High-resolution spectra ($R \sim 86\,000$) covering the wavelength range from 377.0  to 
900.0~nm  were obtained for these objects using the
HERMES spectrograph \citep{Raskin2011} mounted on the 1.2m Mercator telescope at the Roque de los Muchachos Observatory, La Palma, Canary Islands. 
These spectroscopic data were reduced using the standard HERMES pipeline. 
Thanks to the HERMES long-term radial velocity monitoring of these objects \citep{Gorlova2013}, most of them are identified as radial velocity variables  (the two exceptions being HD 12392 and HD 116869, for which no radial velocity monitoring is available), thus
confirming their extrinsic nature.
Among the many observations available covering the period 2009 -- 2016, we selected the spectra with the best signal-to-noise ratio.

In addition to barium stars, our sample also comprises S-type stars
(both extrinsic and intrinsic) and M stars used for comparison, 
all studied by \citet{Neyskens2015} using HERMES spectra as well.
These stars are listed in Table~\ref{Tab:NbZr}.
Moreover, in order to select good, relatively unblended lines with accurate $\log gf$ values, two reference cool giant stars were also included; one is the K1.5 giant Arcturus \citep{Hinkle-2000}, and the other the M3 giant V762~Cas (HERMES spectrum).
Finally, HD 28159 is a K5III star (also classified as M) initially selected to serve as a reference cool K star for our abundance analysis. However, it proved to be slightly $s$-process enriched (Sect.~\ref{Sect: s-process}, Table~\ref{Tab:NbZr}, and Table \ref{Tab:abundances}) and characterised by a variable radial velocity  (Sect.~\ref{Sect:pattern}). Therefore, we include it in our extrinsic star sample.

\section{Derivation of the atmospheric parameters}
\label{Sect:parameters}

Atmospheric parameters for the majority of the programme  stars are derived using the {\sc bacchus} pipeline \citep{Masseron2016}. This pipeline uses the 1D local thermodynamical equilibrium (LTE)
 spectrum-synthesis code Turbospectrum \citep{Alvarez1998,Plez2012}.
The spectroscopic method for deriving the atmospheric parameters is based on ensuring consistency between the abundances derived from  Fe~I and Fe~II lines (for  $\log g$), between Fe lines of various excitation potentials (for $T_{\rm eff}$), and between Fe lines of various reduced equivalent widths (for the microturbulence velocity $\xi$).

For a few of the coolest stars, the spectra are too blended by molecular features and {\sc bacchus} could not converge. A $\chi^2$ method was used instead, selecting models which minimise $\chi^2$ computed from differences between observed and synthetic spectra. The spectra were split into 17 chunks of $\sim 200$ \AA\ (see Table~\ref{Tab:chi2limits}) in order to correctly adjust the pseudo-continuum for each chunk. The derived parameters are presented in Table~\ref{Tab:program_stars} along with the spectra S/N. We assume the model microturbulence of 2~km~s$^{-1}$, a reasonable value for such cool giants. Additional broadening is necessary to correctly reproduce the width of spectral lines; convolution by a Gaussian profile of FWHM typically 6 km/s was applied to account for macroturbulence and spectral resolution. 

The atmospheric parameters for the  S and M stars are not listed in Table~\ref{Tab:program_stars} since they can be found in \citet{Neyskens2015}.

\section{Abundance determination}
\label{Sect:abundances}

Abundances for all the elements are derived by comparing observed spectra with 
synthetic spectra generated by the Turbospectrum radiative transfer code using MARCS model 
atmospheres. We have used the same molecular lines as presented in Table~B.1 of \citet{Merle2016} to derive C and N abundances. The molecular line list  was taken from
\citet{masseron2014} for CH and from \citet{Sneden2014} for CN. The references for the other molecular line lists (TiO, SiO, VO, C$_{2}$,  NH, OH, MgH, SiH, CaH, and FeH) can be found in \citet{Gustafsson2008}. 
The line list used for all the elements is presented in Table~\ref{Tab:linelist}, and is largely based upon the one used for the Gaia/ESO survey \citep{Heiter2015}.  
The lines are carefully selected to keep only those with negligible blending.

\subsection{Elemental abundances in the reference stars Arcturus and V762 Cas}

Table~\ref{Tab:benchmark_abundances} lists the abundances derived in the reference stars Arcturus and V762~Cas, using the atmospheric parameters listed in Table~\ref{Tab:program_stars}. 
Previous abundance analyses  \citep[e.g.][]{Maeckle1975,Ramirez2011,Neyskens2015} have shown that these stars are not enriched in $s$-process elements; they are used as benchmark stars to test the accuracy of our analysis. We confirm that
V762~Cas has a nearly solar metallicity, with all elemental abundances within $\pm0.1$~dex of the solar values, giving confidence in the adopted $\log gf$ of the selected lines. Arcturus is somewhat metal-poorer, and its abundance distribution differs from solar 
\citep[e.g.][]{vanderswaelman2013,Worley2009}.
As shown in Table~\ref{Tab:benchmark_abundances}, the abundances derived in the present paper are in good agreement with previous literature values (differences smaller than 0.1~dex). For the analysis of our target stars, only the lines tested in the reference stars have been used. 

\subsection {Abundances in the programme stars}

The programme star abundances, along 
with their standard errors calculated from the line-to-line dispersion, are presented in Table~\ref{Tab:abundances}.

\subsubsection{C, N, and O}

Abundances of C, N, and O were obtained for all programme stars. Oxygen abundances are derived first from the [\ion{O}{I}] line at 6300.304~\AA, but we also use the \ion{O}{I} resonance triplet at 7770~\AA. This yields abundances that are consistently 
higher by 0.3~dex in all the objects compared to the abundance derived from the [\ion{O}{I}] line. This difference is usually ascribed to non-LTE (NLTE) effects
strongly affecting the resonance line \citep{Asplund2005, Amarsi2016}. Hence, the O abundance 
obtained from the forbidden line is the one finally retained. For a few cool barium stars ($T_\mathrm{eff} \leq 4000$~K) for which we could detect neither the 6300.304~\AA\ [\ion{O}{I}] line nor the high excitation line at 6363.776~\AA, we
used TiO bands instead to get the oxygen abundance. 

For the coolest barium stars in our sample, the carbon abundance is obtained from the 4300~\AA~CH G-band. Once the C abundance was known, we  re-derived the O abundance since in cool-star atmospheres, the O atoms trapped in the CO molecule may have a strong impact on the overall O abundance.
 For the hottest barium stars in our sample,  
the C$_2$ molecular lines \citep[as listed in Table~B.1 of][]{Merle2016} are used to obtain the C abundance,
which is consistent with the value derived from the CH G-band. The $^{12}$C/$^{13}$C  ratio is derived using $^{12}$CN features at 8003.553 and
 8003.910~\AA\ and $^{13}$CN features at 8004.554, 8004.728, and 8004.781~\AA, which are considered to be the most reliable according to \citet{Barbuy1992}. We also used the $^{13}$CN 
features at 8010.458 and 8016.429~\AA, and checked that the abundances are consistent overall.  

Finally, the nitrogen abundance is obtained in all objects from the CN lines available in the 7500 -- 8400~\AA\ range. 
\subsubsection{Na and Mg}
Sodium abundances are mainly obtained from the \ion{Na}{I} lines at 6154.226 and 6160.753~\AA. 
Magnesium is derived using \ion{Mg}{I} lines at 6318.717 and 6319.21~\AA. In some objects, 
we were also able to use the lines at 8717.810~\AA\ and 8736.080~\AA\ in addition to the above two lines.

 \begin{figure}
   \centering
    \includegraphics[width=9cm]{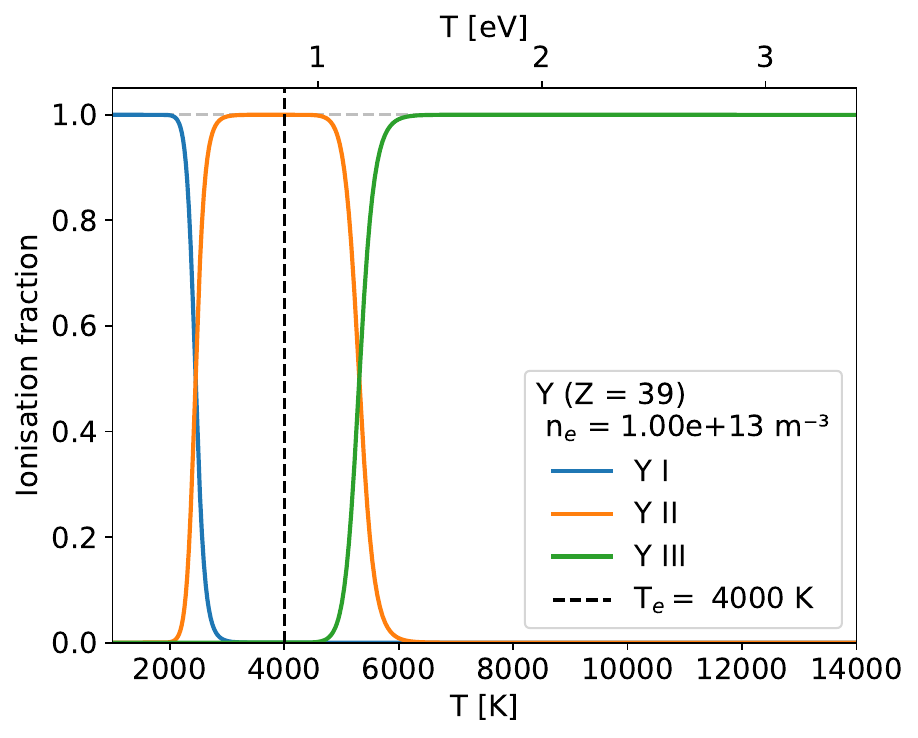}
    \hfill
    \caption{\label{Fig:NLTE}Ionisation fraction vs. temperature for yttrium. }
    
  \end{figure}

\subsubsection{$s$-process}
\label{Sect: s-process}
Abundances for all the $s$-process elements are derived from spectrum synthesis. We also considered the hyperfine splitting (hfs) for Ba, La, and Eu, whose lines are known to be strongly affected by hfs.\\

\noindent{\em Rb}

\noindent We derived Rb abundances in the ten coolest stars in our sample from the two \ion{Rb}{I}
 resonance lines at 7800.26 and 7947.59~\AA. We were not able to detect \ion{Rb}{I} lines in the warmest targets.
\medskip\\

\begin{table}
\caption{Ratios of [Zr/Fe] and [Nb/Fe] in the programme stars
\label{Tab:NbZr}
}
\begin{tabular}{lrccccc}
\hline
\\
Star name & $[\frac{\rm Fe}{\rm H}]$ & $\sigma_{[\frac{\rm Fe}{\rm H}]}$ &  $[\frac{\rm Nb}{\rm Fe}]$ & $\sigma_{[\frac{\rm Nb}{\rm Fe}]}$ &$[\frac{\rm Zr}{\rm Fe}]$& $\sigma_{[\frac{\rm Zr}{\rm Fe}]}$  \\
\\
\hline
\multicolumn{7}{c}{Barium stars} \\
HD~5424    & $-0.43$ &  0.13  &   1.13  & 0.10 &  1.05 &  0.13\\
HD~12392   & $-0.38$ &  0.13  &   1.07  & 0.16 &  1.21 &  0.11 \\  
HD~16458   & $-0.64$ &  0.11  &   1.32  & 0.17 &  1.29 &  0.11\\    
HD~27271   & $-0.07$ &  0.12  &   0.79  & 0.17 &  0.79 &  0.17\\
HD~28159   & $-0.50$ &  0.13  &   0.43  & 0.16 &  0.30 &  0.12\\
HD~31487   & $-0.04$ &  0.13  &   1.15  & 0.16 &  1.11 &  0.11 \\      
HD~43389   & $-0.35$ &  0.13  &   0.64  & 0.16 &  0.32 &  0.12\\
HD~46407   & $-0.35$ &  0.12  &   1.29  & 0.16 &  1.28 &  0.10\\  
HD~50082   & $-0.32$ &  0.11  &   1.14  & 0.17 &  1.04 &  0.10\\
HD~60197   & $-0.60$ &  0.13  &   0.89  & 0.16 &  0.87 &  0.10\\ 
HD~88562   & $-0.53$ &  0.12  &   0.53  & 0.16 &  0.43 &  0.12\\
HD~100503  & $-0.72$ &  0.13  &   1.33  & 0.16 &  1.00 &  0.12\\
HD~116869  & $-0.44$ &  0.12  &   0.98  & 0.21 &  1.01 &  0.17\\  
HD~120620  & $-0.29$ &  0.12  &   1.35  & 0.16 &  1.27 &  0.14\\
HD~121447  & $-0.90$ &  0.13  &   1.92  & 0.16 &  1.57 &  0.14\\
HD~123949  & $-0.23$ &  0.12  &   0.96  & 0.16 &  0.88 &  0.14 \\ 
HD~178717  & $-0.52$ &  0.13  &   0.68  & 0.16 &  0.44 &  0.17\\ 
HD~199939  & $-0.22$ &  0.13  &   1.33  & 0.16 &  1.19 &  0.13\\ 
\hline
\multicolumn{7}{c}{Extrinsic S stars}\\
BD Cam     & $-0.03$ &  0.06  &   0.61  & 0.16 &  0.78 &  0.18\\
HD~119667  & $ 0.01$ &  0.07  &   0.39  & 0.17 &  0.59 &  0.18\\
V613 Mon   & $-0.26$ &  0.15  &   0.40  & 0.22 &  0.66 &  0.21\\
HD~191226  & $-0.28$ &  0.10  &   0.19  & 0.19 &  0.35 &  0.16\\
HD~191589  & $ 0.01$ &  0.06  &   0.20  & 0.17 &  0.29 &  0.15\\
V530 Lyr   & $-0.16$ &  0.09  &   0.20  & 0.23 &  0.29 &  0.20\\
CPD-19 1672& $-0.01$ &  0.06  &   0.28  & 0.19 &  0.41 &  0.21\\
HR 363     & $-0.38$ &  0.12  &   0.82  & 0.19 &  0.83 &  0.22\\
V1261 Ori  & $-0.22$ &  0.15  &   0.80  & 0.23 &  0.85 &  0.22\\
\hline
\multicolumn{7}{c}{Intrinsic S stars}\\
 CSS 454    & $-0.40$ &  0.12  &  -0.07  & 0.24 &  0.80 &  0.21\\
$\sigma$ Ori&$-0.45$ &  0.07  &  -0.04  & 0.17 &  0.45 &  0.20\\
KR CMa     & $-0.34$ &  0.11  &   0.02  & 0.23 &  0.44 &  0.27\\
NQ Pup     & $-0.31$ &  0.13  &  -0.11  & 0.16 &  0.76 &  0.21\\
AD Cyg     & $-0.05$ &  0.14  &   0.05  & 0.24 &  0.90 &  0.20\\
HR Peg     & $ 0.00$ &  0.06  &   0.06  & 0.08 &  0.55 &  0.18\\
AA Cam     & $-0.04$ &  0.05  &   0.02  & 0.24 &  0.15 &  0.25\\
HIP 103476 & $-0.01$ &  0.06  &  -0.03  & 0.16 &  0.61 &  0.18\\
\hline
\multicolumn{7}{c}{M stars}\\
RR UMi     & $-0.06$ &  0.06  &  -0.01  & 0.13 & -0.04 &  0.22\\
V762 Cas$^*$& $-0.08$ &  0.12  &  -0.02  & 0.17 & -0.10 &  0.14\\
V465 Cas   & $ 0.03$ &  0.05  &   0.03  & 0.27 & -0.13 &  0.21\\
$\mu$ Gem  & $-0.15$ &  0.06  &   0.18  & 0.18 &  0.05 &  0.20\\
$\rho$ Per & $-0.06$ &  0.09  &   0.10  & 0.17 & -0.09 &  0.23\\
RZ Ari     & $-0.22$ &  0.20  &  -0.15  & 0.27 &  0.02 &  0.25\\
\hline 
\end{tabular}

Data for intrinsic S stars and M stars are taken from \citet{Neyskens2015}\\
$^*$ Reference object (see Sect.~4)\\
\end{table}

\begin{table}
\begin{center}
\caption{Sensitivity of the abundances ($\Delta \log \epsilon_{X}$) upon variations of the atmospheric parameters.}
\label{Tab:uncertainties}
\begin{tabular}{lrrr}
\hline
\\
Element &   $\Delta T_{\rm eff}$ & $\Delta \log g$ & $\Delta \xi_t$  \\
&   ($-$100) & ($-$0.5) & ($-$0.5) \\
&   (K) & (dex) & (km~s$^{-1}$) \\
\\
\hline
C  & $-$0.15 &$-$0.08& 0.00\\
N &$-$0.30 & 0.30 & $-$0.10\\
O &$-$0.15& $-$0.08&0.00\\
Na I& 0.15 & 0.00 & 0.00\\
Mg I& $-$0.45 & 0.22 & $-$0.05\\
Fe & 0.15 &$-$0.20 &$-$0.07\\
Sr I & $-$0.10 &   0.00 &   0.00  \\
Sr II &  $-$0.15&  $-$0.08&   0.02\\
Y I &   $-$0.23 & $-$0.15  &  0.10\\
Y II&     0.03&  $-$0.17&   0.09\\
Zr I&0.08  &$-$0.13&$-$0.03\\
Zr II&    0.02&   0.10 &  0.30\\
Nb I &$-$0.24 &$-$0.16&$-$0.20\\
La II &    $-$0.04 & $-$0.16 & 0.16\\
Ce II &     0.03&  $-$0.11 &  0.22\\
Pr II &     0.05  & 0.05 &  0.30\\
Nd II &    0.05 & $-$0.07  & 0.20  \\ 
Sm II&     0.15&   0.03 &  0.30\\
Eu II &    $-$0.02 & $-$0.22 &  0.00\\

\hline 
\end{tabular}
\end{center}
\end{table}

\noindent{\em Light $s$-process elements (Sr, Y, Zr)}

\noindent For the three first-peak $s$-process  elements Sr, Y, and Zr, lines are available for both the 
neutral and ionised species. Abundances for the neutral  species  are consistently  lower than those of the ionised species. This 
discrepancy is discussed  by many authors for barium stars \citep{Smith1984,Allen2006,Smiljanic2007} and S stars \citep{Vanture2003}. It is likely caused by NLTE effects affecting the minority species (here the neutral one when $T_{\rm eff} \sim 4500$~K). Figure~\ref{Fig:NLTE} indicates the yttrium dominant species according to temperature; the situation is very similar for strontium and zirconium (the singly ionised species dominating for temperatures in the range 2500K -- 5500K).
 
The situation for Zr is a bit more complex. In addition to the effect discussed above, there is a systematic discrepancy
between abundances derived from the two 7819.37 and 7849.36~\AA~
\ion{Zr}{I} lines, and from many other \ion{Zr}{I} lines.
More precisely, the 7819.37 and 7849.36~\AA~ \ion{Zr}{I} lines systematically lead to
\ion{Zr}{I} abundances $0.30$~dex lower than \ion{Zr}{I} abundances
derived from other lines with transition probabilities
from \citet{Corliss1962}. 
This problem has already been noted by \citet{Neyskens2015}, who decided to rely solely on abundances derived from the first pair of lines which have transition probabilities obtained from laboratory measurements \citep{Biemont1981}. Table~\ref{Tab:benchmark_abundances} reveals that in the benchmark stars, these two \ion{Zr}{I} lines yield a Zr abundance  0.1~dex below the solar abundance. This will be of importance in the discussion of the Zr/Nb thermometer (Sect.~\ref{Sect:thermometer}).
\medskip\\
\noindent{\em Nb}

\noindent Our selected \ion{Nb}{I} lines are listed in Table~\ref{Tab:linelist}. Given the importance of this element for the $s$-process thermometer described in Sect.~\ref{Sect:thermometer}, special care has been applied in testing the adequacy of these lines to derive a reliable Nb abundance in  the two reference stars Arcturus and V762~Cas. In Arcturus, we could not find any line sensitive to a change in the Nb abundance. However,  it was possible to measure the Nb abundance  in V762~Cas, yielding the solar value, similar to the situation encountered for the other $s$-process elements.

To ensure homogeneity and ease the comparison between the present sample of barium stars and the \citet{Neyskens2015} sample of S stars, Nb and Zr abundances  in  S stars from \citet{Neyskens2015} have been re-determined in the present paper, using the parameters of \citet{Neyskens2015} but for exactly the same lines as for the barium stars. The resulting abundances (see Table~\ref{Tab:NbZr}) are found to be in very close agreement ($\sim0.1$~dex) with those of \citet{Neyskens2015}.
\medskip\\

\noindent{\em Heavy $s$-process elements (Ba, La, Ce, Pr, Nd).}

\noindent  
The Ba abundances listed in Table~\ref{Tab:abundances} are derived from the only unsaturated line at 4524.924~\AA. 

Although there are several useful \ion{La}{II} lines which can be used to measure 
the La abundance,  only those with available hyperfine splitting data were considered. 

To derive the Ce abundances,  \ion{Ce}{II} lines in the wavelength range 4300 -- 6500~\AA\ were used; however, these lines yield  abundances that are systematically higher by about 0.3~dex than those from red \ion{Ce}{II} lines (above 7000~\AA).
This may be due to non-LTE effects, which are difficult to quantify without detailed calculations. 

 Not many Pr lines proved useful, so  we  used  four \ion{Pr}{II} lines  to get the Pr abundance (see
Table~\ref{Tab:linelist}). 
 
 Since most of the violet Nd lines are blended, we  used \ion{Nd}{II} lines in the range 5200 -- 5400~\AA.
For a few objects, we were  also able to use clean, unblended lines at 4797.150, 4947.020, 4961.387, 5089.832, and 5132.328~\AA. All these lines give consistent abundances.

\noindent{\em $r$-process elements (Sm, Eu).}

Good Sm lines are found in the bluer part of the spectrum, and are well fitted by the synthetic spectrum. 

The Eu abundance in our objects were   measured  from six \ion{Eu}{II} lines that yield consistent abundances,  but the final abundances which are listed in Table~\ref{Tab:abundances} are derived from the  lines at 6437.640~\AA\ and  6645.134~\AA,\ which yield quasi-solar abundances in the reference stars Arcturus and V762~Cas.

\subsection{Uncertainties on the abundances}

We  estimated the uncertainties on all the elemental abundances $\log \epsilon$ following Eq.~2 from \citet{Johnson2002}:
\begin{equation}
\label{Eq:Johnson}
\sigma^{2}_{\rm \log \epsilon}=\sigma^{2}_{\rm ran} 
\;+\; \left(\frac{\partial \log\epsilon}{\partial T}\right)^{2}\sigma^{2}_{T}  \;+\; \left(\frac{\partial \log \epsilon}{\partial \log g}\right)^{2}\;\sigma^{2}_{\log g} 
\;+\; \left(\frac{\partial \log \epsilon}{\partial \xi }\right)^{2}\;\sigma^{2}_{\xi}.
\end{equation}
In Eq.~\ref{Eq:Johnson}  $\sigma_{T}$, $\sigma_{\log g}$, and $\sigma_{\xi}$  are
the typical uncertainties on the atmospheric parameters, which are estimated as $\sigma_{T}$ = 50~K, $\sigma_{\log g}$ = 0.2~dex, and $\sigma_{\xi}$ = 0.05~km/s. 
 Since uncertainties on the abundances due to the uncertainties on [Fe/H] are very 
 small  compared to uncertainties arising from other sources, we do not consider the $\sigma_{\rm [Fe/H]}$ term  in our error calculations. 

The partial derivatives appearing in Eq.~\ref{Eq:Johnson} were evaluated in the specific case of HD~43389, a cool barium star with $T_{\rm eff} = 4000$~K, varying the  
atmospheric parameters $T_{\rm eff}$, $\log g$, and microturbulence $\xi$ by $100$~K,  $0.5$~dex, and $0.5$~km/s, respectively. 
The resulting changes in the abundances are presented in Table~\ref{Tab:uncertainties}. Since Eq.~\ref{Eq:Johnson} assumes that the uncertainties due to the different parameters are uncorrelated, the resulting abundance uncertainties are likely overestimated.

Finally, the random error $\sigma_{\rm ran}$ is the line-to-line scatter.
For most of the elements, we were able to use more than four lines to derive the abundances. In that case, we  adopted $\sigma_\mathrm{ran} = \sigma_{s}/N^{1/2}$, where $\sigma_{s}$ is the standard deviation of the abundances derived from all the $N$ lines of the considered element. It involves uncertainties caused by factors like line blending, continuum normalisation, and
oscillator strength. 
However, for elements like Zr and Eu only two or three lines could be used, so that the above method had to be adapted. We first calculated  $\sigma_\mathrm{Zr,avg}$, which is the average standard deviation on Zr abundances as measured on all the stars in our sample.
Then $\sigma_\mathrm{Zr,ran}$ = $\sigma_\mathrm{Zr,avg}$/$N^{1/2}$, where $N$ is the number of considered Zr lines. We used a similar procedure for $\sigma_\mathrm{Eu,ran}$.
Finally the error on  [X/Fe] was calculated from
$\sigma^{2}_{\rm [X/Fe]} =\sigma^{2}_{\rm X} + \sigma^{2}_{\rm Fe}$.

\section{Light-element abundance analysis: C, $^{12}$C/$^{13}$C, N, Na, and Mg }
\label{Sect:light}

In this section we discuss the physical conditions required for the production of the light elements C, N, F, Na, and Mg.

\begin{figure}
   \includegraphics[width=9cm]{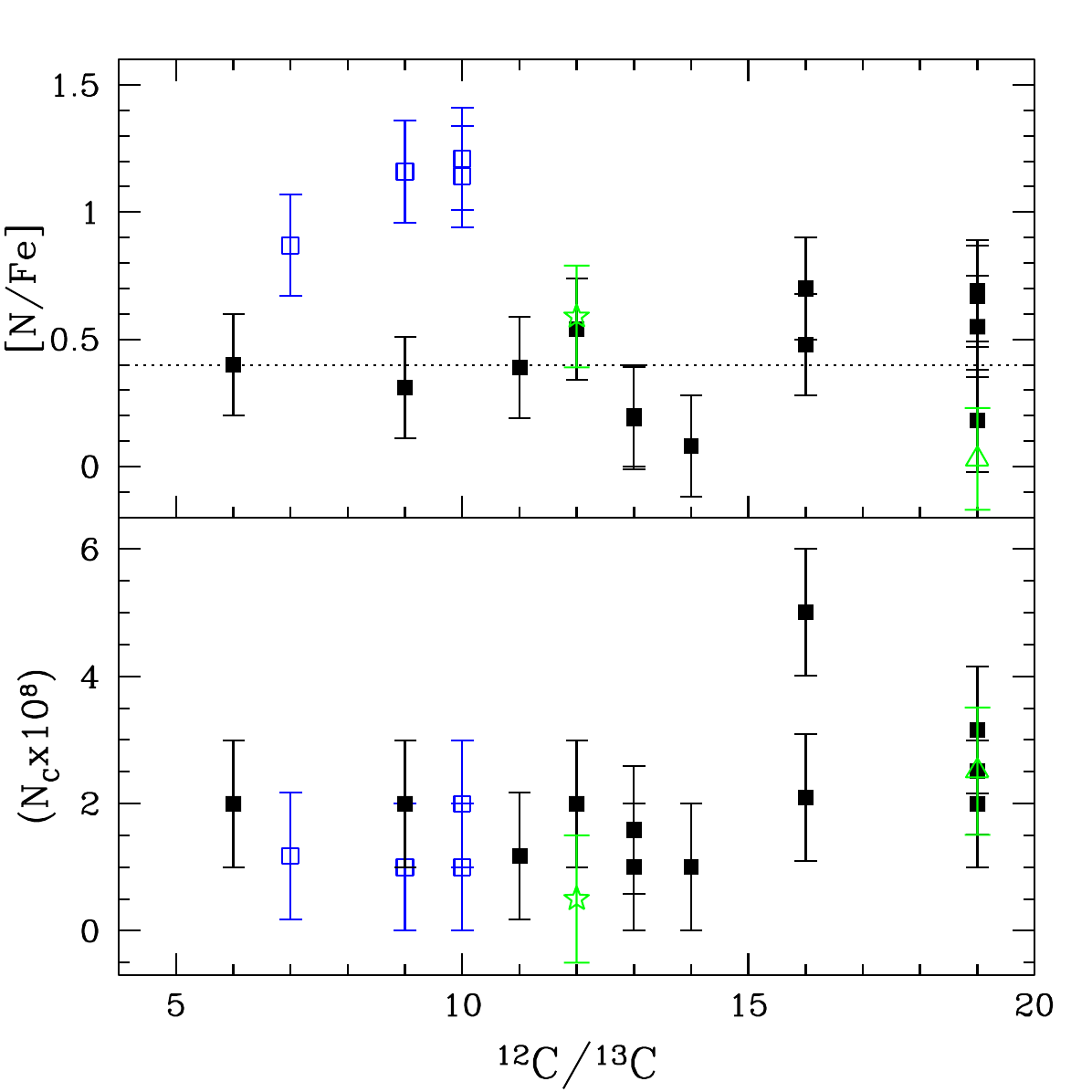}
\caption{{\em Top panel:}  [N/Fe]  as a function of  $^{12}$C/$^{13}$C, represented for our programme stars with full and open blue squares; open blue squares  specifically identify the N-enriched objects (HD~43389, HD~60197, HD~100503, and to a lesser extent HD~121447). Arcturus (green star) and V762~Cas (green triangle) are also shown for comparison. 
The horizontal dotted line shows the abundance ratio expected in red giant stars after the mixing processes occurring along the giant branch \citep{Eggleton2008}. {\em Bottom panel:} Same as the top panel for the C abundance (by number, in the scale where $\log \epsilon_H = 12$). 
              }
         \label{Fig:N}
   \end{figure}

Except for HD~120620 (with $^{12}$C/$^{13}$C~=~90), all barium stars have $^{12}$C/$^{13}$C values lower than 20 (Fig.~\protect\ref{Fig:N}), a feature already noted by \citet{Tomkin1979}, \citet{Sneden1981}, \citet{Smith1984}, \citet{Harris1985}, and \citet{Barbuy1992}. This is indeed expected if matter processed by the CN cycle and brought to the surface by the first dredge-up dominates over the 
$^{12}$C pollution from the AGB companion.
In most of the programme stars, we find a nitrogen overabundance [N/Fe]~$ \sim 0.4$~dex (top panel of Fig.~\ref{Fig:N}), in agreement with the target stars being subject to efficient mixing \citep{Lambert1981}.
The situation observed in barium stars (namely $^{12}$C/$^{13}$C$< 20$ and 
[N/Fe]~$ \sim 0.4$~dex)  matches the predictions of models combining the first dredge-up on the red giant branch with the subsequent thermohaline mixing occurring at the  `bump' \citep{Eggleton2008}.

In some barium stars, however (HD~43389, HD~60197, HD~100503, and to a lesser extent HD~121447), N overabundances as high as
$\sim$1.2~dex are measured.
Nitrogen enrichments in barium stars have already been  reported by \citet{Barbuy1992}, \citet{Allen2006}, and \citet{Merle2016}, although generally not to such extreme levels. 
This situation may still be accounted for by the previous models, provided the barium star was formerly enriched  by $^{14}$N from a massive AGB companion experiencing
hot-bottom burning \citep[HBB; e.g.][and references therein]{Boothroyd1995} where hydrogen burns at the bottom of the convective envelope.


   \begin{figure}
\includegraphics[width=9cm]{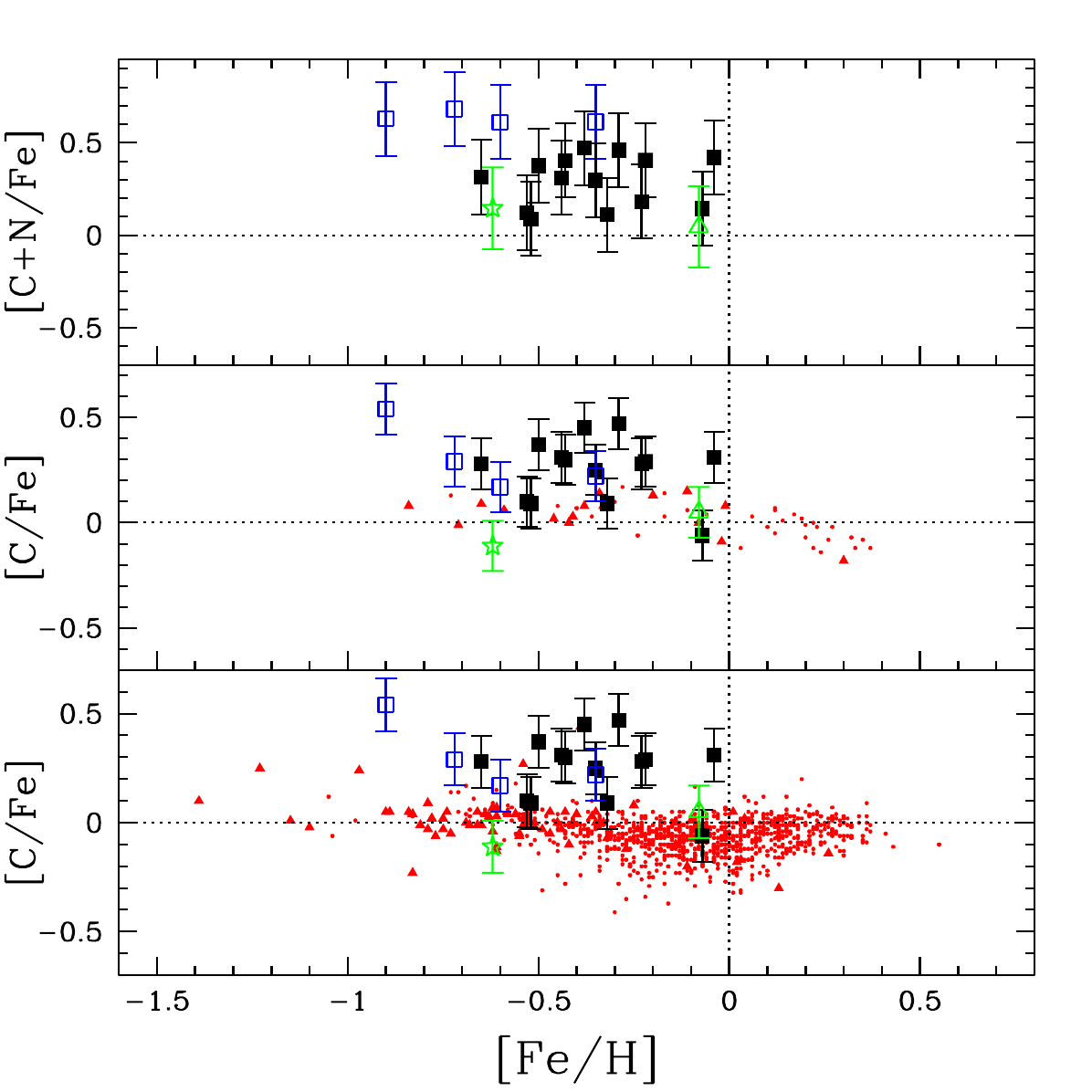}

      \caption{Abundances of  C ({\em bottom and middle panels}) and C+N ({\em top panel})  in our programme stars, as a function of metallicity (symbols  defined  in Fig.~\ref{Fig:N}). The C abundances of thick disc (red triangles) and thin disc (red dots) stars from \citet{Bensby2006} ({\em middle panel}) and  
      \citet{suarez2017} ({\em bottom panel})
      are also shown to outline the Galactic trend.
              }
         \label{Fig:C+N}
   \end{figure}

   \begin{figure}
\includegraphics[width=9cm]{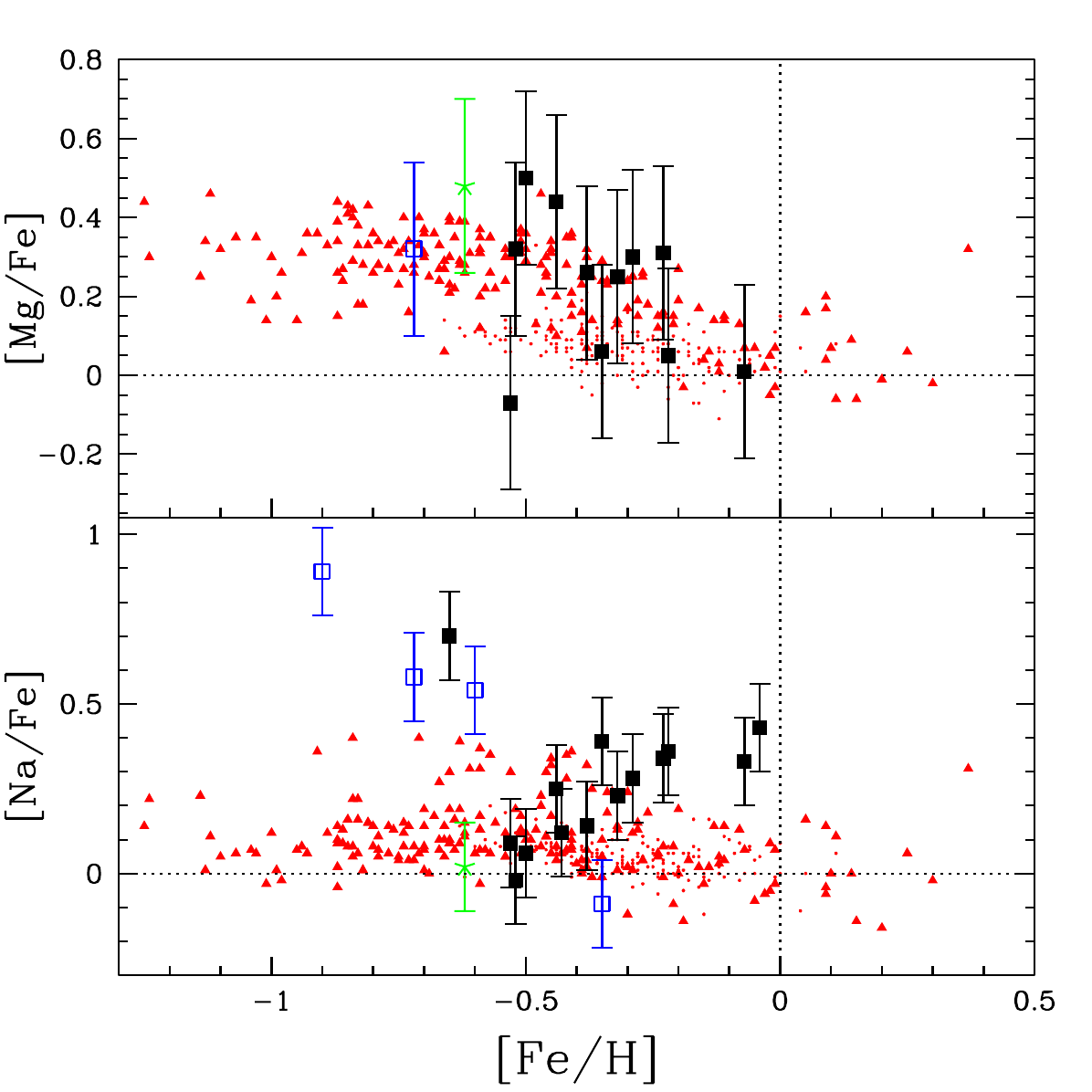}
\caption{Abundances of  Na and Mg  of our programme stars (symbols  defined  in Fig.~\ref{Fig:N}) along with Arcturus (green star), compared with thin disc stars \citep[red dots;][]{bensby2005,reddy2003}. Thick disc stars \citep[red triangles;][]{bensby2005,reddy2006} are also shown for comparison.
\label{Fig:NaMg_Galactic}     }
   \end{figure}
      
   \begin{figure}
\includegraphics[width=9cm]{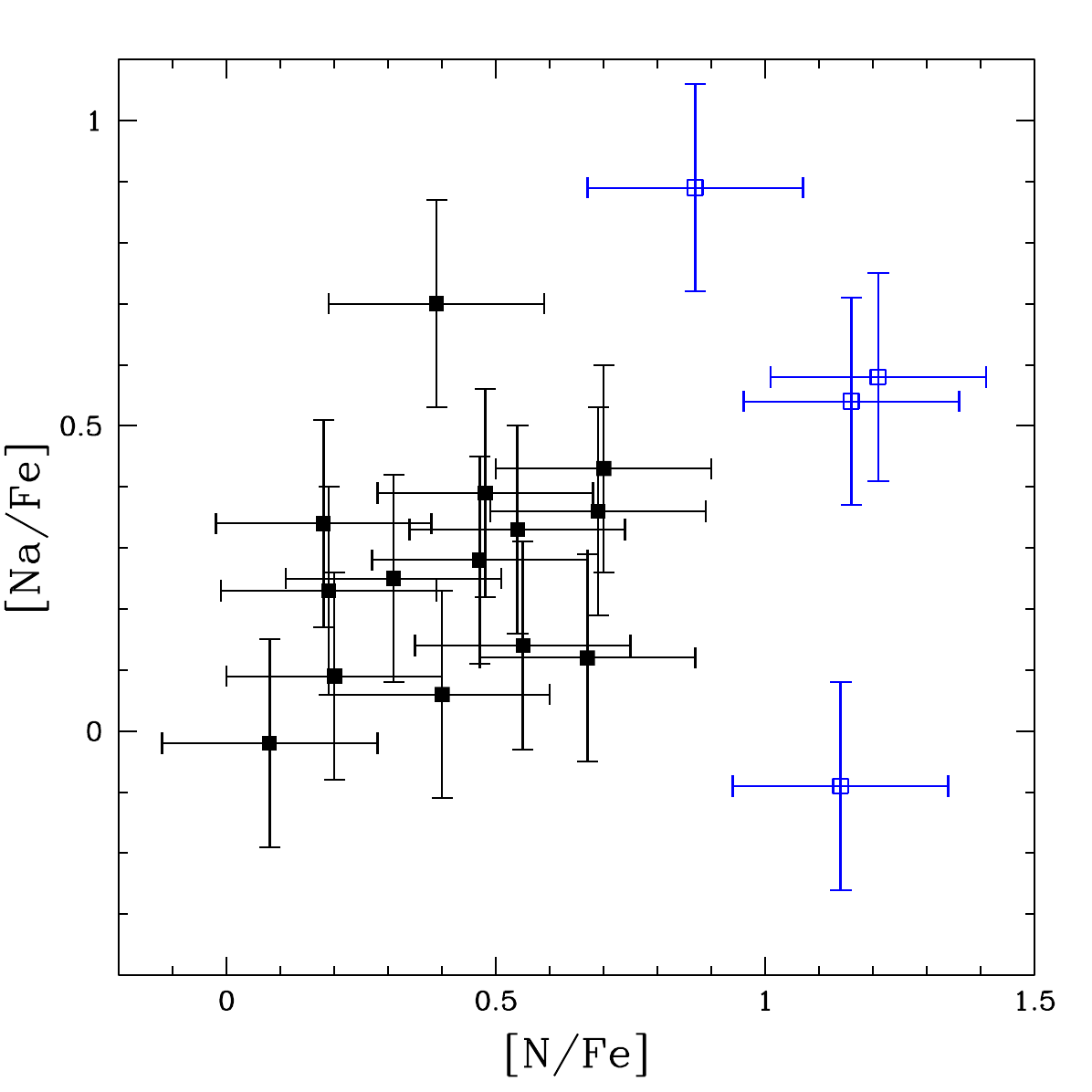}
      \caption{Barium stars in the ([Na/Fe], [N/Fe]) plane. The four N-rich objects HD~43389, HD~60197, HD~100503, and HD~121447 are shown as blue open squares. 
      }
         \label{Fig:N+Na}
   \end{figure}

A further confirmation of this $^{14}$N pollution is given in the top panel of Fig.~\ref{Fig:C+N}, which reveals that the four N-enriched objects have distinct C abundances (see bottom panel of Fig.~\ref{Fig:C+N}) but similar C+N abundances,
 revealing that CN-processing might indeed have operated. Interestingly, the C+N abundances in these four peculiar objects are globally higher than those of the bulk of barium stars (top panel of Fig.~\ref{Fig:C+N}).

The bottom panels of Fig.~\ref{Fig:C+N} compare the C abundance in barium stars with the trend present in the Galactic discs derived by \citet{Bensby2006} and \citet{suarez2017}. 
Both Galactic determinations of [C/Fe] as a function of [Fe/H] yield a broadly consistent picture. We note, however, that \citet{reddy2006} find a significant increase in [C/Fe] with decreasing [Fe/H], contrarily to \citet{Bensby2006} and \citet{suarez2017}. The majority of barium stars in our sample clearly exhibit carbon overabundances with respect to the Galactic trends.
This is consistent with stellar evolution considerations since the matter accreted by the barium star was processed by the third dredge-up in the donor star.
 



The partial mixing of protons inside the C-rich layers of the H-He intershell region of AGB stars can be responsible not only for an efficient $s$-process nucleosynthesis but also for a significant enrichment in $^{19}$F  \citep{goriely2000}. 
Fluorine abundances have been determined in only three barium stars
so far
\citep[HD~121447, HD~123396, and HD~178717,  the first and the last belong to our sample;][]{Jorissen1992,Alves-Brito2011};
they all reveal  fluorine enrichments (at the level $\rm 0.4 \le [F/O] \le 0.7$). Even though the very high F abundances found in carbon stars in the early work of \citet{Jorissen1992} were later revised downwards by \citet{Abia2009,Abia2010,Abia2015} (due to blends of CN and C$_2$ lines with the HF lines), these stars still remain F-rich \citep[see Fig.~3 from][]{Pilachowski2015}.

A sodium enrichment may also be expected in AGB stars (and thus in barium stars) through the NeNa chain of proton captures 
\citep{goriely2000}.
Such a Na enrichment in barium stars has already been noted by \citet{DeCastro2016}, and is observed in some of our barium stars as well (Fig.~\ref{Fig:NaMg_Galactic}). It is  explained by our models (Figs.~\ref{Fig:N+Na+Mg} and ~\ref{Fig:N+Na+Mg_3M}) except for HD~121447 whose interpretation is quite challenging (see discussion below). 

The four lowest metallicity ([Fe/H] $\le-0.6$) barium stars in our sample (HD~16458, HD~60197,  HD~100503,  and HD~121447) are all very Na-enriched ([Na/Fe]~$>0.5$), as shown in Figs.~\ref{Fig:NaMg_Galactic} and \ref{Fig:N+Na}.
The last three  
are also enriched in nitrogen, along with HD~43389 (of higher metallicity: [Fe/H]=-0.35). 
Non-LTE effects are known to affect the Na lines;
the corrections (to be applied on LTE abundances) are smaller than $\Delta_{NLTE}= -0.15$ dex considering the barium star parameter range \citep[][and the INSPECT project]{Lind-2011}. This correction would not be sufficient to reconcile the most-enriched Na abundances in barium stars with those of disc stars.

Interestingly, at the very high temperatures achieved at the base of the convective envelope of super-AGB stars ($T_\mathrm{env} > 10^8$~K), N  can be substantially produced by HBB \citep{doherty2014}.
These four N-enriched objects  show variations in the C abundance from star to star 
(see bottom panel of Fig.~\ref{Fig:C+N}), but similar C+N abundances (top panel of Fig.~\ref{Fig:C+N}), revealing that CN processing has probably occurred.

Above [Fe/H]~$\sim -0.6$, the predictions of \cite{doherty2014} for super-AGB stars agree with the observed decrease in Na abundance with metallicity,  but they remain at odds with the trend shown by the four most metal-poor stars in our sample (with enhanced Na). As noted by \citet{Doherty2014b}, a very high mass-loss rate during the AGB phase would minimise the $^{23}$Na destruction [through $^{23}$Na(p,$\alpha$)$^{20}$Ne and $^{23}$Na(p,$\gamma$)$^{24}$Mg], thus preserving a relatively high Na abundance. Interestingly enough, we note that in the present sample, the star with the highest [Na/Fe] abundance is HD~121447, which is also the barium star with the third shortest orbital period (185.7~d). If the orbital period was as short at the end of the TP-AGB phase of the donor star, this star could well have suffered from a strongly enhanced mass transfer, as proposed for example by \cite{Tout1988}, thus shortening the time spent on the AGB. Combined with the \citet{Doherty2014b} argument above, this could possibly account for the high Na abundance observed in this star.

Finally, if the $s$-process nucleosynthesis is taking place in hot thermal pulses, Mg is expected to be overproduced as a consequence of $\alpha$-captures by $^{22}$Ne. This signature is not observed in our barium stars since they show a level of Mg enrichment compatible with the Galactic trend 
(Fig.~\ref{Fig:NaMg_Galactic}), which further discards 
$^{22}$Ne($\alpha$,n)$^{25}$Mg as the main neutron source for the $s$-process. 
\citet{DeCastro2016} report measurements in barium stars of [Mg/Fe] (and more generally of [$\alpha$/Fe])  higher than local field giant stars;
these barium stars could be transition objects between 
the thin and thick disc, or be thick disc stars.

To conclude, the most salient feature emerging from the present study of light  elements in barium stars is the suggestion offered by the N and Na overabundances that some may have been polluted by rather massive AGB stars undergoing HBB. However, we  see  in Sect.~\ref{Sect:pattern} that this conclusion is not supported by the $s$-process abundance pattern, which forbids AGB stars more massive than 3~M$_\odot$ from being at the origin of the pollution of barium stars.

In Sect.~\ref{Sect:heavylight} we  connect
 the light-element abundances discussed so far with abundances of $s$-process elements and compare the results with stellar model and nucleosynthesis predictions that are described in the next section.

\begin{figure*}
 \includegraphics[clip, trim=3.5cm 0cm 0cm 0cm,height=19cm, width=22cm]{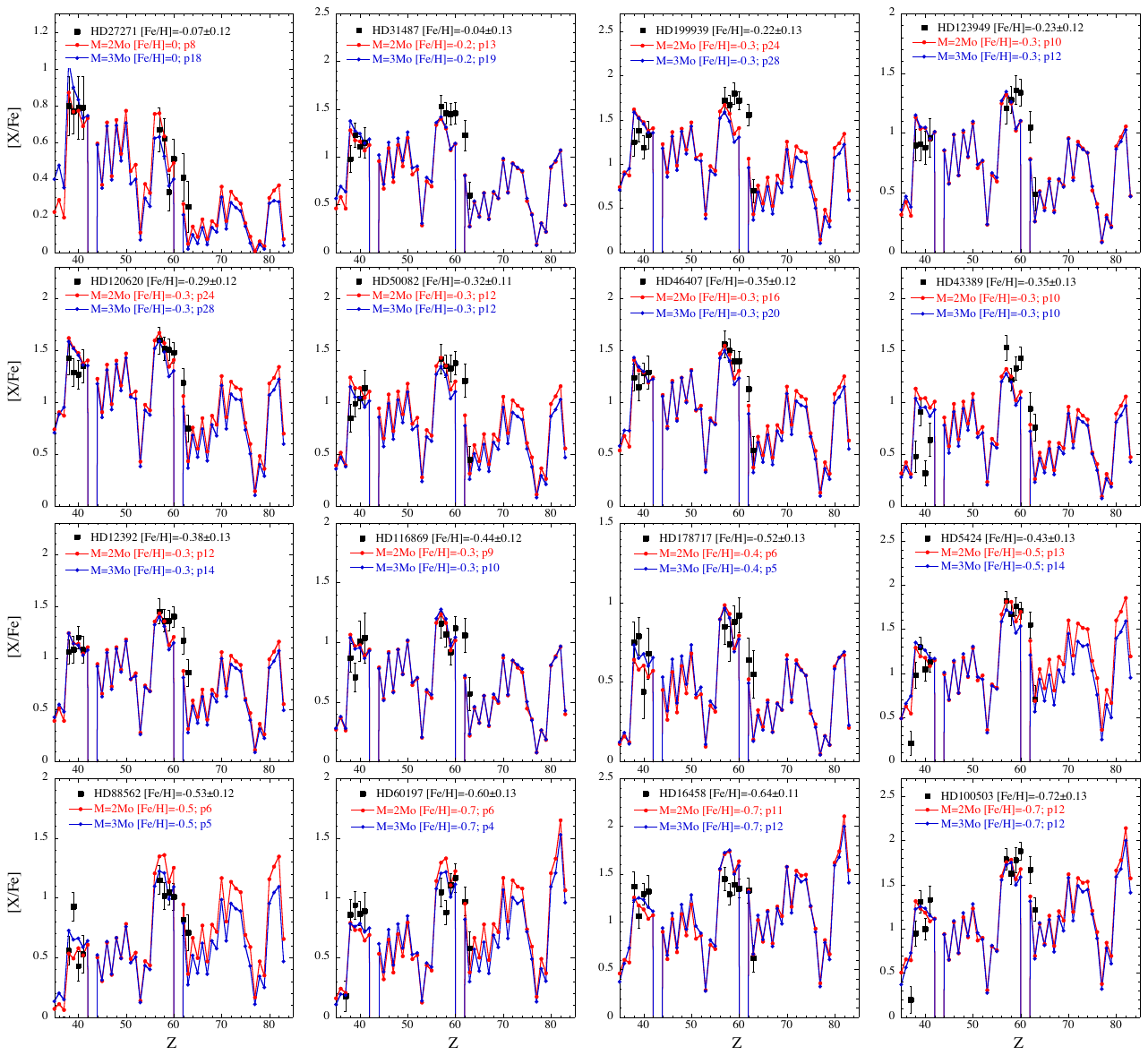}
      \caption{Observed $s$-process abundance patterns (black squares with error bars) compared with the predictions for 2~\Msun\  (red line) and 3~\Msun\   (blue line) models with the diffusive mixing scheme. The pulse number is quoted in the label as `p'. The stars are ordered by decreasing model metallicity.
              }
         \label{Fig:abundances}
\end{figure*}

\section{STAREVOL models}
\label{Sect:STAREVOL}

Asymptotic giant branch  models have been computed with the STAREVOL code \citep{Siess2000,Siess2006} using an extended $s$-process reaction network of 411 species and the same input physics as described in \cite{Goriely2017}. 
In our computations, we use the standard mixing length theory with $\alpha = 1.75$. We take into account the change in  opacity due to the formation of molecules when the star becomes carbon rich, as prescribed by \cite{Marigo2002}. The reference solar composition is given by \cite{Asplund2009} which  corresponds to a metallicity $Z = 0.0134$. 
 For the lower metallicity models, we assume a solar-scaled composition without $\alpha$-element enhancement.
For the mass-loss rate, we consider the \cite{Reimers1975} prescription with $\eta_R = 0.4$ from the main sequence up to the beginning of the AGB and then switch to the \cite{Vassiliadis1993} rate.   Dedicated models have been computed for  [Fe/H] between -1 and 0, with steps of 0.1,  for masses of 2 and 3~\Msun.

In the present simulations, a diffusion equation is used to simulate the partial mixing of protons in the C-rich layers at the time of the third dredge-up. 
Following the formalism of Eq.~9 of \citet{Goriely2017},
the diffusive mixing (DM) parameters adopted in our calculations are $D_{\rm min}=10^9~{\rm cm^2/s}$ and $p=5$,
where $D_{\rm min}$ is the value of the diffusion coefficient at the innermost boundary of the diffusive region
and $p$ is an additional free parameter defining the slope of the exponential decrease in 
the overshoot diffusion coefficient with depth.
This formalism is actually used for 2 and 3~\Msun\  stars only.
In 4 and 5~\Msun~stars, the high temperature at the base of the convective envelope requires a coupling of mixing and burning, which is not yet included in our simulations. Nevertheless, it was shown in \citet{goriely-siess-2004} that 
when HBB is effective, the protons are burnt on-the-fly leading to an overlap of $^{13}$C with (neutron poison) $^{14}$N, resulting in a very inefficient $s$-process nucleosynthesis. We note, however, that an $s$-production by $^{22}$Ne can still occur in such stars.


\section{Heavy-element abundance analysis} 
\label{Sect:heavy}

\subsection{$s$-process abundance pattern}
\label{Sect:pattern}

Comparison between the observed and predicted $s$-process abundance patterns for various stellar masses is shown in Figs.~\ref{Fig:abundances} and \ref{Fig:abundances121447}. 
The pulse number corresponding to the model that best reproduces the overall surface enrichment is indicated for each observed star.
A good agreement is systematically obtained with the predictions corresponding to  2~\Msun{} (red line) or 3~\Msun{}  (blue line)  stars, with metallicities matching the measured values. 
This initial mass of the polluting AGB stars has to be higher than the mass of the current Ba star (because the polluting star evolved first), but this effect might be softened because the current barium star accreted mass (up to a few tenths of~\Msun{} from the simulations). In any case, the masses of the current barium stars  derived from their position in the Hertzsprung--Russell (HR) diagram ($M=1.0 - 2.7$ \Msun, Table~\ref{Tab:masses} and Sect.~\ref{Sect:Masses}) is compatible with a 2--3~\Msun{} companion.

Abundances from more massive AGB stars (4--5 \Msun) have systematically lower [hs/ls] ratios than those displayed in Fig.~\ref{Fig:abundances}, where ls and hs denote the average abundance of $s$-process elements belonging to first peak (Sr, Y, Zr) and second peak (Ba, La, Ce),  respectively. This results from the fact that the $s$-process is taking place in the convective pulse rather in a radiative layer \citep[see e.g.][]{Goriely2017}.

\begin{figure}
 \includegraphics[clip, trim=1cm 10cm 1cm 1cm,height=9cm, width=9cm]{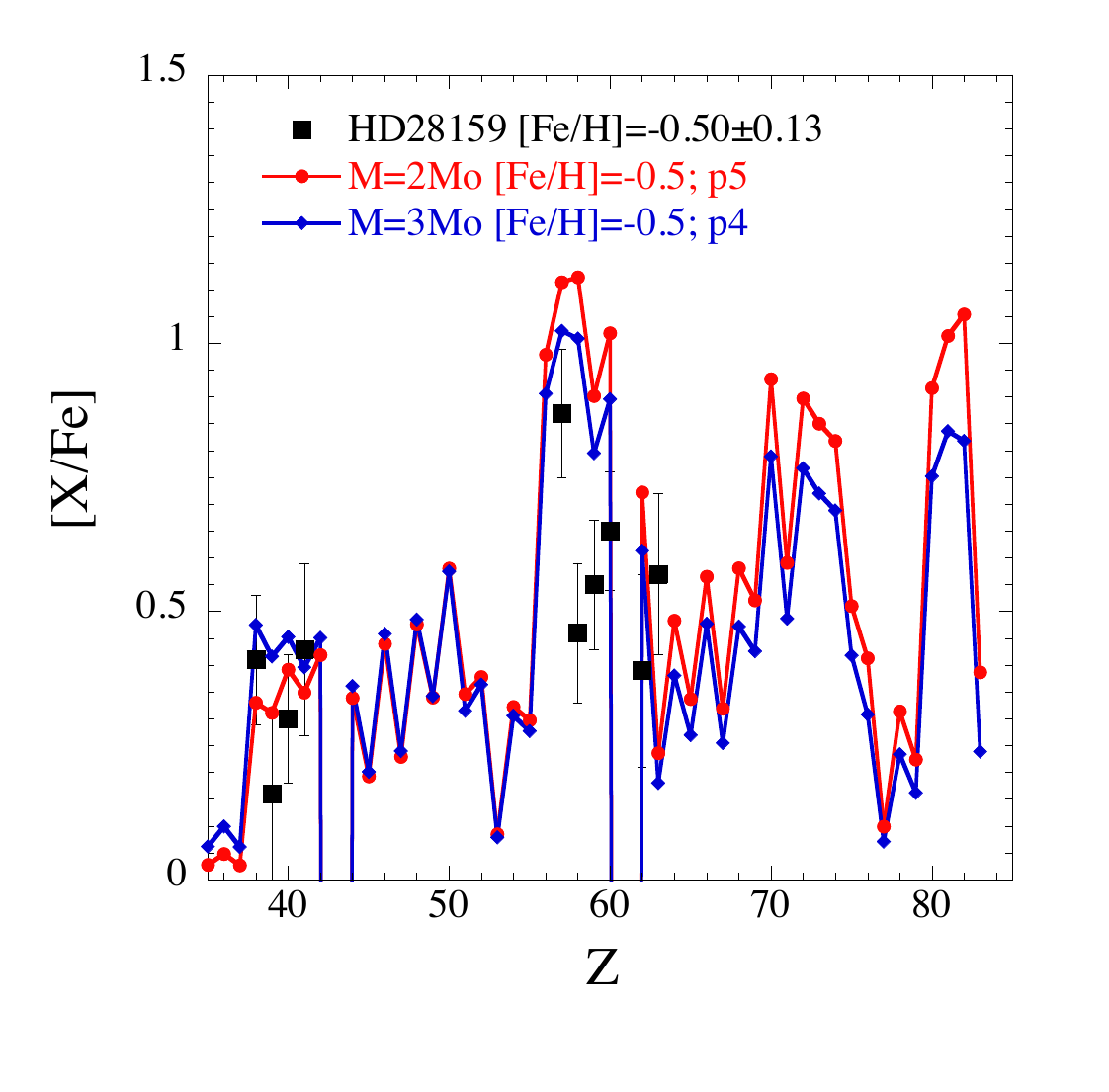}
      \caption{Same as Fig.~\ref{Fig:abundances}, but for HD~28159. Our analysis reveals that this object, initially classified as a normal giant (K5III), is slightly $s$-process enriched. The agreement with the predictions for a 2~\Msun\ (red line) and 3~\Msun\ (blue line) model with the diffusive mixing scheme is not satisfactory. 
              }
         \label{Fig:abundances28159}
\end{figure}
\begin{figure}
 \includegraphics[clip, trim=1cm 10cm 1cm 2cm,height=9cm, width=9cm]{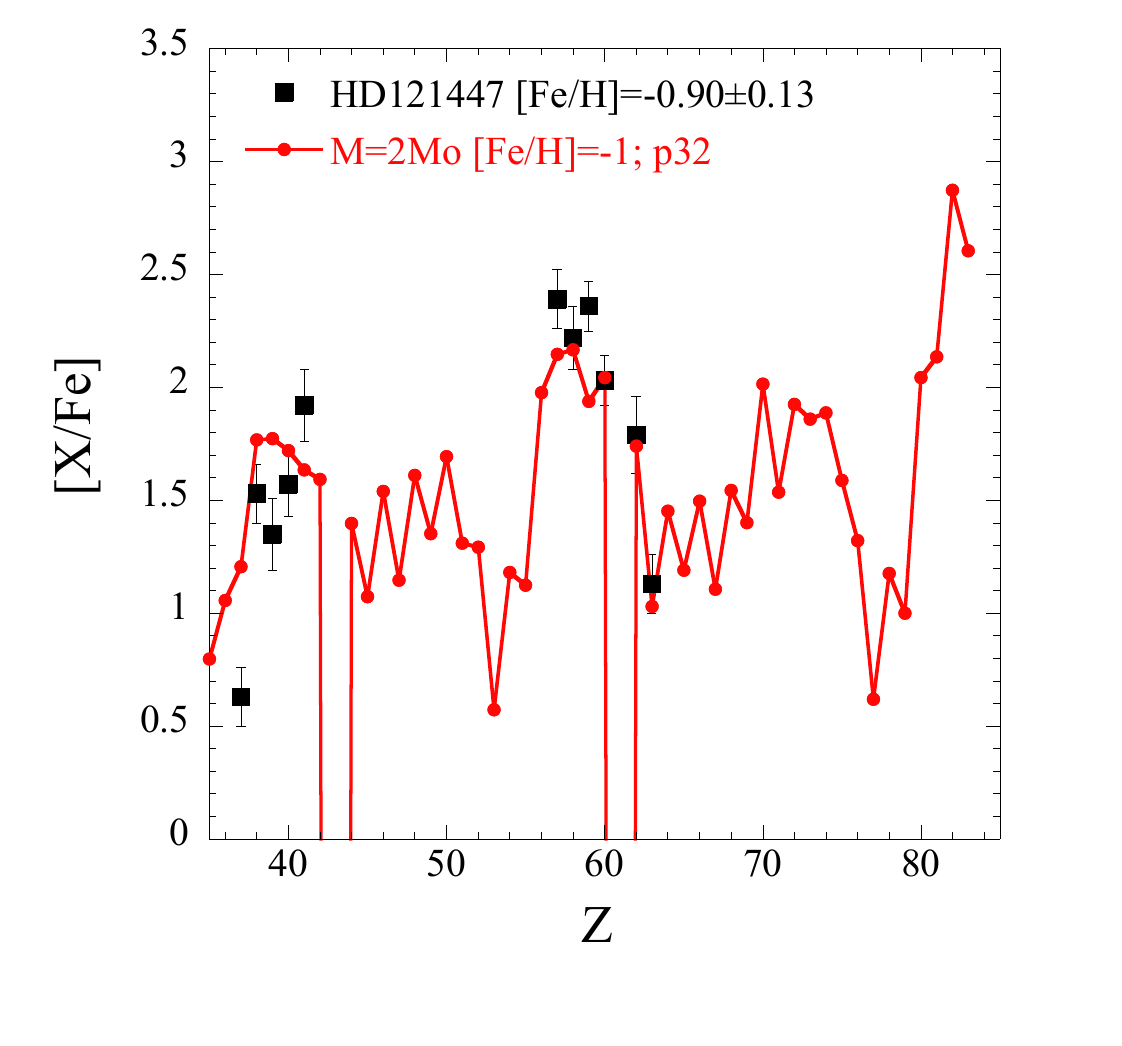}
      \caption{Same as Fig.~\ref{Fig:abundances}, but for HD~121447, the most extreme barium star in our sample, compared with the predictions for a 2~\Msun\ (red line) model with the diffusive mixing scheme.
              }
         \label{Fig:abundances121447}
\end{figure}

Two stars deserve a special discussion. The first, 
HD~28159, is the least $s$-process enriched star in our sample. As already mentioned in Sect.~\ref{Sect:observations}, this K5III giant was initially selected to serve as a solar-scaled composition reference object. 
Its abundance distribution (Fig.~\ref{Fig:abundances28159}) reveals that it is slightly enriched in La ([La/Fe] = $0.87 \pm 0.12$) and to a lesser extent in Zr, Ce, Pr, Nd, Sm, and Eu. 
The La abundance can hardly be reproduced, even considering its error bar, with the 3~\Msun~ model predictions, which moreover overestimates by $\sim 0.4$ dex the abundances of Ce, Pr, and Nd (Fig.~\ref{Fig:abundances28159}).
To reconcile such a modest [hs/ls] ratio with the metallicity of the star ([Fe/H]$=-0.5$) is difficult.
Interestingly enough, this object is probably a binary. 
This conclusion is not drawn from the radial velocity of -6.5 ($\pm 3$) km/s quoted (without associated date) by \citet{Wilson1953}, but rather from the two available HERMES spectra, which lead to the following radial velocities:
$-7.745 \pm 0.007$ km/s on JD $2\,455\,508.70$; 
 $-8.177 \pm 0.014$ km/s on JD $2\,456\,619.57$.
This object is thus a newly discovered $s$-process enriched, probably extrinsic star. Its hs abundance profile is difficult to reconcile with that of the other extrinsic stars in our sample and with the model predictions, but  its Zr and Nb abundances perfectly follow the trend expected for extrinsic stars (see discussion in Sect~\ref{Sect:thermometer} and Fig.~\ref{Fig:NbZr}).

The second star, HD~121447, on the contrary, is the most metal$-$poor ([Fe/H]~$=-0.9$) and the most $s$-process enriched barium star in our sample. It is also the coolest known barium star with the third shortest orbital period ($P=185.7$~d). It
shows several peculiarities including ellipsoidal photometric variations and a rotation rate not synchronised with the orbital motion, though orbital circularisation has been achieved
\citep{Jorissen1995}. Scattering on some obscuring material, possibly trapped around the WD companion, could also potentially account for the photometric variability.
From the analysis of its light and velocity curves, \citet{Jorissen1995} estimate a mass of 1.5 -- 1.7~\Msun\, 
consistent with the range we derive in Sect.~\ref{Sect:Masses} from
its location in the HR diagram.
The abundance distribution (probing the initial mass of the primary, i.e. of the current white dwarf) is  reproduced well by a 2~\Msun~model, as shown in Fig.~\ref{Fig:abundances121447}, 
and is consistent with the primary being initially more massive than the secondary, and evolving more quickly.

\subsection{Trends between light and heavy elements}
\label{Sect:heavylight}

\begin{figure}
\includegraphics[width=9cm]{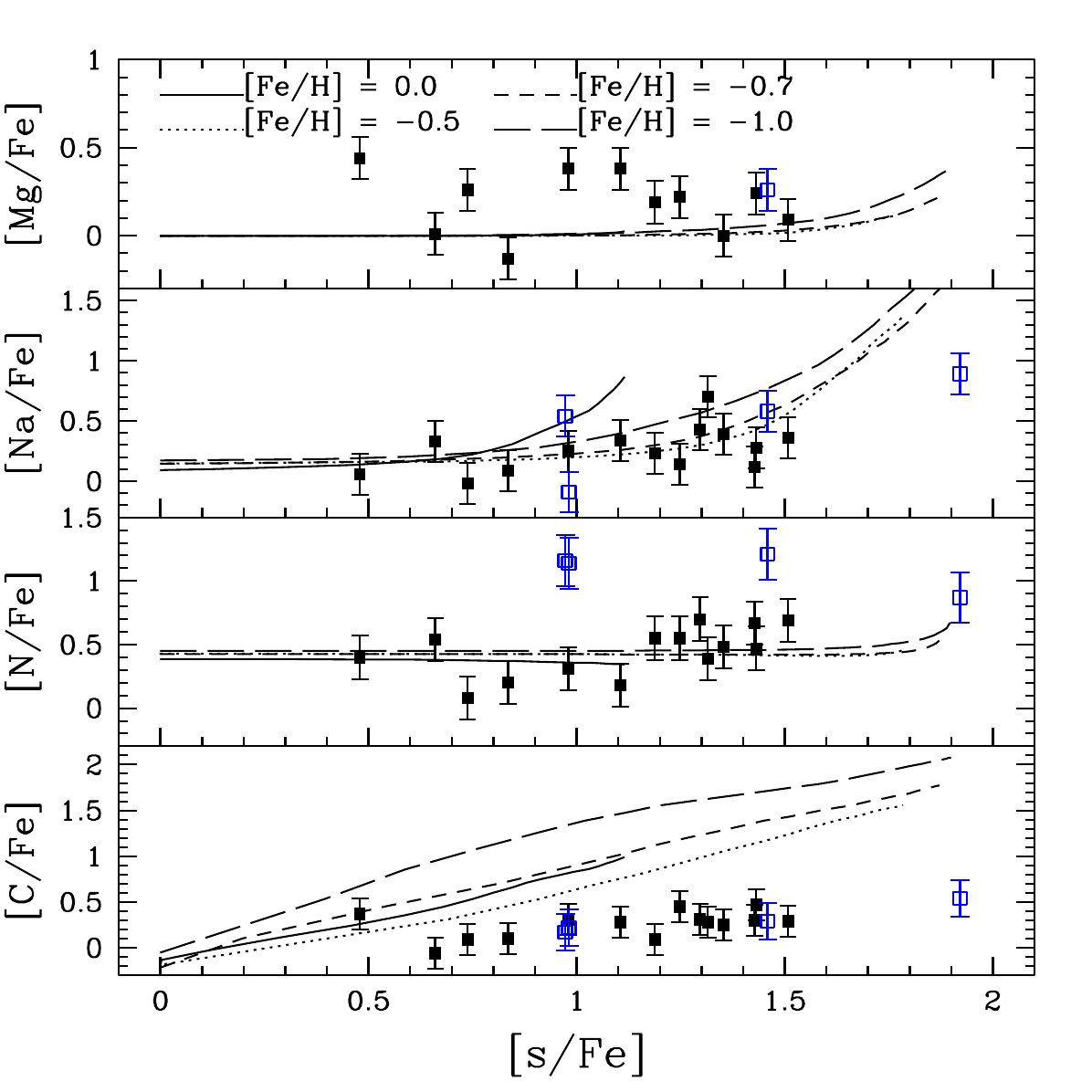}
      \caption{Abundances of  C, N, Na, and Mg (from bottom to top)   in our programme stars are shown as a function of the $s$-process abundances [s/Fe], where [s/Fe] = ([Sr/Fe] + [Y/Fe] + [Zr/Fe] + [Nb/Fe] + [La/Fe] + [Ce/Fe] + [Pr/Fe] + [Nd/Fe])/8. Predicted values from  2~\Msun\  models for different metallicities  are shown for comparison (symbols  defined  in Fig.~\ref{Fig:N}).
              }
         \label{Fig:N+Na+Mg}
   \end{figure}

\begin{figure}
\includegraphics[width=9cm]{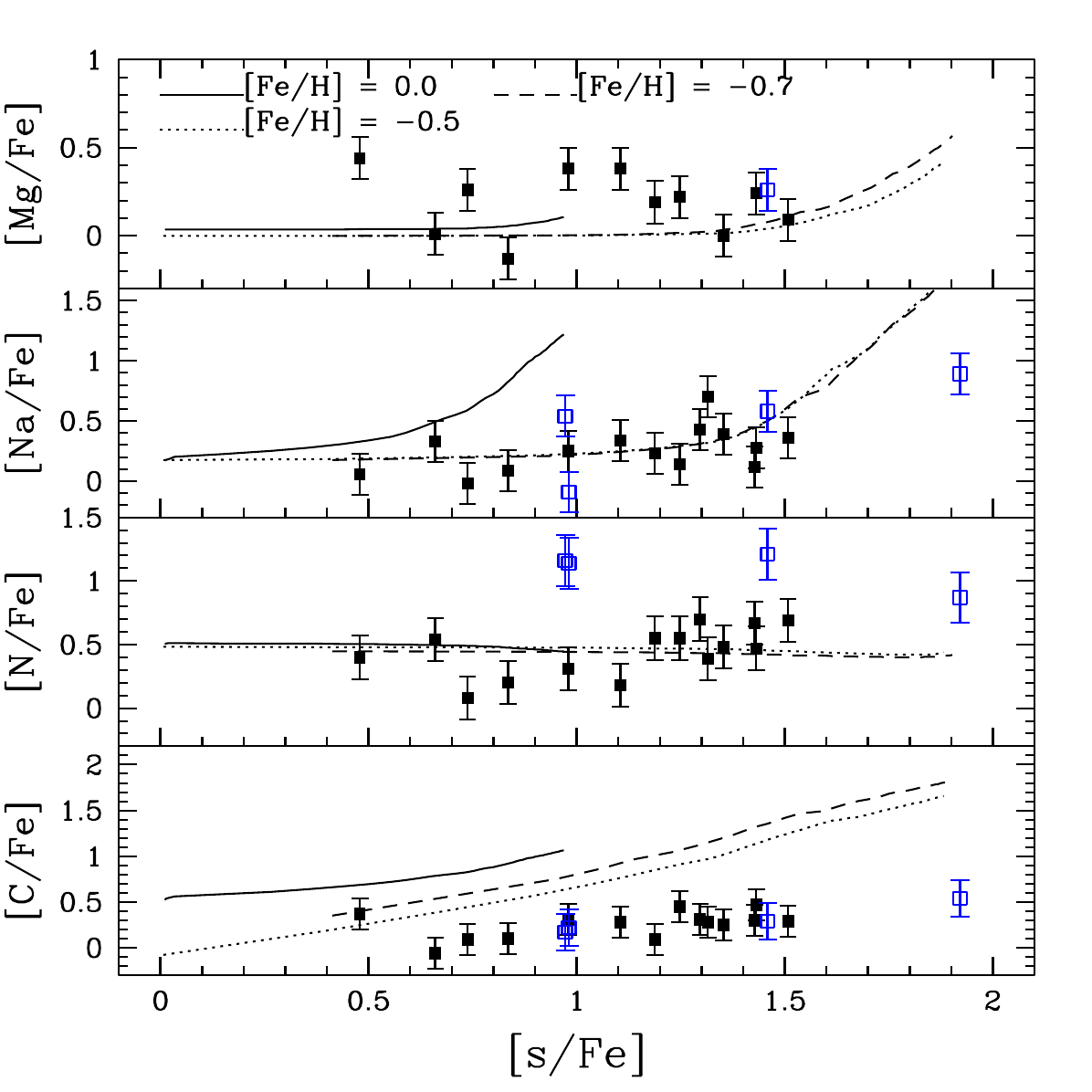}
      \caption{Same as Fig.~\ref{Fig:N+Na+Mg}, but for 3~\Msun\ models.
              }
         \label{Fig:N+Na+Mg_3M}
   \end{figure}

Figures~\ref{Fig:N+Na+Mg} and \ref{Fig:N+Na+Mg_3M}  compare the measured average $s$-process abundance [s/Fe] and the light elements C, N, Na, and Mg with the predictions from 2 and 3~\Msun\ models of various metallicities. 

The predicted carbon abundances are generally too high with respect to the observed ones. 
This may simply be due to the fact that dilution of the AGB matter in the barium star envelope has not been considered in Figs.~\ref{Fig:N+Na+Mg} and \ref{Fig:N+Na+Mg_3M}, and that barium stars are post-first dredge-up. Both processes will bring  the predicted carbon abundance down,  in better agreement with the observed value. 

\begin{figure}
   \includegraphics[width=9cm]{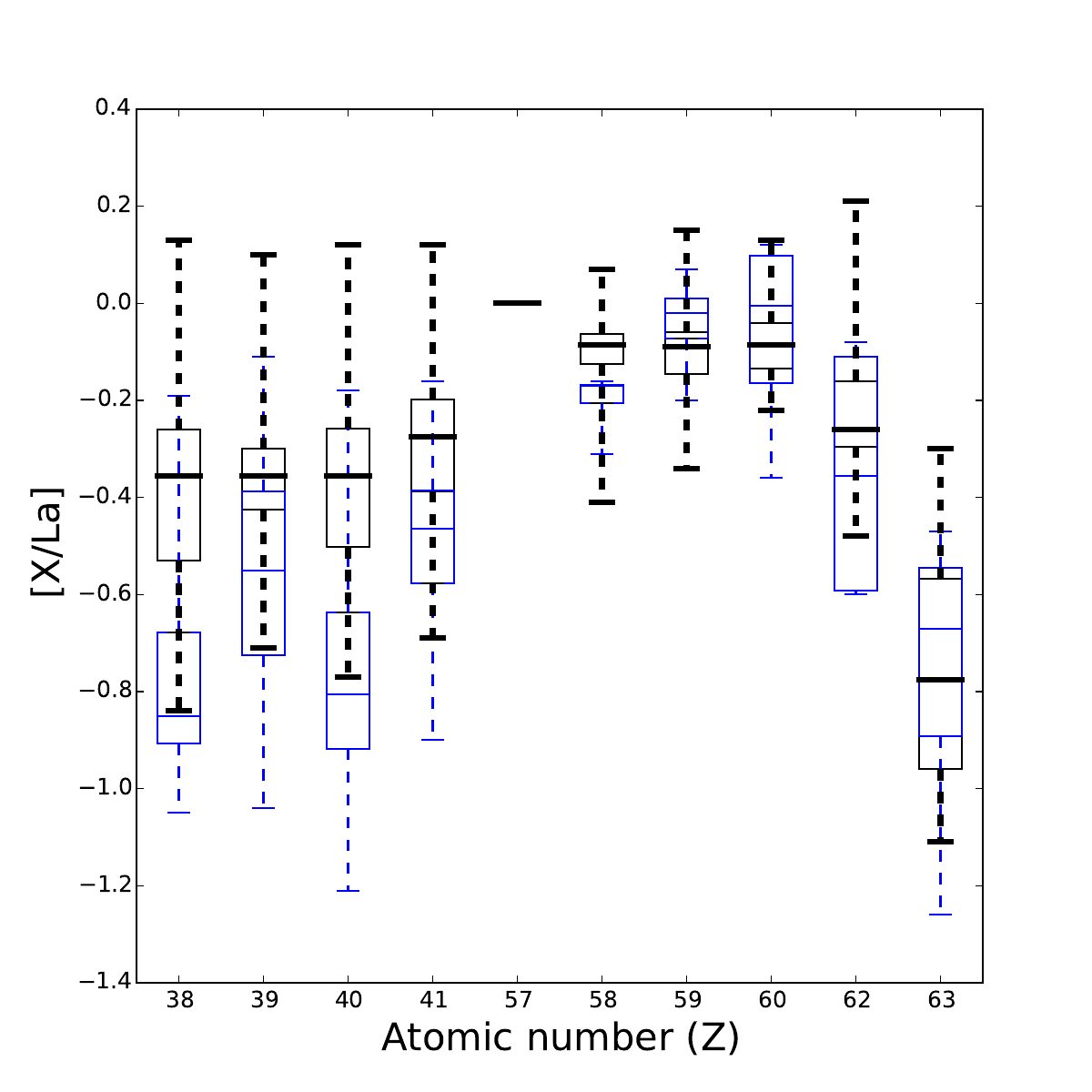}
      \caption{Heavy-element abundances in the programme stars normalised to their La abundance ($Z = 57$). The blue boxes refer to the four N-rich stars, whereas black boxes depict the range covered by all the other stars. The bottom and top of the boxes refer to the  25th and 75th percentiles in the considered elemental abundance. The light  and heavy  $s$-elements (ls and hs) are delineated by the atomic number $Z=57$.  
              }
         \label{Fig:X/La}
   \end{figure}

The above arguments, however, cannot explain the highest measured N overabundances. Figure~\ref{Fig:X/La} reveals that all N-rich stars are characterised by  high [hs/ls] ratios pointing toward an efficient $s$-process. Therefore, the N enrichment cannot be attributed to hot-bottom burning occurring in massive AGB stars, since as discussed in Sects.~\ref{Sect:pattern} and \ref{Sect:Rb}, they are characterised by low [hs/ls] ratios and high Rb abundances. Similarly, large envelope pollutions in carbon and nitrogen are expected from low-mass metal-poor AGB stars  experiencing H-ingestion events \citep{Campbell-2008}, but the required metallicity is too low and incompatible with those measured in the N-rich barium stars, whose origin thus remains a mystery.

Finally, the observed  Na abundances are well reproduced globally, except for the star most enriched in Na. Again, the dilution factor has not been considered.

The Mg abundances are not well reproduced by the predictions likely because the corresponding models did not start from initial abundances matching the Galactic trend displayed in  Fig.~\ref{Fig:NaMg_Galactic}.

\begin{figure}
\includegraphics[width=9cm]{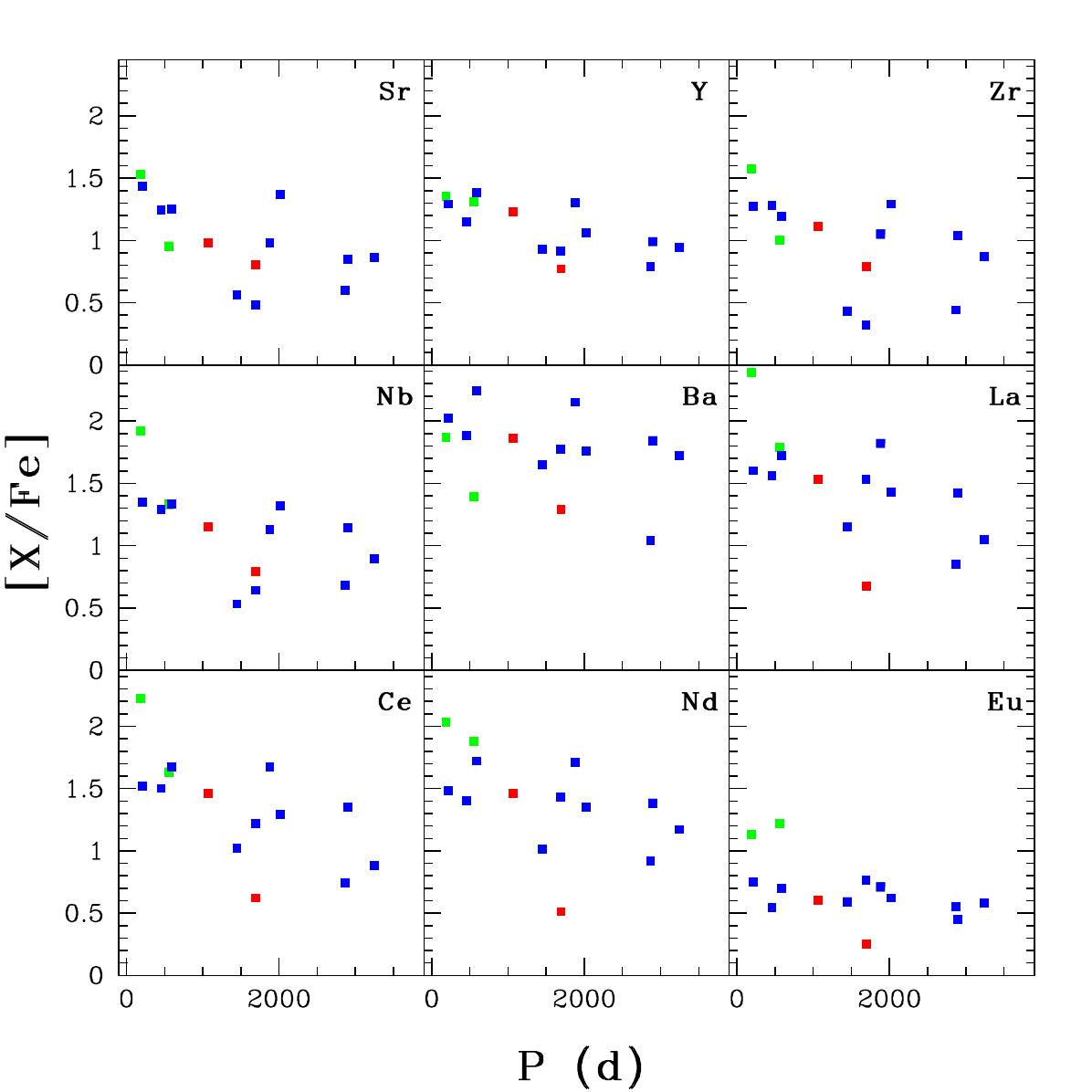}
\caption{Abundance of heavy elements (as labelled) as a function of the orbital period. The metallicity is coded as follows: green: [Fe/H] $\le -0.65$; blue: $-0.65 <$ [Fe/H] $\le -0.1$; red: [Fe/H] $> -0.1$. 
\label{Fig:allabundP}   }
\end{figure}
   
\begin{figure}
\includegraphics[width=9cm]{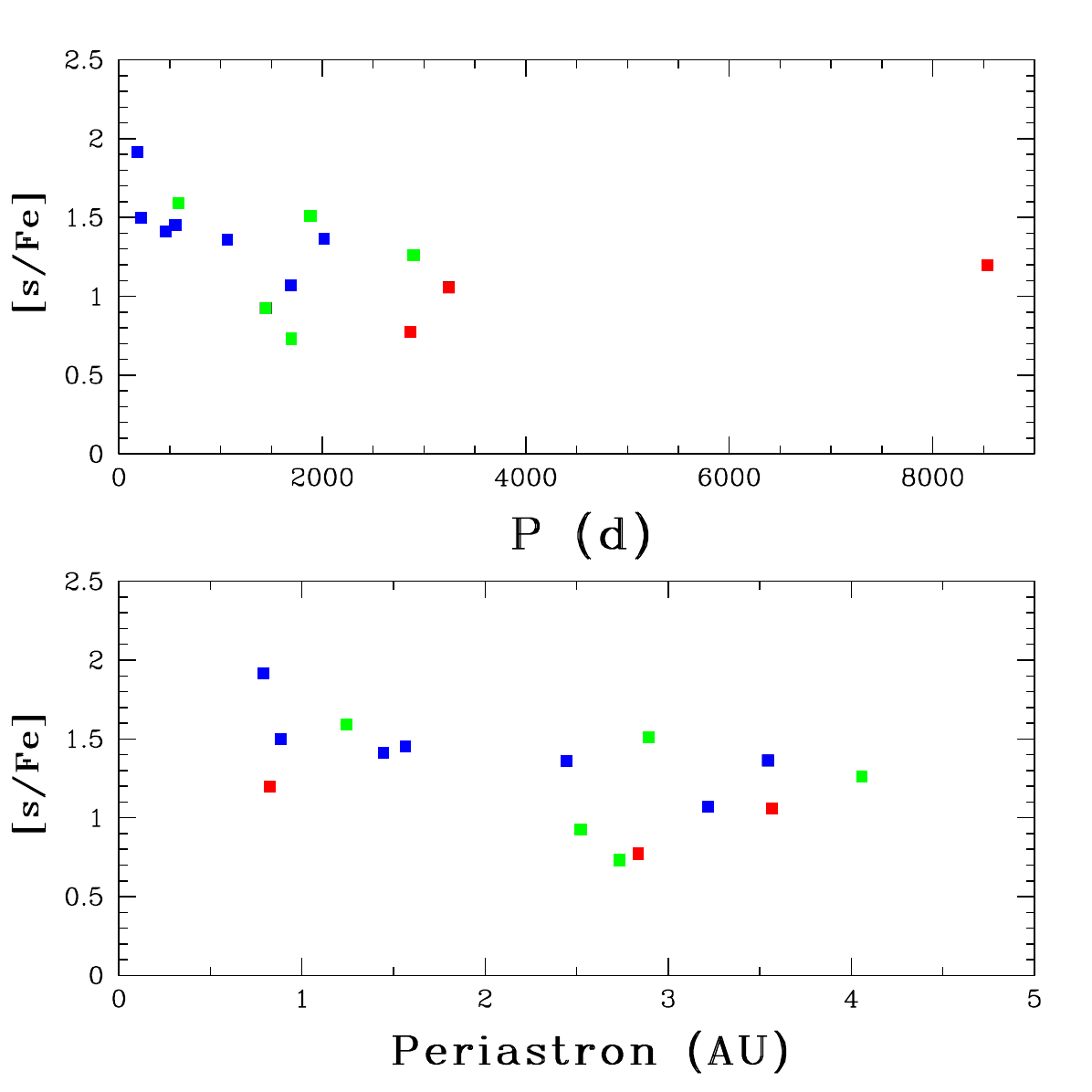}
\caption{  
\label{Fig:sP} 
Mean $s$-process abundance (s is an average of the Sr, Y, Zr, Nb, Ba, La, Ce, Pr, Nd abundances) as a function of the orbital period (top) and of the periastron distance (bottom), assuming the total mass of the binary system is 2~\Msun. The  eccentricity is coded as follows: red: $e>0.3$; green: $0.1< e\le 0.3$; blue: $e\le 0.1$. }
\end{figure}

\begin{figure}
   \includegraphics[width=9cm]{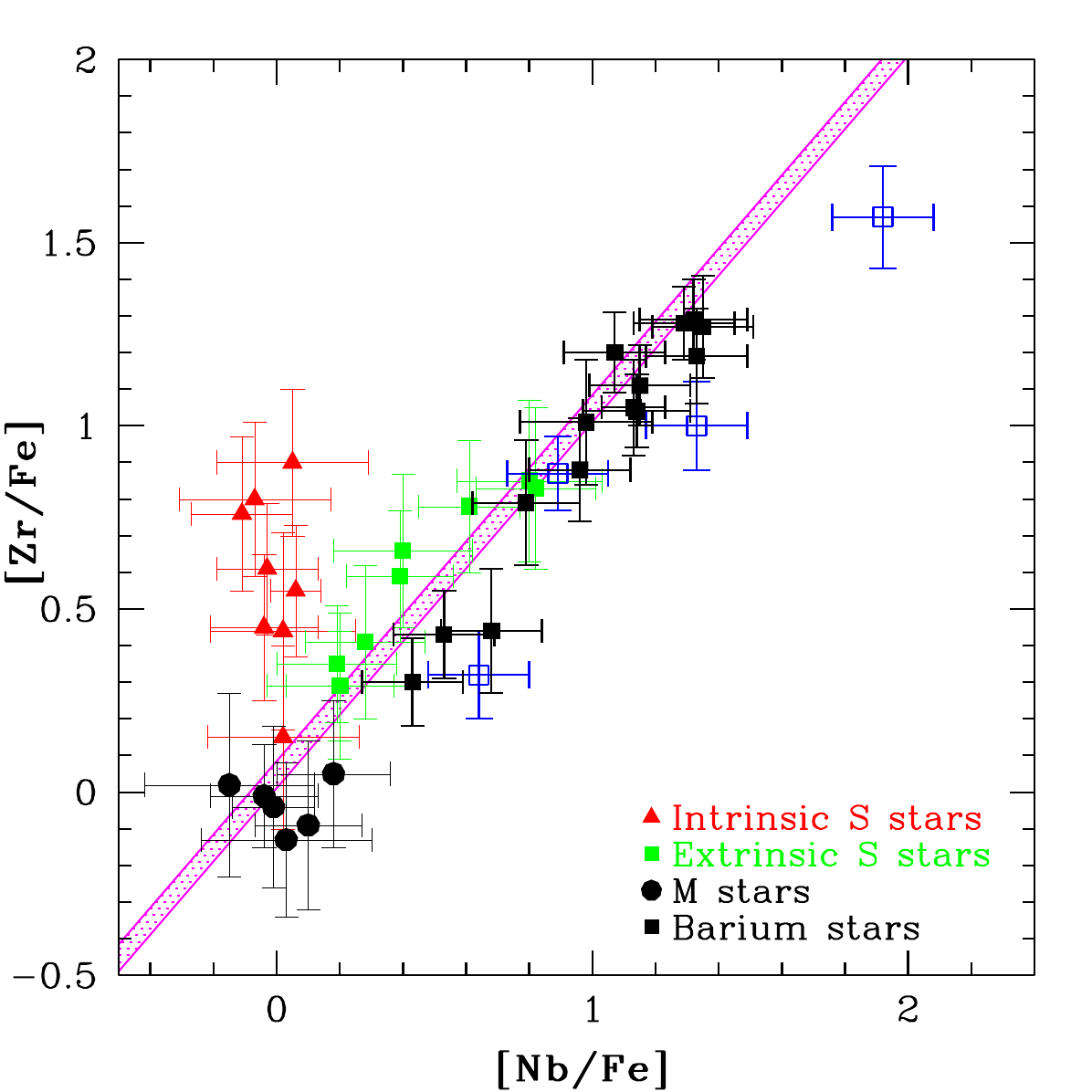}
    \caption{\label{Fig:NbZr} Distribution of [Zr/Fe] vs. [Nb/Fe] for our target stars. Abundances of intrinsic S stars and M stars  are from \citet{Neyskens2015}. The abundances of extrinsic S stars and barium stars are from this work. The open blue squares refer to the four N-enriched barium stars (HD~43389, HD~60197, HD~100503, and HD~121447 in order of increasing value of [Nb/Fe]). 
    HD~28159, a star previously known as K0III and which proved to be slightly enriched in $s$-process (see Sect.~\ref{Sect:pattern}) is located at ([Nb/Fe],[Zr/Fe])= (0.43, 0.30).
    The magenta shaded band indicates the expected location of stars polluted by material resulting from the $s$-process  operating at temperatures between 1.0 (upper line) and 3.0$\times 10^8$~K (lower line). These simple predictions assume that  the chain of neutron captures along the Zr isotopes is in local equilibrium. } 
\end{figure} 

\subsection{Chemical overabundances and orbital elements} 

Mass transfer being responsible for the chemical peculiarities of barium stars, a correlation is expected between the orbital period and the overabundance level.
Figure~\ref{Fig:allabundP} displays the barium-star abundances of $s$- and $r$-process elements  derived in the present study as a function of their orbital period \citep[Appendix A of][]{VanderSwaelmen-2017}. A loose correlation is indeed present, maybe more clearly seen for Zr, Ba, and  Nd, pointing to a lower enrichment when the orbital period is longer. No clear  metallicity modulation  (colour-coded in Fig.~\ref{Fig:allabundP}) is observed, though the small number of investigated stars prevents us from drawing any firm conclusion on a possible impact of the stellar metallicity on the $s$-process overabundance level.

The overabundance--period trend is somewhat clearer when considering an average of eight measured $s$-process elements as indicator (top panel of Fig.~\ref{Fig:sP}).
This correlation remains when considering the periastron distance instead of the orbital period. Another interesting aspect to be discussed is the impact of the orbital eccentricity. 
The most eccentric barium stars ($e>0.3$) tend to be among the least $s$-process enriched (red points in Fig.~\ref{Fig:sP}). 
When occurring in a system characterised by an eccentric orbit, the efficiency of the mass transfer, being probably intermittent, could be less than when occurring continuously along a circular orbit.

\subsection{The Zr-Nb pair as $s$-process thermometer} 
\label{Sect:thermometer}

\subsubsection{Method}
\label{Sect:ZrNb}

The aim of this section is to apply the method described in \citet{Neyskens2015} to a larger sample of stars characterised by high levels of $s$-process enrichments.
We  show that for low-mass stars (M$\le$3~\Msun) a simple straight line drawn in the ([Zr/Fe],[Nb/Fe]) diagram provides an estimate of the $s$-process operation temperature, whereas for higher mass stars (M$\ge$4~\Msun), this diagram keeps a predictive power for the $s$-process operation temperature, but the relation is more complex 
(see in Sect.~\ref{Sect:NbZrSTAREVOL}).

The method compares the [Zr/Fe] and  [Nb/Fe] abundances in intrinsic stars (red symbols in Fig.~\ref{Fig:NbZr}), which are still on the AGB producing $^{93}$Zr, with abundances in the older extrinsic stars that accreted the AGB material. In these extrinsic stars, $^{93}$Zr had time to decay into $^{93}$Nb, since the $\beta$-decay half-life of $^{93}$Zr is $1.53\times10^6$~ yr.
 
These polluted stars thus appear as either barium stars, CH stars, or extrinsic S stars (depending on their $T_{\rm eff}$ and metallicity), enriched in Nb. They are located along the diagonal in Fig.~\ref{Fig:NbZr} (blue, black, and green symbols, respectively). 
The ratio Zr/Nb  is thus a powerful tool that can  distinguish intrinsic from extrinsic stars \citep{Bisterzo2010,Neyskens2015}.

The barium stars in the present sample have been deliberately selected to be highly enriched in $s$-process elements: their maximum overabundance is around 
[Zr/Fe]~$\sim~1.6$.
Intrinsic S stars, in turn, show a lower maximum overabundance level (around [Zr/Fe]~$\sim 0.9$).  This is not surprising, since intrinsic S stars spend only a small portion of their TPAGB lifetime being classified as S stars.
As they evolved along the AGB, third dredge-ups raise the surface C/O ratio and rapidly turn the S star into a carbon star 
(with C/O$>1$). The latter are expected and known to be more $s$-process enriched indeed \citep[see Extended Data Fig.~2 in][]{Neyskens2015}.

For the sake of clarity, we reproduce here the main predictions from \citet{Neyskens2015}, based on the {\em assumption of local equilibrium}, i.e., the product $\langle \sigma_i \rangle N_i$ is constant along an isotopic chain (where $\langle \sigma_i \rangle$ and  $N_i$ are the Maxwellian-averaged neutron-capture cross section and abundance of isotope $i$, respectively). The observed ([Zr/Fe], [Nb/Fe]) ratios are then expected to fall on the relation
\begin{equation}
\label{Eq:Zr/Fe}
\left[\frac{\rm Zr}{\rm Fe}\right] = \left[\frac{\rm Nb}{\rm Fe}\right] + \log \omega^* - \log\frac{N_\odot({\rm Zr})}{N_\odot({\rm Nb})},  
\end{equation}
where 
\begin{equation}
\label{Eq:omega}
\omega^* = \langle \sigma_{93} \rangle \left[\frac{1}{\langle \sigma_{90}\rangle} + \frac{1}{\langle\sigma_{91}\rangle} + \frac{1}{\langle\sigma_{92}\rangle} + \frac{1}{\langle\sigma_{94}\rangle}\right].
\end{equation}
The above relation further assumes that the neutron density is low enough for the branching at $^{95}$Zr not to produce $^{96}$Zr.
Thus, the $y$-intercept of the straight line of slope 1 defined by Eq.~\ref{Eq:Zr/Fe} in the ([Zr/Fe], [Nb/Fe]) plane depends on the neutron-capture cross sections of the Zr isotopes,  through the quantity $\omega^*$. Since these cross sections are temperature dependent, so is $\omega^*$, and hence the $y$-intercept of the diagonal. The location of the diagonal should thus give insights into the operation temperature of the $s$-process. To illustrate these principles, the band corresponding to $s$-process operating temperatures of $1$ to $3\times10^8$~K is displayed in Fig.~\ref{Fig:NbZr}; this theoretical band assumes local nuclear equilibrium and is therefore independent of the metallicity (Eqs.~\ref{Eq:Zr/Fe} and \ref{Eq:omega}).
The sensitivity of the method remains modest because the error bars on our data points prevent a precise determination of the temperature. 
Furthermore, detailed nucleosynthesis calculations in hot thermal pulses of AGB stars show that the assumption of local equilibrium is not necessarily valid. The situation is thus more complex than that described by  Eqs.~\ref{Eq:Zr/Fe} and \ref{Eq:omega}, as we discuss in Sect.~\ref{Sect:NbZrSTAREVOL}.

\begin{figure}
\includegraphics[width=9cm]{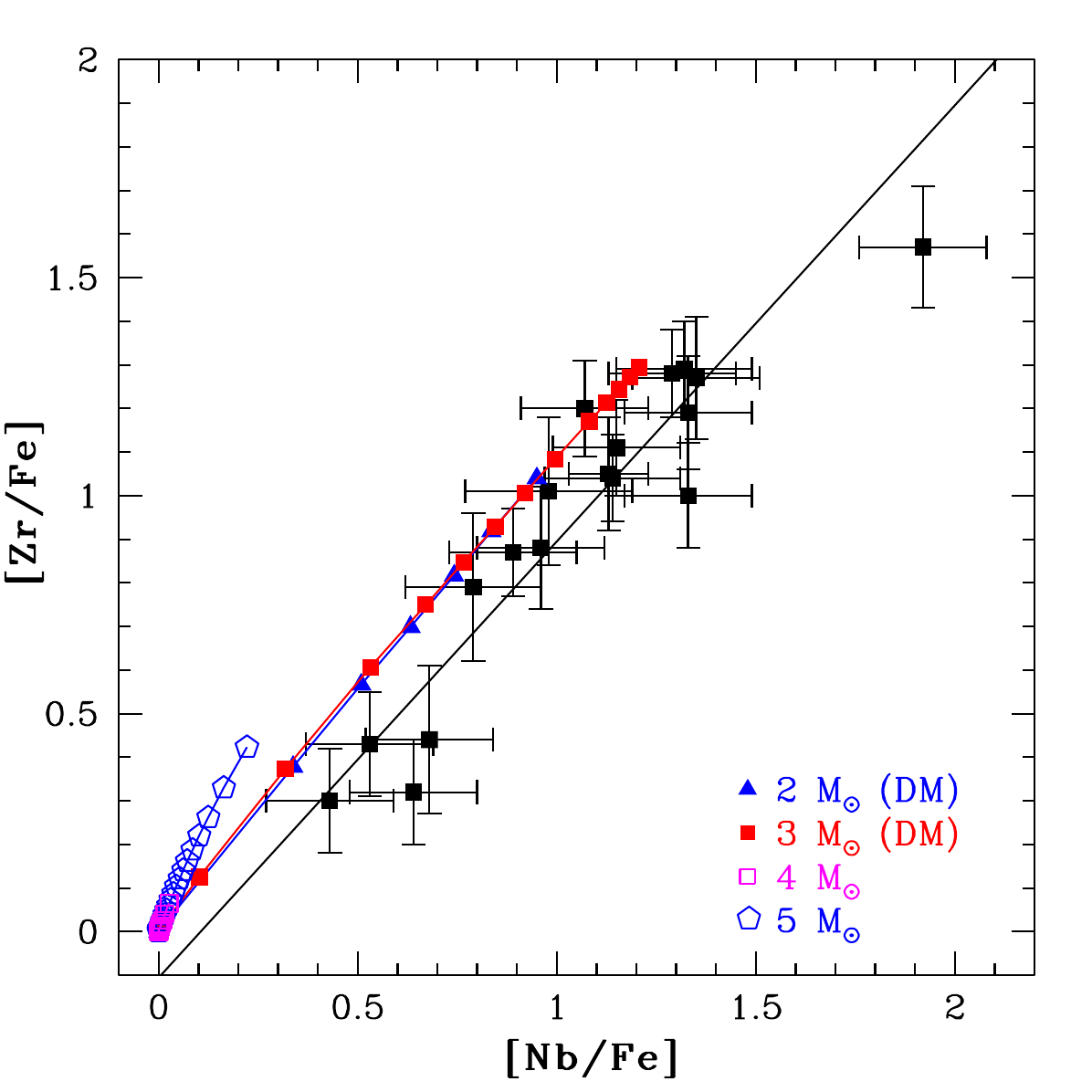}
\caption{Predictions from the STAREVOL code for extrinsic stars polluted by a 2, 3, 4, or 5~\Msun~AGB star with  [Fe/H]~=~-\nobreak0.5.
The 4~\Msun{} predictions are  superimposed on the 5~\Msun{} values, but their track stops at 
([Nb/Fe], [Zr/Fe])~$\sim (0.03, 0.07)$.
For the 2 and 3~\Msun{}  models, diffusive mixing (DM) of protons in the $^{12}$C pocket during the interpulse is modelled, whereas for the
4 and 5~\Msun{} models, the $s$-process is assumed to be purely convective in the pulses. Each symbol corresponds to a thermal pulse enrichment.  The black line corresponds to the fit to the barium stars (black squares). }
\label{Fig:NbZrSTAREVOL}
\end{figure}

\subsubsection{Predictions from STAREVOL}
\label{Sect:NbZrSTAREVOL}


The STAREVOL predictions for [Fe/H]=-0.5 (not assuming local equilibrium and not neglecting the 
$^{95}$Zr$-^{96}$Zr branching)
relative to Nb and Zr are displayed in Fig.~\ref{Fig:NbZrSTAREVOL}, where 
all $^{93}$Zr is assumed to have decayed into Nb; therefore,  only extrinsic stars are  plotted.

The 2 and 3~\Msun{} models follow a diagonal of slope 1 with a $\omega^*$ value of 15.8, consistent with a $s$-process temperature of $10^8$~K 
typical of the operation of the $^{13}$C($\alpha$,n)$^{16}$O neutron source.
However, for AGB stars more massive than $\sim 4$~\Msun, the situation becomes more complex. First, given the greater envelope mass and the small pulse extent, many more third dredge-ups are required to reach a given surface overabundance level, and at the end of the computations the enrichment remains modest. 
Second, it does not comply any longer with the picture drawn under the hypothesis of local equilibrium
since the Nb and Zr abundances for these more massive stars (with hot pulses thus) fall above the predictions for the cooler pulses in lower-mass AGB stars. This result is in contradiction with the temperature dependence predicted by the local-equilibrium analysis (Fig.~\ref{Fig:NbZr} and Sect.~\ref{Sect:ZrNb}).

There are three reasons responsible for the breakdown of the local-equilibrium assumption. First, local equilibrium does not have enough time to establish since the duration of thermal pulses 
is short compared to the neutron-capture timescale for the Zr isotopes, especially in view of the rapidly varying profile of neutrons in the convective pulse and the small region of the thermal pulse subject to the neutron irradiation.
Second, the neutron density in a pulse of the 4~\Msun\  star (with [Fe/H]=-0.5) with base temperature of $3.7\times10^8$~K is high enough to bypass $\beta$-decay of $^{95}$Zr and produce  $^{96}$Zr. This results in values of Zr/Nb as high as 27, to be compared with Zr/Nb  in the range  12 -- 17 (or more exactly 15.8, as found above for the 2 and 3~\Msun{} stars) as predicted by the local equilibrium analysis (without $^{96}$Zr; see Fig.~1 of Neyskens et al. 2015).  
Third, with the high neutron densities found at the bottom of hot thermal pulses ($N_n \approx 10^{12}$~cm$^{-3}$), an important amount of the unstable 
$^{90}$Sr (with a half-life of $t_{1/2}=28.8$~y), $^{90}$Y ($t_{1/2}=2.7$~d), and 
$^{91}$Y ($t_{1/2}=58.5$~d)
is produced. During the neutron irradiation, such long-lived branchings contribute to the production of $^{91,92}$Zr isotopes but bypass $^{90}$Zr. After the neutron irradiation, their $\beta$-decays lead to a delayed increase in the Zr abundance without affecting the $^{93}$Zr (thus $^{93}$Nb) yield, hence a higher value of the Zr/Nb ratio.   

All these reasons explain why the ([Zr/Fe], [Nb/Fe]) curves corresponding to hot pulses in 4 -- 5~\Msun{}  AGB stars are located {\em above} the  $^{13}$C($\alpha$, n)$^{16}$O locus in Fig.~\ref{Fig:NbZrSTAREVOL}. Because of this behaviour, the discriminating power of the ([Zr/Fe], [Nb/Fe]) diagram is restored, and in fact, turns out to be better than initially inferred from the simple local-equilibrium analysis (Fig.~\ref{Fig:NbZr}). Figure~\ref{Fig:NbZrSTAREVOL} clearly shows that all the extrinsic stars analysed so far are consistent with an $s$-process triggered by the
$^{13}$C($\alpha$,n)$^{16}$O neutron source operating in 2 -- 3~\Msun{}  AGB stars (see, however, a caveat below) rather than in more massive ones, a conclusion in agreement with that reached in Sect.~\ref{Sect:pattern} on the basis of the $s$-process abundance pattern. 
This confirms conclusions from earlier studies \citep[e.g.][]{Busso-2001} that the abundance pattern, as probed specifically by the [hs/ls] ratio, is compatible with the $^{13}$C($\alpha$,n)$^{16}$O neutron source. 

It is noteworthy that no extrinsic stars seem to originate from more massive AGB progenitors. This could be the result of a combined effect of the lower probability of occurrence of more massive stars, and of the moderate levels of pollution expected from these stars (Fig.~\ref{Fig:NbZrSTAREVOL}) due to large dilution factors. Therefore, extrinsic stars polluted by massive AGB stars, should they exist, would exhibit very modest enhancement levels, and might even not be detected as $s$-process-rich stars.

Many extrinsic stars in Fig.~\ref{Fig:NbZrSTAREVOL} are, however, off the STAREVOL  2 -- 3~\Msun\  predictions by a  few tenths of a dex. 
Given the careful abundance calibration performed in the present paper, 
there are no obvious reasons why such a bias would affect our Nb abundances.
However, as stressed in Sect.~\ref{Sect: s-process}, the oscillator strengths for the two lines used to derive the Zr abundance have a tendency to yield abundances that are about 0.1~dex too low in the benchmark stars. This slight shift would be enough to reconcile observed  ([Zr/Fe], [Nb/Fe]) abundances with those predicted in 2 -- 3~\Msun\  stars (Fig.~\ref{Fig:NbZrSTAREVOL}). The uncertainty on the $^{93}$Zr neutron-capture cross section \citep{Tagliente2013} could also modify somewhat the STAREVOL predictions \citep[see Fig.~1 of][for an illustration of the impact of the cross-section uncertainties on $\omega^*$]{Neyskens2015}.


\subsubsection{Gaia parallaxes and determination of the mass of the extrinsic stars }
\label{Sect:Masses}

\begin{table*}
\centering
\caption{\label{Tab:masses}
Luminosities, masses, and metallicities for the programme stars.}
\begin{tabular}{llllllll}
\hline
\\
Name & [Fe/H] & $\log L_{\rm min}$ & $\log L$&  $\log L_{\rm max}$ &$M_{\rm min}$  & $M$ & $M_{\rm max}$ \\
 & & & ($L_{\odot}$) & & & (\Msun)\\
\hline 
HD~5424   & $-0.43$ & 1.47  & 1.56  & 1.66   &  1.0  & 1.1  & 1.2  \\  
HD~12392  & $-0.38$ & 1.28  & 1.36  & 1.45   &  1.0  & 1.1  & 1.2 \\ 
HD~16458  & $-0.64$ & 2.31   & 2.39  & 2.47   &  1.9  & 2.2  & 2.5\\  
HD~27271  & $-0.07$ & 1.44  & 1.49 &  1.55 &   2.3 &  2.5 &  2.7  \\  
HD~31487  & $-0.04$ & 1.50 &  1.54  & 1.59  &   2.3 &  2.5 &  2.7  \\ 
HD~43389  & $-0.35$ & 2.23 &  2.34  & 2.47  &   -    &  -    & 0.9  \\  
HD~46407  & $-0.35$ & 1.86 &  2.01  & 2.20  &   2.0 &  2.4  & 2.9 \\    
HD~50082  & $-0.32$ & 1.70  & 1.76  & 1.82  &   1.5 &  1.7  & 2.0 \\    
HD~88562  & $-0.53$ & 2.41 &  2.57  & 2.76   &  0.8  & 1.0   &1.2  \\ 
HD~120620 & $-0.29$ & 0.29 &  0.39  & 0.50   &   -    &  -  & 0.9  \\ 
HD~121447 & $-0.90$ & 2.53  & 2.86   &3.40    & 1.0  & 1.4  & 2.0 \\   
HD~123949 & $-0.23$ & 2.01  & 2.16  & 2.34   &  1.0 &  1.3 &  1.7  \\ 
HD~178717 & $-0.52$ & 3.15  & 3.32  & 3.53   &  1.4  & 1.9  & 2.5 \\   
HD~199939 & $-0.22$ & 2.00 &  2.19  & 2.44  &   2.0  & 2.7  & 3.5  \\  
\hline 
\end{tabular}
\end{table*}
   
 \begin{figure}

    \includegraphics[width=8cm]{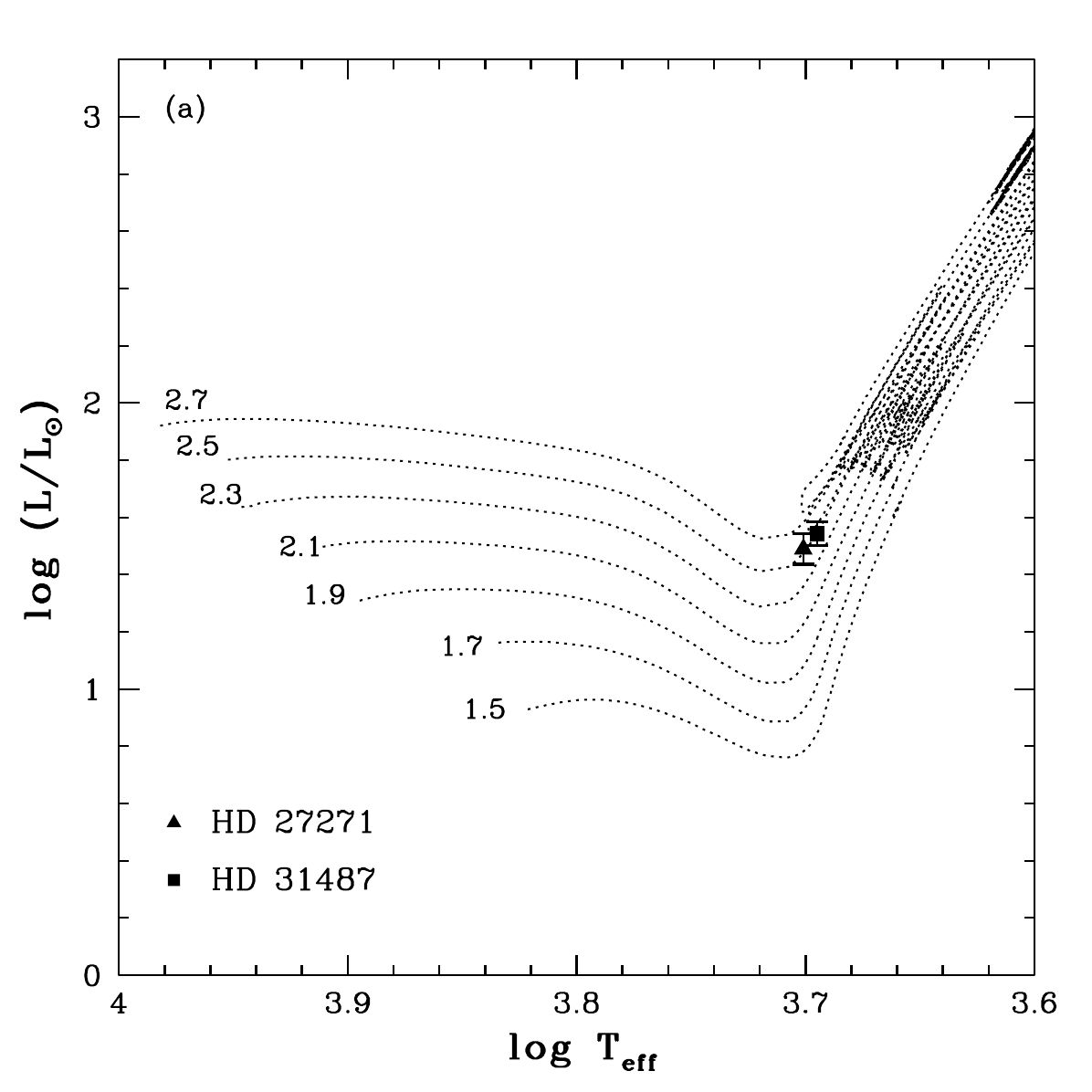}
    \includegraphics[width=8cm]{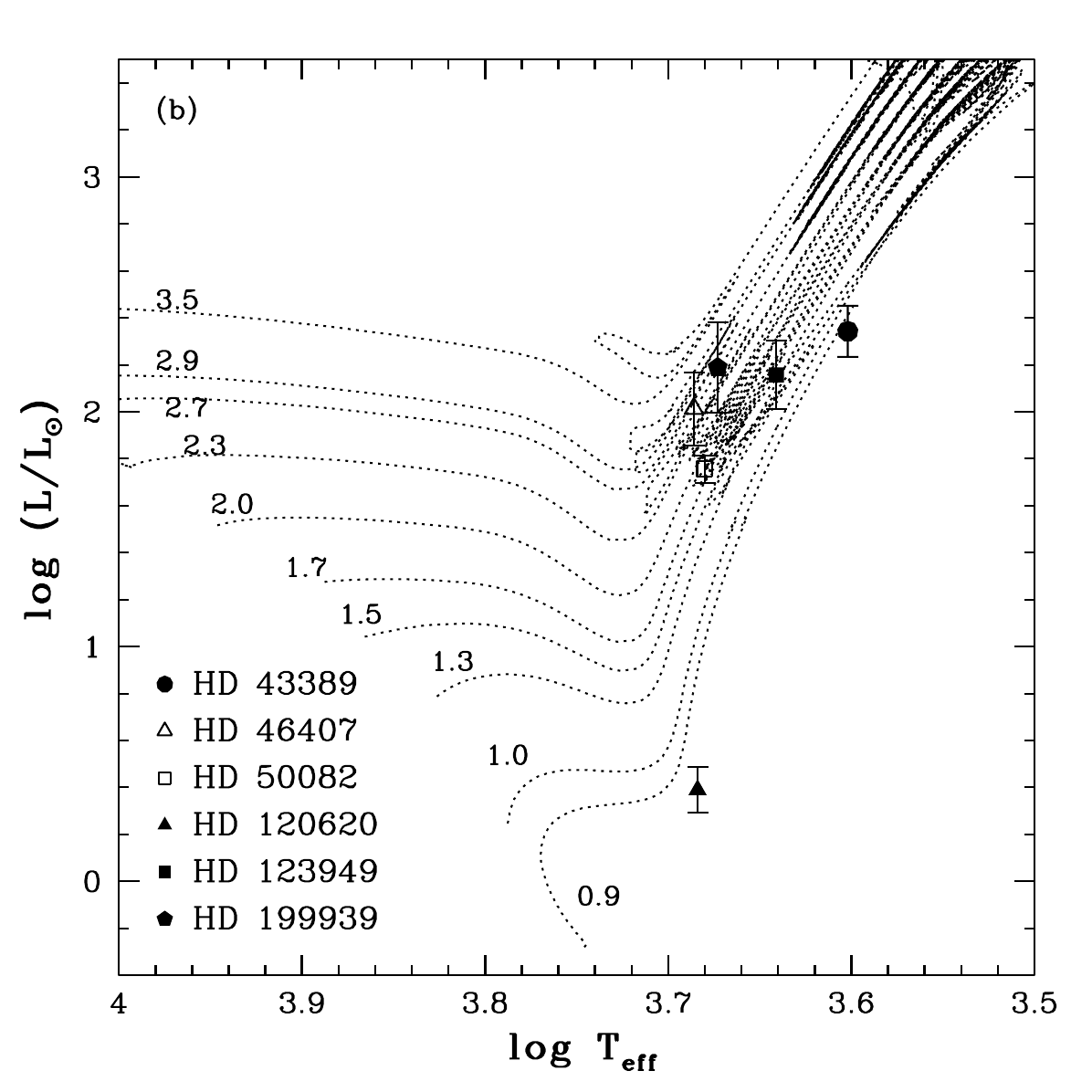}
    \includegraphics[width=8cm]{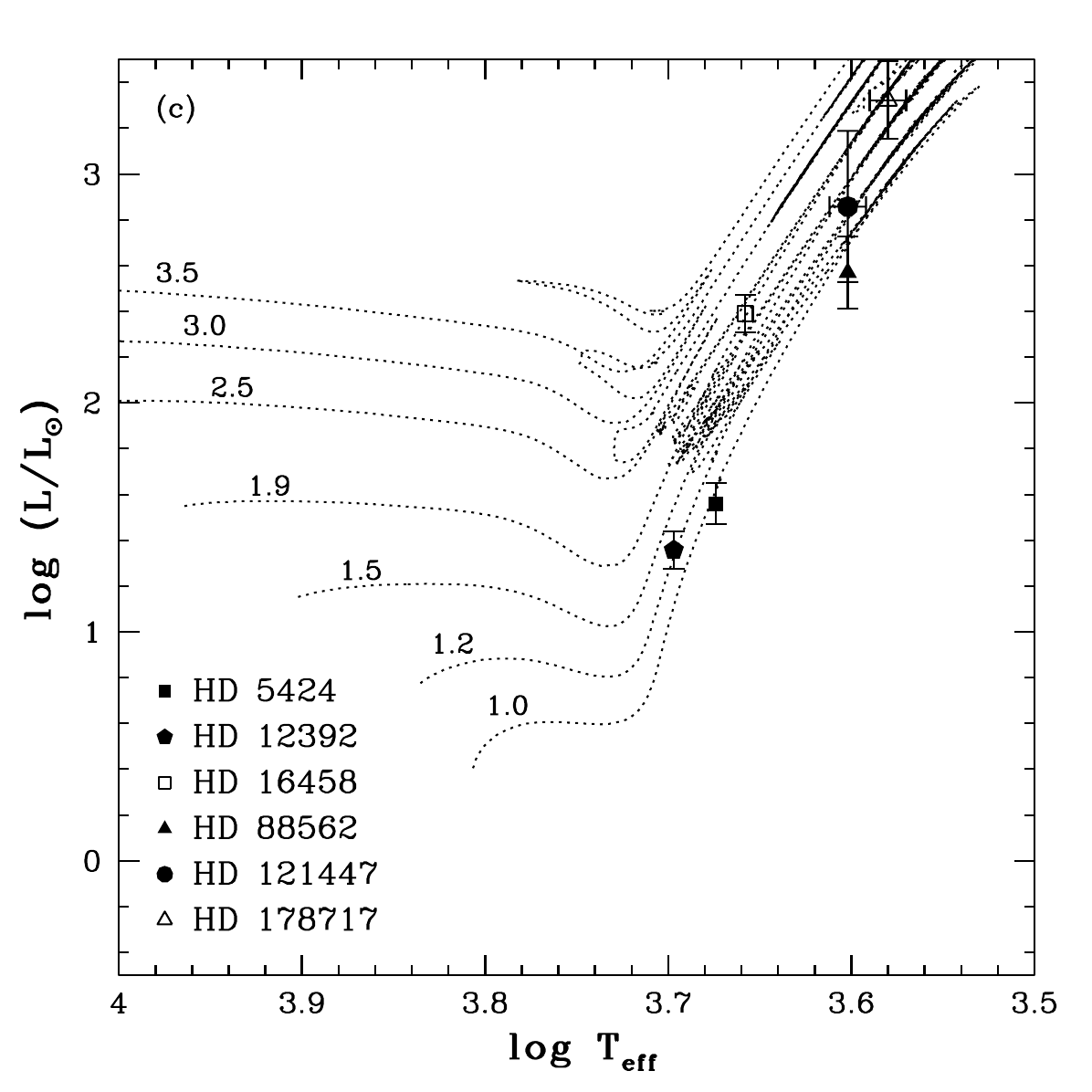}
    \caption{\label{Fig:masses} HR diagram for the programme stars, ordered according to their metallicities: (a) [Fe/H] $\approx 0.0$; (b) [Fe/H] $\approx -0.25$; (c) [Fe/H] $\approx -0.5$. The tracks are labelled according to the mass (in~\Msun).}
 \end{figure} 
   
     \begin{figure}
\includegraphics[width=9cm]{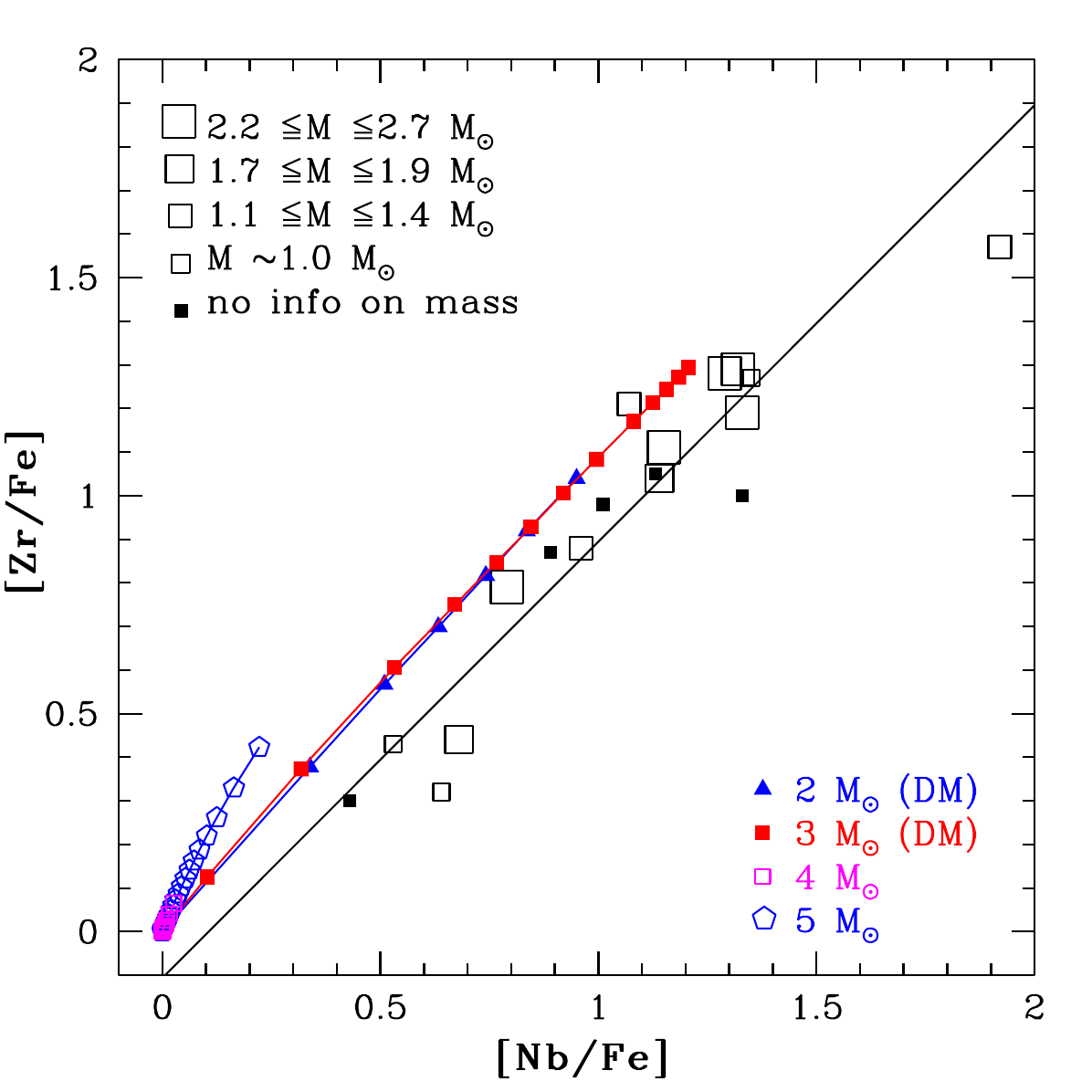}
\caption{Same as Fig~\ref{Fig:NbZrSTAREVOL}, but coded according to the mass of the extrinsic star whenever available.  
\label{Fig:NbZrSTAREVOL_mass}   }
   \end{figure}


The positions of the stars in the HR diagram was derived using the procedure described by \citet{Escorza2017}: the bolometric flux is computed by integrating the spectral energy distribution corresponding to a MARCS model atmosphere with the stellar parameters (effective temperatures, gravities, and metallicities) derived from the spectroscopic analysis (Table~\ref{Tab:program_stars}). 
Then the distance modulus corresponding to the TGAS parallax \citep[The Tycho-Gaia astrometric solution;][]{Michalik2015} available for 14 (out of 18) of our sample stars was computed to obtain the luminosity. Once the luminosity and effective temperature were known, the stellar mass was estimated by finding the STAREVOL evolutionary track of the corresponding metallicity that passes through this point in the HR diagram (Fig.~\ref{Fig:masses}).
The resulting luminosities and masses are listed in Table~\ref{Tab:masses}, as are the uncertainties derived from the uncertainties on the TGAS parallaxes. 
Two objects (HD~43389 and HD~120620, see middle panel of Fig.~\ref{Fig:masses}) have luminosities fainter than the 0.9~\Msun\ evolutionary track corresponding to their metallicity (though considering its error bar HD~43389 fits on the 0.9~\Msun{} track). Only an upper limit for their mass is therefore listed in Table~\ref{Tab:masses}. We note that the position in the HR diagram of HD~120620 is consistent with its unusually high $^{12}$C$/^{13}$C ($\approx 90$), indicative of a pre-first dredge-up composition.

The inferred masses of the barium stars range between 0.9 and 3~\Msun{} and are compatible with the conclusions that the $s$-process material currently observed in the atmospheres of the extrinsic stars must have originated from a low-mass AGB companion.
Because the Zr/Nb ratio depends on the operation temperature of the $s$-process, thus on the mass of the AGB polluter, we might expect that the position of the barium star in the ([Zr/Fe], [Nb/Fe]) plane depends on its mass as well.
Figure~\ref{Fig:NbZrSTAREVOL_mass} reveals no such correlation, however,
perhaps because, as discussed above, stars polluted by more massive AGB stars
never display overabundance levels as high as those in barium stars.


\subsection{Rubidium as a diagnostic for the operation of the $^{22}$Ne neutron source }
\label{Sect:Rb}
   
\begin{figure}
\includegraphics[width=9cm]{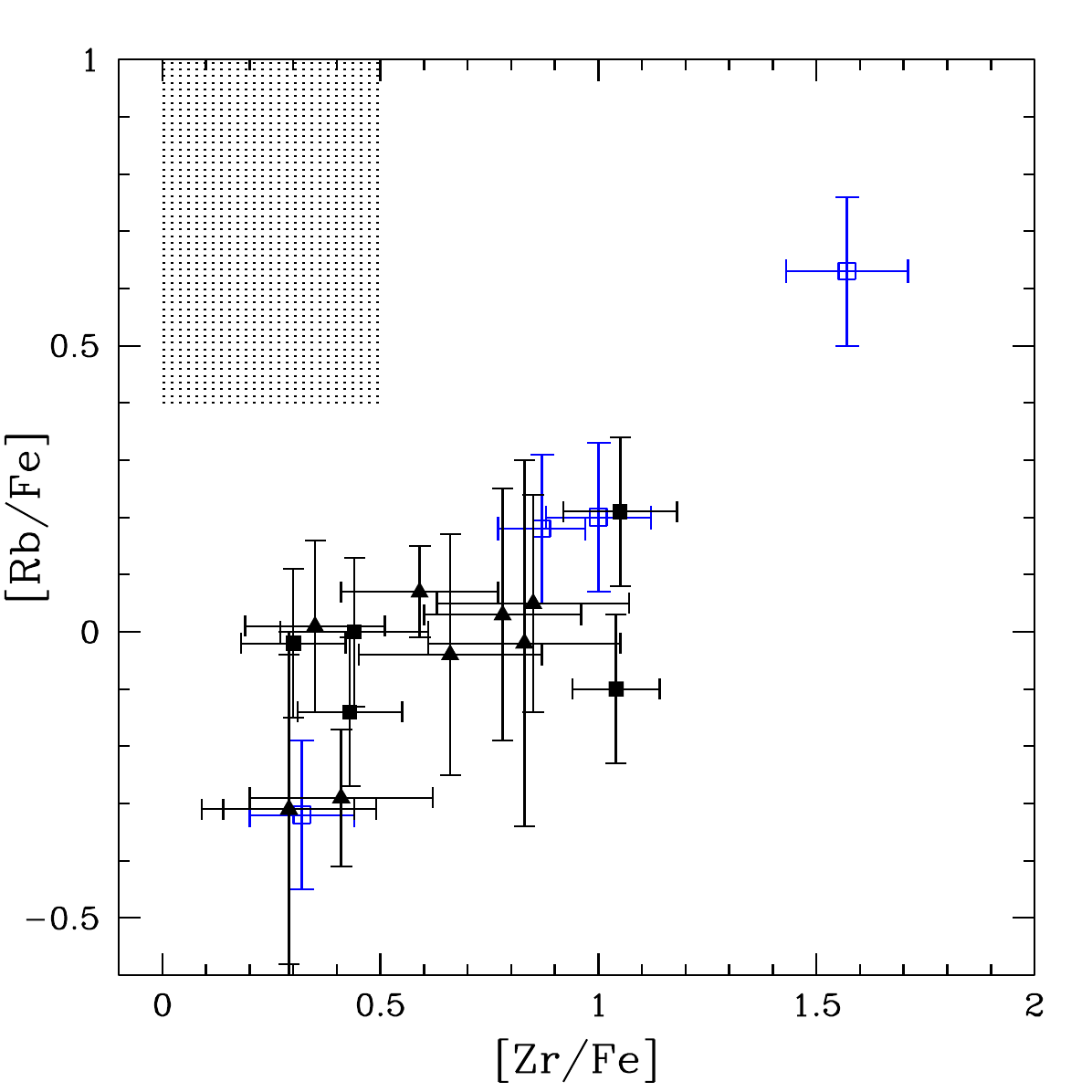}
\caption{Abundances of  Rb vs. Zr  for barium (squares) and extrinsic S stars (filled triangles). The four usual N- and Nb-rich barium stars are shown as open blue squares. Among these, HD~121447 is the most enriched in Zr. The shaded area represents the observed range of Rb and Zr abundances in intermediate-mass AGB stars \citep{vanRaai2012}, clearly incompatible with the range measured in barium stars.
\label{Fig:Rb}              }
   \end{figure}

Models predict higher Rb abundances in AGB stars where neutrons are produced by the $^{22}$Ne($\alpha$,n)$^{25}$Mg reaction \citep{Abia2001,vanRaai2012}.
The amount of produced Rb depends on the ability of $^{85}$Kr and $^{86}$Rb to capture a neutron before
decaying, since they act as branching points.  The  probability  of  this  happening  depends on the local neutron density \citep{Beer1989}.
The $^{22}$Ne($\alpha$,n)$^{25}$Mg neutron source favours the production of $^{87}$Rb. Hence, the determination of the $^{87}$Rb/$^{85}$Rb isotopic ratio in AGB stars could inform us about the neutron source but that ratio is impossible to measure in stellar spectra, even with a very high spectral resolution. 
Nevertheless, models predict that the ratio of Rb with other light $s$-process elements like Sr, Y, or Zr can be used for the same purpose. 
As shown in Fig.~\ref{Fig:Rb}, derived [Rb/Zr] ratios in AGB stars of different masses both in our Galaxy and the Magellanic Clouds vary from $-1$ to 1.2~dex \citep{Abia2001,Garcia2006a,vanRaai2012,Zamora2014,Perez-Mesa2017}, the highest values (shaded region in Fig.~\ref{Fig:Rb}) corresponding to the most massive AGB stars (4 to 6~\Msun), and negative values to lower mass AGB stars where $^{13}$C($\alpha$,n)$^{16}$O operates.

While HD~121447 shows a substantial Rb overabundance ([Rb/Fe] = 0.63), this is a sequel to the very high overall $s$-process overabundances  for this star. In terms of [Rb/Zr] ($= -0.94$), HD~121447 is not different from the other stars, which all have [Rb/Zr]$ < 0$ as expected when $^{13}$C($\alpha$,n)$^{16}$O operates \citep[e.g.][]{vanRaai2012}.
In fact, none of the measured [Rb/Zr] values in our barium stars
agrees with the [Rb/Zr] values
measured in intermediate-mass AGB stars (shaded region in Fig.~\ref{Fig:Rb})
nor with the expectations for the operation of $^{22}$Ne($\alpha$,n)$^{25}$Mg in 
thermal pulses occurring in more massive AGB stars \citep{vanRaai2012, Karakas-2016}.


\subsection{HD 100503: a CEMP-$rs$ analogue  at a relatively high metallicity ([Fe/H]=-0.7)?}

\begin{figure} 
    \includegraphics[width=9cm]{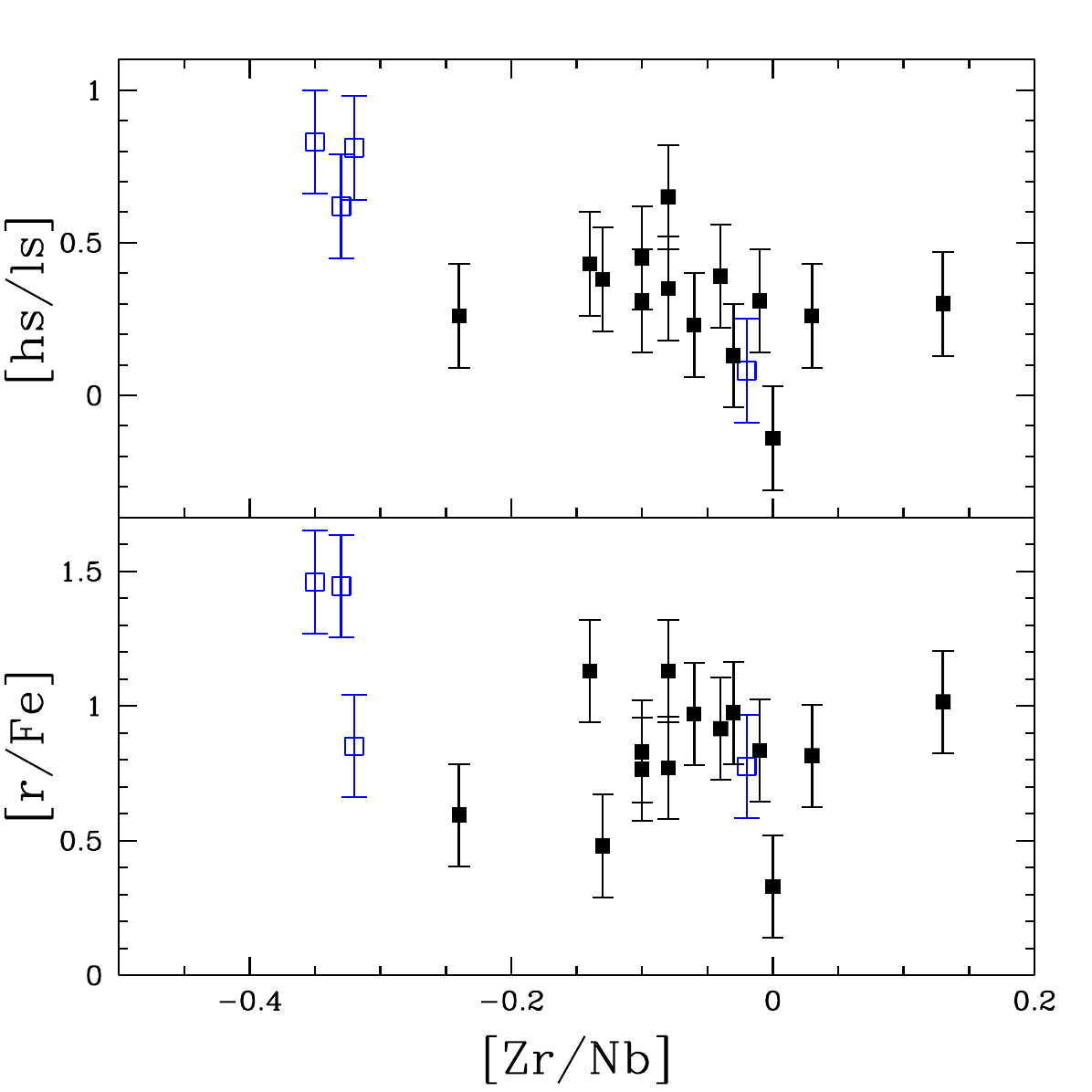}  
\caption{Ratios of  [hs/ls]  (top panel, where $ls$ is the average of Sr, Y, and Zr, and $hs$ the average of La and Ce) and [r/Fe]  (bottom panel, where `r' stands for the average of the [Sm/Fe] and [Eu/Fe] ratios) are shown as a function of [Zr/Nb]. The four N-rich barium stars are shown as open blue squares, of which  HD~100503 is found to be highly enriched in $r$-process elements. 
              }
     \label{Fig:hsls}
   \end{figure}

\begin{figure} 
\includegraphics[width=9cm]{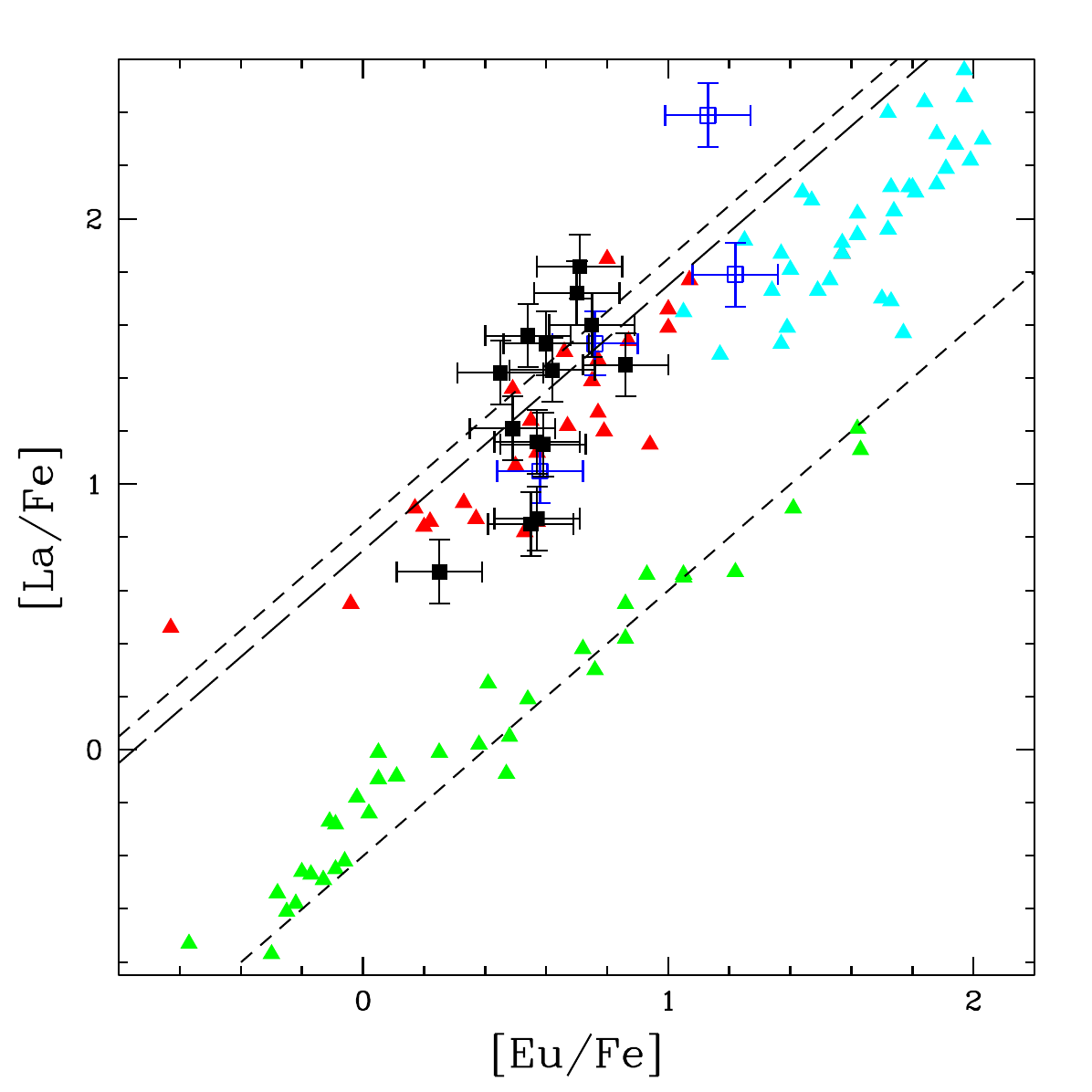}  
\caption{Abundances of [La/Fe] vs. [Eu/Fe]  in barium stars (squares) and in CEMP-$s$  (red triangles), CEMP-$r$ (green triangles), and CEMP-$rs$ (cyan triangles) stars. Data for CEMP stars are from \citet{Masseron2010}. 
The four usual N-rich barium stars are shown as open blue squares.
The short-dashed line through the CEMP-$r$ stars corresponds to pure solar $r$-process \citep{Goriely1999}, whereas the short-dashed line through the CEMP-$s$ stars corresponds  to pure $s$-process nucleosynthesis  predictions for a 0.8~M$_{\odot}$ star \citep{Masseron2006} and a 3~M$_{\odot}$  \citep[long-dashed line;][]{Goriely2005} metal-poor AGB star. 
\label{Fig:CEMP} }
\end{figure}

Most of the N-rich stars show enrichments in Na (Fig.~\ref{Fig:N+Na}), Nb (Fig.~\ref{Fig:NbZr}), and exhibit  high [hs/ls] ratios (Fig.~\ref{Fig:X/La} and top panel of Fig.~\ref{Fig:hsls} with data from Table~\ref{Tab:hsls}).  A further important property, not yet discussed, is their enrichment in the $r$-process elements Sm and Eu (bottom panel of Fig.~\ref{Fig:hsls}).
Figure~\ref{Fig:CEMP} compares the abundances of s and r elements, in a  way similar to that done for CEMP stars to distinguish CEMP-$s$ stars from CEMP-$r$ and CEMP-$rs$  stars \citep{Masseron2010}. One of our target stars clearly stands out, namely HD~100503, which falls in the region occupied by CEMP-$rs$ stars.
This similarity suggests that a common nucleosynthetic process may be responsible for the chemical anomalies of CEMP-$rs$ stars and the barium star HD~100503. 

We note that HD~121447 is characterised by a high [Eu/Fe] ratio, but it has an extremely high [La/Fe] ratio, locating it on the $s$-process side in the ([La/Fe], [Eu/Fe]) plane (Fig.~\ref{Fig:CEMP}). Therefore its connection with the CEMP-$rs$ objects is less obvious.  

The scenarios invoked to explain the $r$-process overabundances of CEMP-$rs$ stars involve either an external pollution from a nearby supernova  or the  $i$-process \citep[see references in][]{Masseron2010}, i.e. a process leading to neutron densities of the order of
$N_n \approx 10^{15}$~cm$^{-3}$,
intermediate between those of the $s$-process
($N_n \approx 10^{8}$~cm$^{-3}$)
and those of the $r$-process
($N_n \ge 10^{20}$~cm$^{-3}$).

The operation of $^{22}$Ne($\alpha$,n)$^{25}$Mg in the hot thermal pulses of intermediate-mass AGB stars \citep{Goriely2005} leads to 
such intermediate neutron densities ($N_n \approx 10^{11}- 10^{14}$ cm$^{-3}$).
Another possibility for the $i$-process operation would be the ingestion of protons in a He-burning convective region, as first described by \citet{Cowan1977}.
In this situation, protons are captured by $^{12}$C and lead to the formation of $^{13}$C which activates the $^{13}$C($\alpha$,n)$^{16}$O neutron source.
The typical neutron densities in such conditions are of the order of $N_n \approx 10^{15}$~cm$^{-3}$, substantially higher than the $s$-process neutron densities when protons are mixed in the intershell region to form the $^{13}$C pocket. 
Different sites have been proposed for this $i$-process, including (i) proton ingestion in very metal-poor low-mass stars during the core He-flash or at the beginning of the TP-AGB phase,
(ii) the dredge-out in super-AGB stars, (iii) the late pulses in post-AGB stars, and 
(iv) He-shell flashes in rapidly accreting white dwarfs \citep[see][and references therein]{Karakas2014, Denissenkov-2017}. We note that three-dimensional simulations show that turbulent entrainment of material could produce proton ingestion episodes at metallicities close to solar during the core He-flash \citep{Mocak2008, Mocak2009,Mocak2010}, and this could affect stars of masses $M\le 2.25$~\Msun.

The $i$-process has been recently invoked to explain not only CEMP-$rs$ stars \citep[e.g.][]{Hampel-2016}, but also low-metallicity
post-AGB stars showing low lead and other heavy-s abundances that cannot be reconciled with standard $s$-process model predictions. 
More precisely, below [Fe/H] $\sim -0.7$, a very wide range of neutron irradiations is needed to explain the spread in heavy-s abundances observed in these objects \citep{Lugaro-2015, DeSmedt-2016, Hampel-2016}. These authors suggested that the $i$-process could meet this challenge.
If HD~100503 turns out to be a genuine analogue of CEMP-$rs$ stars,
we note that it would be the first case found at such metallicity ([Fe/H] $= -0.7$).

\section{Conclusions}

Our analysis of the abundances of light (C, N, Na, Mg) and heavy elements in barium stars has shown that they are consistent with the operation of the  $^{13}$C($\alpha$,n)$^{16}$O neutron source in 2 -- 3~\Msun\  AGB stars. We show in particular 
that the Nb/Zr ratio offers a powerful diagnostic in that respect. 

We have identified four N-rich stars with [N/Fe]~$>0.7$ (HD~43389, HD~60197, HD~100503, and
HD~121447). All of them, except HD~60197, have  [Nb/Zr] and [hs/ls] ratios higher than any of the barium stars studied. Moreover, HD~100503 and HD~121447 
are enriched in Eu, and the latter is also the most Na-rich star in our sample. The origin of these peculiarities remains a mystery; the high N abundance  points towards a rather massive AGB companion undergoing hot-bottom burning, but this conclusion is at odds with the high [hs/ls] ratios that cannot be obtained from the activation of the $^{22}$Ne($\alpha$,n)$^{25}$Mg source
in those AGB stars. 
We conclude from the observed $s$-process abundance distributions that all the presently studied barium stars were polluted by low-mass (M$<4$\Msun) AGB stars. 

Of the four N-rich stars, HD~100503 has been identified as a possible analogue of CEMP-$rs$ stars with the highest metallicity ([Fe/H] $= -0.7$) known so far.



\begin{table}
\caption[]{
[ls/Fe] and [hs/Fe] in the programme stars.
\label{Tab:hsls}}

\begin{tabular}{lcccc}
\hline
\\
HD&  [Fe/H]& [ls/Fe] &[hs/Fe]& [hs/ls]\\
\\
\hline
5424    & $-0.43 $ &   1.11 &1.75&0.64  \\
12392   & $-0.38 $ &   1.11 &1.41& 0.30     \\
16458   & $-0.64 $ &   1.24 &1.37&0.13     \\
27271   & $-0.07 $ &   0.79 &0.65&-0.14     \\
28159   & $-0.50 $ &   0.29 &0.67&0.38\\
31487   & $-0.04 $ &   1.11 &1.50& 0.39      \\
43389   & $-0.35 $ &   0.57 &1.38&0.81      \\
46407   & $-0.35 $ &   1.22 &1.53&0.31     \\
50082   & $-0.32 $ &   0.96 &1.39&0.43      \\
60197   & $-0.60 $ &   0.89 &0.97&0.08     \\
88562   & $-0.53 $ &   0.64 &1.09&0.45     \\
100503  & $-0.72 $ &   1.09& 1.71& 0.62    \\
116869  & $-0.44 $ &   0.86 &1.12& 0.26     \\
120620  & $-0.29 $ &   1.33& 1.56& 0.23   \\
121447  & $-0.90 $ &   1.48 &2.31& 0.83    \\ 
123949  & $-0.23 $ &   0.90 &1.25& 0.35     \\
178717  & $-0.52 $ &   0.66 &0.80&0.26   \\
199939  & $-0.22 $ &   1.27 &1.70&0.43    \\
\hline 
\end{tabular}
\end{table}

\begin{acknowledgements}
D.K. acknowledges the support from Belgian Science Policy Office (BELSPO). This research has been partly funded by the Belgian
Science Policy Office under contract BR/143/A2/STARLAB.
S.V.E, T.M., and M.V.d.S. are thankful to "Fondation ULB" for its support.
L.S. and S.G. are senior FRS-FNRS research associates. T.M. acknowledges support provided by the Spanish Ministry of
Economy and Competitiveness (MINECO) under grant AYA-2017-88254-P.
This research is based on observations performed with the
Mercator
telescope and the HERMES spectrograph. The Mercator telescope  is operated  thanks to grant number  G.0C31.13 of the FWO under  the `Big  Science'  initiative  of the Flemish
government. The HERMES spectrograph is
supported by the Fund for Scientific Research of Flanders (FWO), the Research
Council of K.U.Leuven, the Fonds National de la Recherche Scientifique (F.R.S.-
FNRS),  Belgium, the  Royal  Observatory  of  Belgium;  the  Observatoire  de
Gen\`eve, Switzerland; and the Th\"{u}ringer Landessternwarte Tautenburg, Germany.
This work has made use of data from the European Space Agency (ESA)
mission {\it Gaia} (\url{https://www.cosmos.esa.int/gaia}), processed by
the {\it Gaia} Data Processing and Analysis Consortium (DPAC,
\url{https://www.cosmos.esa.int/web/gaia/dpac/consortium}). Funding
for the DPAC has been provided by national institutions, in particular
the institutions participating in the {\it Gaia} Multilateral Agreement. The authors thank the referee for the very constructive comments.
\end{acknowledgements}
\bibliographystyle{aa}
\bibliography{references}

\begin{appendix}

\section{$\chi^2$ fitting regions}

\begin{table}[h!]
\caption{
Boundaries (expressed in nm) of the continuum (C) and band (B) spectral regions used in the $\chi^2$ fitting (Sect.~\ref{Sect:parameters}).}
\label{Tab:chi2limits}
\begin{tabular}{llll}
\hline\\
  $\lambda_{B,i}$ & $\lambda_{B,f}$ & $\lambda_{C,i}$ & $\lambda_{C,f}$ \\ 
  \hline
          400.0&420.0&408.0& 412.0 \\
          420.0&440.0&432.0&434.0\\
          440.0&458.0&456.0&458.0\\      
          458.0&480.0&472.0&476.0\\      
          480.0&490.0&489.4&489.9\\
          490.0&495.0&494.0&495.0\\       
          495.0&516.0&515.5&516.0\\       
          516.0&530.0&516.0&516.3\\     
          530.0&544.0&543.0&544.5\\     
          544.0&565.0&560.5&562.0\\
          565.0&575.5&573.0&573.7\\
          575.5&588.0&580.0&582.0\\
          588.0&615.0&613.0&615.0\\
          615.0&647.0&645.0&647.0\\
          647.0&671.5&670.0&671.5\\
          671.5&685.0&683.0&685.0\\
          703.0&720.0&703.0&705.0\\
          \hline
\end{tabular}
\end{table}

\section{Line list}
Table~\ref{Tab:linelist} presents the lines used in the present abundance analysis.\\
\topcaption{Lines used in the abundance analysis.\label{Tab:linelist}}\\
\begin{supertabular}{llr}
\hline\\
$\lambda$ & $\chi_{\rm low}$ & $\log gf$ \\
(\AA) & (eV) & \\
\hline\\
\tablehead{
\hline\\
$\lambda$ & $\chi_{\rm low}$ & $\log gf$ \\
(\AA) & (eV) & \\
\hline\\}

C I&&\\
5380.337&7.685&-1.615\\
5553.174&8.643&-2.370\\
\hline\\
O I&      &      \\
6300.304& 0.000 &-9.715\\
6363.776 &0.020&-10.190\\
7771.941&9.146&0.369\\
7774.161&9.146&0.223\\
7775.338&9.146&0.001\\
\hline\\
Na I&&\\
6154.226& 2.102 &-1.547 \\
6160.747&  2.104 &-1.246 \\ 
\hline\\
Mg I &&\\
6318.717 & 5.108 & -2.103\\
6319.237 & 5.108 &-2.324\\
8717.810 & 5.933 &-1.850\\
8736.006 & 5.946 &-1.930 \\
8736.019 & 5.946 &-0.690 \\
8736.019 & 5.946 &-1.970\\
8736.029 & 5.946 &-1.020 \\
\hline\\
Rb I      &     &      \\ 
7800.259 & 0.000  &   0.137 \\ 
7947.597 & 0.000  &  -0.167 \\ 
\hline\\
Sr I     &            &         \\ 
4962.259 & 1.847  &   0.200 \\
6408.459 & 2.271  &   0.510 \\
7070.070 & 1.847  &  -0.030 \\
Sr II    &        &         \\
4872.494 & 1.798  &  -0.060 \\
\hline\\
Y I      &        &         \\
6435.004 & 0.066  &   -0.82\\  
5630.134 & 1.356  &   0.211 \\
Y II     &        &         \\
4883.684 &  1.084 &   0.265 \\   
4900.124 &  1.033 &   0.103 \\   
5087.416 &  1.084 &  -0.170 \\   
5200.406 &  0.992 &  -0.570 \\  
5289.815 &  1.033 &  -1.850 \\   
5320.782 &  1.084 &  -1.950 \\   
5402.774 &  1.839 &  -0.630 \\
5544.611 &  1.738 &  -1.090\\   
5728.890 &  1.839 &  -1.120 \\   
6795.414 &  1.738 &  -1.030 \\
7881.880 &  1.840 &  -0.570 \\
\hline\\
Zr I     &        &         \\
7819.374  & 1.820  & -0.380 \\ 
7849.374  & 0.690  & -1.300 \\
Zr II     &        &        \\
4379.742  & 1.532  & -0.356 \\
5112.270  & 1.665  & -0.850 \\
5350.089  & 1.827  & -1.240\\
5350.350  & 1.773  & -1.160 \\ 
\hline\\
Nb I      &        &        \\
4667.221  & 0.267  & -1.160 \\  
5017.736  & 0.049  & -1.320 \\ 
5095.293  & 0.086  & -1.048 \\ 
5100.153  & 0.049  & -1.440 \\ 
5134.737  & 0.019  & -1.390 \\ 
5189.186  & 0.130  & -1.394 \\ 
5271.524  & 0.142  & -1.240 \\ 
5318.598  & 0.197  & -1.085\\ 
5350.722  & 0.267  & -0.910 \\            
5642.096  & 0.739  & -1.060 \\           
5729.185  & 0.197  & -1.530 \\            
5787.512  & 0.267  & -1.650 \\               
5983.206  & 0.657  & -1.250 \\
\hline\\
La II     &        &        \\
4662.478  & 0.000  & -2.952 \\
4662.482  & 0.000  & -2.511 \\
4662.486  & 0.000  & -2.240 \\
4662.491  & 0.000  & -2.253\\
4662.492  & 0.000  & -2.137 \\
4662.493  & 0.000  & -2.256 \\
4662.503  & 0.000  & -2.511 \\
4662.505  & 0.000  & -2.056 \\
4662.507  & 0.000  & -1.763 \\
4748.726  & 0.927  & -0.540 \\ 
4920.965  & 0.126  & -2.261 \\ 
4920.965  & 0.126  & -2.407 \\ 
4920.966  & 0.126  & -2.065 \\ 
4920.966  & 0.126  & -2.078\\ 
4920.966  & 0.126  & -2.738 \\ 
4920.968  & 0.126  & -1.831 \\ 
4920.968  & 0.126  & -1.956 \\ 
4920.968  & 0.126  & -2.629 \\ 
4920.971  & 0.126  & -1.646 \\ 
4920.971  & 0.126  & -1.895 \\ 
4920.971  & 0.126  & -2.650 \\ 
4920.975  & 0.126  & -1.490 \\ 
4920.975  & 0.126  & -1.891 \\ 
4920.975  & 0.126  & -2.760\\ 
4920.979  & 0.126  & -1.354 \\ 
4920.979  & 0.126  & -1.957 \\ 
4920.979  & 0.126  & -2.972 \\ 
4920.985  & 0.126  & -1.233 \\ 
4920.985  & 0.126  & -2.162 \\ 
4920.985  & 0.126  & -3.375 \\   
5290.818  & 0.000  & -1.650 \\
5301.845  & 0.403  & -2.587 \\
5301.857  & 0.403  & -2.684 \\
5301.860  & 0.403  & -2.508\\
5301.878  & 0.403  & -2.830 \\
5301.882  & 0.403  & -2.325 \\
5301.885  & 0.403  & -2.809 \\
5301.908  & 0.403  & -3.065 \\
5301.913  & 0.403  & -2.266 \\
5301.917  & 0.403  & -2.391 \\
5301.946  & 0.403  & -3.483 \\
5301.953  & 0.403  & -2.300 \\
5301.958  & 0.403  & -2.120 \\
5302.001  & 0.403  & -2.483\\
5302.008  & 0.403  & -1.913 \\
5302.067  & 0.403  & -1.742 \\
5302.582  & 2.058  & -0.435 \\
5303.513  & 0.321  & -1.874 \\
5303.513  & 0.321  & -2.363 \\
5303.514  & 0.321  & -3.062 \\
5303.531  & 0.321  & -2.167 \\
5303.532  & 0.321  & -2.247 \\
5303.532  & 0.321  & -2.622 \\
5303.546  & 0.321  & -2.366\\
5303.546  & 0.321  & -2.622 \\
5303.547  & 0.321  & -2.351 \\ 
6172.721  & 0.126  & -2.253 \\    
6262.113  & 0.403  & -3.047 \\ 
6262.114  & 0.403  & -2.901 \\ 
6262.132  & 0.403  & -2.705 \\ 
6262.134  & 0.403  & -2.718 \\ 
6262.135  & 0.403  & -3.378 \\ 
6262.164  & 0.403  & -2.471 \\ 
6262.166  & 0.403  & -2.596\\ 
6262.169  & 0.403  & -3.269 \\ 
6262.208  & 0.403  & -2.286 \\ 
6262.212  & 0.403  & -2.535 \\ 
6262.215  & 0.403  & -3.290 \\ 
6262.266  & 0.403  & -2.130 \\ 
6262.271  & 0.403  & -2.531 \\ 
6262.275  & 0.403  & -3.400 \\ 
6262.338  & 0.403  & -1.994 \\ 
6262.343  & 0.403  & -2.597 \\ 
6262.348  & 0.403  & -3.612\\ 
6262.422  & 0.403  & -1.873 \\ 
6262.429  & 0.403  & -2.802 \\ 
6262.434  & 0.403  & -4.015 \\ 
6390.455  & 0.321  & -2.012 \\ 
6390.468  & 0.321  & -2.183 \\ 
6390.468  & 0.321  & -2.752 \\ 
6390.479  & 0.321  & -2.570 \\ 
6390.479  & 0.321  & -3.752 \\ 
6390.480  & 0.321  & -2.390 \\ 
6390.489  & 0.321  & -2.536\\ 
6390.489  & 0.321  & -3.334 \\ 
6390.490  & 0.321  & -2.661 \\ 
6390.496  & 0.321  & -3.100 \\ 
6390.497  & 0.321  & -2.595 \\ 
6390.498  & 0.321  & -3.079 \\ 
6390.502  & 0.321  & -2.954 \\ 
6390.503  & 0.321  & -2.778 \\ 
6390.506  & 0.321  & -2.857 \\
\hline\\
Ce II     &        &        \\
4515.848  & 1.058  & -0.240\\ 
4562.359  & 0.478  &  0.230 \\
4628.169  & 0.516  &  0.200 \\
4943.441  & 1.206  & -0.360 \\
5274.229  & 1.044  &  0.130 \\ 
5330.556  & 0.869  & -0.400 \\
5472.279  & 1.247  & -0.100 \\
5975.818  & 1.327  & -0.460 \\
6043.373  & 1.206  & -0.480 \\
6051.815  & 3.486  & -0.210 \\
\hline\\
Nd II     &        &       \\
4797.150  & 0.559  & -0.690 \\
4947.020  & 0.559  & -1.130 \\
4961.387  & 0.631  & -0.710 \\ 
5089.832  & 0.205  & -1.160 \\
5092.788  & 0.380  & -0.610 \\
5132.328  & 0.559  & -0.710 \\
5212.360  & 0.205  & -0.960 \\ 
5276.869  & 0.859  & -0.440 \\ 
5293.160  & 0.823  &  0.100 \\ 
5311.450  & 0.986  & -0.420\\
5319.810  & 0.550  & -0.140 \\
5356.967  & 1.264  & -0.280 \\
5361.165  & 0.559  & -1.480 \\ 
5361.467  & 0.680  & -0.370 \\
\hline\\
Pr II     &        &        \\
5219.045  & 0.795  & -0.053 \\ 
5220.108  & 0.796  &  0.298 \\ 
5259.728  & 0.633  &  0.114 \\
5322.772  & 0.483  & -0.141 \\
\hline\\
Sm II     &        &       \\
4318.926  & 0.277  & -0.250 \\
4499.475  & 0.248  & -0.870 \\
4519.630  & 0.544  & -0.350 \\
4566.200  & 0.333  & -0.590 \\ 
4577.688  & 0.248  & -0.650 \\
4615.440  & 0.544  & -0.690 \\
4726.026  & 0.333  & -1.250 \\
\hline\\
Eu II     &        &        \\
6437.601  & 1.320  & -0.960\\ 
6437.603  & 1.320  & -0.960 \\ 
6437.606  & 1.320  & -2.191 \\ 
6437.609  & 1.320  & -2.191 \\ 
6437.617  & 1.320  & -2.191 \\ 
6437.619  & 1.320  & -2.191 \\ 
6437.620  & 1.320  & -1.070 \\ 
6437.623  & 1.320  & -1.998 \\ 
6437.627  & 1.320  & -1.070 \\ 
6437.627  & 1.320  & -1.998 \\ 
6437.630  & 1.320  & -1.181\\ 
6437.633  & 1.320  & -1.956 \\ 
6437.633  & 1.320  & -1.956 \\ 
6437.633  & 1.320  & -1.998 \\ 
6437.635  & 1.320  & -1.287 \\ 
6437.635  & 1.320  & -2.010 \\ 
6437.635  & 1.320  & -2.206 \\ 
6437.637  & 1.320  & -1.377 \\ 
6437.637  & 1.320  & -1.428 \\ 
6437.637  & 1.320  & -2.010 \\ 
6437.639  & 1.320  & -2.206\\ 
6437.640  & 1.320  & -1.998 \\ 
6437.647  & 1.320  & -1.181 \\ 
6437.652  & 1.320  & -1.956 \\ 
6437.657  & 1.320  & -1.956 \\ 
6437.662  & 1.320  & -1.287 \\ 
6437.667  & 1.320  & -2.010 \\ 
6437.669  & 1.320  & -2.010 \\ 
6437.674  & 1.320  & -1.377 \\ 
6437.677  & 1.320  & -2.206 \\ 
6437.679  & 1.320  & -2.206\\ 
6437.682  & 1.320  & -1.428 \\ 
6645.055  & 1.380  & -1.823 \\
6645.057  & 1.380  & -0.516 \\
6645.058  & 1.380  & -3.466 \\
6645.061  & 1.380  & -0.516 \\
6645.067  & 1.380  & -1.823 \\
6645.070  & 1.380  & -0.592 \\
6645.073  & 1.380  & -1.628 \\
6645.075  & 1.380  & -3.466 \\
6645.077  & 1.380  & -3.149\\
6645.080  & 1.380  & -0.672 \\
6645.085  & 1.380  & -1.583 \\
6645.086  & 1.380  & -0.592 \\
6645.087  & 1.380  & -0.754 \\
6645.091  & 1.380  & -3.076 \\
6645.093  & 1.380  & -0.838 \\
6645.093  & 1.380  & -1.634 \\
6645.094  & 1.380  & -1.628 \\
6645.097  & 1.380  & -0.921 \\
6645.099  & 1.380  & -1.829\\
6645.100  & 1.380  & -3.244 \\
6645.101  & 1.380  & -3.149 \\
6645.108  & 1.380  & -0.672 \\
6645.116  & 1.380  & -1.583 \\
6645.123  & 1.380  & -3.076 \\
6645.127  & 1.380  & -0.754 \\
6645.134  & 1.380  & -1.634 \\
6645.140  & 1.380  & -3.244 \\
6645.141  & 1.380  & -0.838 \\
6645.148  & 1.380  & -1.829\\
6645.153  & 1.380  & -0.921\\ 
\hline
\end{supertabular}

\section{Elemental abundances for the programme stars}
{\footnotesize
\begin{table*}
\caption{Elemental abundances.
\label{Tab:abundances}}
\begin{tabular}{lllccccccccc}
\hline
\\
\multicolumn{3}{c}{}& \multicolumn{4}{c}{HD~5424} && \multicolumn{4}{c}{HD~12392} \\
\cline{4-7}\cline{9-12}\\
 &    Z  &    log$_{\odot}{\epsilon}^a$ & log${\epsilon}$&$\sigma_{s}$ (N)& [X/H]  & [X/Fe]$\pm \sigma_{[X/Fe]}$ & & log${\epsilon}$&$\sigma_{s}$(N)& [X/H]  & [X/Fe]$\pm \sigma_{[X/Fe]}$  \\
 &       &                         &                  &        &        \\
\hline

C     &  6  & 8.43  & 8.30  &  0.08(8)& -0.13 & 0.30$\pm$0.12 && 8.50  & 0.05(10) &  0.07 & 0.45$\pm$0.12 \\
$^{12}$C/$^{13}$C & &&  19 &         &       &               &&   19  &          &        &              \\
N     &  7  & 7.83  & 8.07 & 0.07(55)&  0.24 & 0.67$\pm$0.20 && 8.00  & 0.08(50) &  0.17 & 0.55$\pm$0.21 \\
O     &  8  & 8.69  & 8.60 & -(1)    & -0.09 & 0.34$\pm$0.18 && 8.70  & -(2)     &  0.01 & 0.38$\pm$0.18 \\
Na I  & 11  & 6.24  & 5.93 & 0.04(2) & -0.31 & 0.12$\pm$0.13 && 6.00  & 0.05(3)  & -0.24 & 0.14$\pm$0.16 \\
Mg I  & 12  & 7.60  & -    & -       & -     &               && 7.50  & 0.17(3)  & -0.10 & 0.26$\pm$0.24 \\
Fe    & 26  & 7.50  & 7.07 & 0.19(43)& -0.43 & -             && 7.12  & 0.14(44) & -0.38 & 0.09          \\
Rb I  & 37  & 2.52  & 2.3: &         & -0.22 & 0.21$\pm$0.14 &&    -  & -        & -     &               \\
Sr I  & 38  & 2.87  & 3.35 & 0.07(2) &  0.48 & 0.91$\pm$0.11 && 3.42  & 0.03(2)  &  0.55 & 0.93$\pm$0.11 \\
Sr II & 38  & 2.87  & 3.42 & 0.15(3) &  0.55 & 0.98$\pm$0.15 && 3.55  & 0.05(3)  &  0.68 & 1.06$\pm$0.12 \\
Y I   & 39  & 2.21  & 2.58 & 0.14(4) &  0.37 & 0.80$\pm$0.17 && 2.73  & 0.05(4)  &  0.52 & 0.90$\pm$0.16 \\
Y II  & 39  & 2.21  & 3.08 & 0.13(6) &  0.87 & 1.30$\pm$0.11 && 2.91  & 0.13(6)  &  0.70 & 1.08$\pm$0.13 \\
Zr I  & 40  & 2.58  & 3.20 & 0.06(2) &  0.62 & 1.05$\pm$0.13 && 3.40  & 0.05(2)  &  0.82 & 1.20$\pm$0.11 \\
Zr II & 40  & 2.58  & 3.78 & 0.15(3) &  1.20 & 1.63$\pm$0.17 && 3.55  & 0.15     &  0.97 & 1.35$\pm$0.14 \\
Nb I  & 41  & 1.46  & 2.16 & 0.15(4) &  0.70 & 1.13$\pm$0.10 && 2.15  & 0.02(1)  &  0.69 & 1.07$\pm$0.16 \\
Ba II & 56  & 2.18  & 3.90:&         &  1.72 & 2.15$\pm$0.14 && 3.3:  &    -     &  1.12 & 1.50$\pm$0.10 \\
La II & 57  & 1.10  & 2.49 & 0.18(5) &  1.39 & 1.82$\pm$0.11 && 2.17  & 0.09(8)  &  1.07 & 1.45$\pm$0.12 \\
Ce II & 58  & 1.58  & 2.82 & 0.11(7) &  1.24 & 1.67$\pm$0.13 && 2.56  & 0.07(7)  &  0.98 & 1.36$\pm$0.11 \\
Pr II & 59  & 0.72  & 2.05 & 0.16(4) &  1.33 & 1.76$\pm$0.10 && 1.70  & 0.06(4)  &  0.98 & 1.36$\pm$0.10 \\
Nd II & 60  & 1.42  & 2.70 & 0.11(10)&  1.28 & 1.71$\pm$0.10 && 2.44  & 0.09(11) &  1.02 & 1.40$\pm$0.10 \\
Sm II & 62  & 0.96  & 2.08 & 0.14(3) &  1.12 &1.55 $\pm$0.15 && 1.75  & 0.07(2)  &  0.79 & 1.17$\pm$0.13 \\
Eu II & 63  & 0.52  & 0.80 & 0.10(2) &  0.28 &0.71 $\pm$0.15 && 1.00  & 0.03(4)  &  0.48 & 0.86$\pm$0.13 \\
\hline
\\
\multicolumn{3}{c}{}& \multicolumn{4}{c}{HD~16458} && \multicolumn{4}{c}{HD~27271} \\
\cline{4-7}\cline{9-12}\\
\\
 &    Z  &    log$_{\odot}{\epsilon}^a$ & log${\epsilon}$&$\sigma_{s}$(N)& [X/H]  & [X/Fe]$\pm \sigma_{[X/Fe]}$ && log${\epsilon}$&$\sigma_{s}$(N)& [X/H]  & [X/Fe]$\pm \sigma_{[X/Fe]}$  \\
 &       &                         &                  &        &        \\
\hline
C  &  6   &  8.43 &  8.07 & 0.05(9) & -0.36 & 0.28$\pm$0.12 &&   8.30 & 0.06(5)  & -0.13  & -0.06$\pm$0.08  \\
$^{12}$C/$^{13}$C & &&  11 &         &       &                 &&   12  &          &        &              \\
N &  7   & 7.83  &  7.58 & 0.05(61)& -0.25 & 0.39$\pm$0.20 &&   8.30 & 0.03(50) &  0.47  & 0.54$\pm$0.21   \\
O &  8   & 8.69  &  8.10 & 0.14(2) & -0.59 & 0.05$\pm$0.18 &&   8.70 &   -(2)   &  0.01  & 0.08 $\pm$0.17  \\
Na I&11   & 6.24  &  6.30 & 0.14(2) &  0.06 & 0.70$\pm$0.15 &&   6.60 & 0.17(3)  & 0.26   & 0.33$\pm$0.15   \\
Mg I&12   & 7.60  &   -   &         & -     &-              &&   7.60 & (1)      & -0.06  & 0.01 $\pm$0.26  \\
Fe & 26   & 7.50  &  6.86 & 0.15(35)& -0.64 &               &&   7.43 & 0.16(44) & -0.07  &   -             \\
Rb I& 37  & 2.52  &   -   &         &-      & -             &&   1.7: &      -   & -0.82  & -0.75$\pm$0.15  \\
Sr I& 38  & 2.87  &  3.48 & 0.18(2) &  0.61 & 1.25$\pm$0.16 &&   3.45 & 0.10(2)  &  0.58  &  0.65$\pm$0.12  \\
Sr II& 38 & 2.87  &  3.60 & 0.17(3) &  0.73 & 1.37$\pm$0.16 &&   3.60 & 0.10(1)  &  0.73  &  0.80$\pm$0.16  \\
Y I & 39  & 2.21  &  2.64 & 0.05(4) &  0.43 & 1.07$\pm$0.16 &&   2.80 & 0.00(2)  &  0.59  &  0.66$\pm$0.16  \\
Y II& 39  & 2.21  &  2.70 & 0.05(4) &  0.49 & 1.06$\pm$0.12 &&   2.91 & 0.13(8)  &  0.70  &  0.77$\pm$0.12  \\
Zr I & 40 & 2.58  &  3.23 & 0.10(2) &  0.65 & 1.29$\pm$0.11 &&   3.30 & 0.20(2)  &  0.72  &  0.79$\pm$0.17  \\
Zr II & 40& 2.58  &  3.25 & 0.08(3) &  0.67 & 1.31$\pm$0.11 &&   3.70 & 0.10(3)  &  1.12  &  1.19$\pm$0.12  \\
Nb I  & 41& 1.46  &  2.14 & 0.16    &  0.68 & 1.32$\pm$0.17 &&   2.18 & 0.10(2)  &  0.72  &  0.79$\pm$0.17  \\
Ba II & 56 & 2.18 &  3.3: &        &  1.12 & 1.76$\pm$0.11 &&   3.4: &          &  1.22  &  1.29$\pm$0.10  \\
La II & 57 & 1.10 &  1.89 & 0.13(10)&  0.79 & 1.43$\pm$0.12 &&   1.70 & 0.11(7)  &  0.60  &  0.67$\pm$0.12  \\
Ce II & 58 & 1.58 &  2.23 & 0.10(8) &  0.65 & 1.29$\pm$0.11 &&   2.13 & 0.11(8)  &  0.55  &  0.62$\pm$0.11  \\
Pr II & 59& 0.72  &  1.47 & 0.11(4) &  0.75 & 1.39$\pm$0.11 &&   0.98 & 0.05(4)  &  0.26  &  0.33$\pm$0.10  \\
Nd II & 60& 1.42  &  2.13 & 0.10(12)&  0.71 & 1.35$\pm$0.10 &&   1.86 & 0.12(12) &  0.44  &  0.51$\pm$0.11  \\
Sm II & 62& 0.96  &  1.65 & 0.08(2) &  0.69 & 1.33$\pm$0.13 &&   1.30 & 0.10(3)  &  0.34  &  0.41$\pm$0.13  \\
EuII  & 63& 0.52  &  0.50 & 0.10(4) & -0.02 & 0.62$\pm$0.14 &&   0.7  & 0.10(4)  &  0.18  &  0.25$\pm$0.14  \\
\hline
\end{tabular}

$^{a}$ Asplund et al. (2009) \\
\end{table*}
}
\addtocounter{table}{-1}
{\footnotesize
\begin{table*}
\caption{Continued.}
\begin{tabular}{lllccccccccc}
\hline
\\
\multicolumn{3}{c}{}& \multicolumn{4}{c}{HD~28159} && \multicolumn{4}{c}{HD~31487} \\
\cline{4-7}\cline{9-12}\\
 &    Z  &    log$_{\odot}{\epsilon}^a$ & log${\epsilon}$&$\sigma_{s}$ (N)& [X/H]  & [X/Fe]$\pm \sigma_{[X/Fe]}$ & &log${\epsilon}$&$\sigma_{s}$(N)& [X/H]  & [X/Fe]$\pm \sigma_{[X/Fe]}$  \\
 &       &                         &                  &        &        \\
\hline
C     &  6  & 8.43  &  8.30 & -(1)    & -0.13 & 0.37$\pm$0.13  && 8.7  & 0.03(11)&  0.27  & 0.31$\pm$0.08  \\
$^{12}$C/$^{13}$C & &&  6 &         &       &               &&   16  &          &        &              \\
N     &  7  & 7.83  &  7.73 & 0.09(45)& -0.10 & 0.40$\pm$0.21  && 8.49 &0.04(52) &  0.66  & 0.70$\pm$0.21  \\
O     &  8  & 8.69  &  8.60 & -(2)    & -0.09 & 0.41$\pm$0.17  && 8.95 & -(2)    &  0.26  & 0.30$\pm$0.17  \\
Na I  & 11  & 6.24  &  5.80:& -(1)    & -0.44 & 0.06$\pm$0.13  && 6.63 &0.07(4)  &  0.39  & 0.43$\pm$0.12  \\
Mg I  & 12  & 7.60  &  7.60:& -(1)    &  0.00 & 0.50$\pm$0.26  && -    &         & -      &-               \\
Fe    & 26  & 7.50  &  7.00 & 0.15(35)& -0.50 &                && 7.46 & 0.19(45)& -0.04  &                \\
Rb I  & 37  & 2.52  &  2.00 & 0.00(2) & -0.52 & -0.02$\pm$0.13 &&    - &         &     -  & -              \\
Sr I  & 38  & 2.87  &  2.25 & 0.07(2) & -0.62 &-0.12$\pm$0.11  && 3.70 & 0.05(2) &  0.83  & 0.87$\pm$0.11  \\
Sr II & 38  & 2.87  &  2.78 & 0.04(1) & -0.09 & 0.41$\pm$0.12  && 3.81 & 0.11(3) &  0.94  & 0.98$\pm$0.14  \\
Y I   & 39  & 2.21  &  1.45 & 0.08(3) & -0.76 &-0.26$\pm$0.16  && 3.03 & 0.09(4) &  0.82  & 0.86$\pm$0.16  \\
Y II  & 39  & 2.21  &  1.87 & 0.20(3) & -0.34 & 0.16$\pm$0.16  && 3.40 & 0.13(5) &  1.19  & 1.23$\pm$0.13  \\
Zr I  & 40  & 2.58  &  2.38 & 0.10(2) & -0.20 & 0.30$\pm$0.12  && 3.65 & 0.07(2) &  1.07  & 1.11$\pm$0.11  \\
Zr II & 40  & 2.58  &  2.80 & 0.00(2) &  0.22 & 0.72$\pm$0.10  && 4.27 & 0.10(3) &  1.69  & 1.73$\pm$0.12  \\
Nb I  & 41  & 1.46  &  1.39 & 0.14    & -0.07 & 0.43$\pm$0.16  && 2.57 & 0.06(4) &  1.11  & 1.15$\pm$0.16  \\
Ba II & 56  & 2.18  &  1.28:&         & -0.90 &-0.40$\pm$0.10  && 4.00:&         &  1.82  & 1.86$\pm$0.11  \\
La II & 57  & 1.10  &  1.47 & 0.07(7) &  0.37 & 0.87$\pm$0.12  && 2.59 & 0.13(7) &  1.49  & 1.53$\pm$0.12  \\
Ce II & 58  & 1.58  &  1.54 & 0.19(6) & -0.04 & 0.46$\pm$0.13  && 3.00 & 0.13(10)&  1.42  & 1.46$\pm$0.11  \\
Pr II & 59  & 0.72  &  0.77 & 0.14(4) &  0.05 & 0.55$\pm$0.12  && 2.13 & 0.10(4) &  1.41  & 1.45$\pm$0.11  \\
Nd II & 60  & 1.42  &  1.57 & 0.15(7) &  0.15 & 0.65$\pm$0.11  && 2.84 & 0.12(8) &  1.42  & 1.46$\pm$0.11  \\
Sm II & 62  & 0.96  &  0.85 & 0.20(2) & -0.11 & 0.39$\pm$0.18  && 2.15 & 0.10(1) &  1.19  & 1.23$\pm$0.16  \\
Eu II & 63  & 0.52  &  0.59 & 0.14(3) &  0.07 & 0.57$\pm$0.15  && 1.08 & 0.08(3) &  0.56  & 0.60$\pm$0.14  \\
\hline
\\
\multicolumn{3}{c}{}& \multicolumn{4}{c}{HD~43389} && \multicolumn{4}{c}{HD~46407} \\
\cline{4-7}\cline{9-12}\\
 &    Z  &    log$_{\odot}{\epsilon}^a$ & log${\epsilon}$&$\sigma_{s}$(N)& [X/H]  & [X/Fe]$\pm \sigma_{[X/Fe]}$ & &log${\epsilon}$&$\sigma_{s}$(N)& [X/H]  & [X/Fe]$\pm \sigma_{[X/Fe]}$  \\
 &       &                         &                  &        &        \\
\hline
C     &  6  & 8.43  &  8.30  & -(1)     & -0.13 & 0.22$\pm$0.13 &&  8.32 & 0.03(10)& -0.11  & 0.25$\pm$0.12 \\
$^{12}$C/$^{13}$C & &&  10&         &       &               &&   16  &          &        &              \\
N     &  7  & 7.83  &  8.62  & 0.15(48) &  0.79 & 1.14$\pm$0.22 &&  7.95 & 0.10(48)&  0.12  & 0.48$\pm$0.20 \\
O     &  8  & 8.69  &  8.40  & -    (2) & -0.29 & 0.06$\pm$0.17 &&  8.40 &  -(1)   & -0.29  & 0.07$\pm$0.18 \\
Na I  & 11  & 6.24  &  5.80  & 0.07(2)  & -0.44 &-0.09$\pm$0.13 &&  6.27 & 0.12(3) &  0.03  & 0.39$\pm$0.17 \\
Mg I  & 12  & 7.60  & -      & -        & -     &               &&  7.30 & -(2)    & -0.30  & 0.06$\pm$0.26 \\
Fe    & 26  & 7.50  &  7.15  & 0.15(30) & -0.35 &               &&  7.14 & 0.15(39)& -0.36  &              \\
Rb I  & 37  & 2.52  &  1.85  & 0.00(2)  & -0.67 &-0.32$\pm$0.12 &&  -    &         &  -     & -         \\ 
Sr I  & 38  & 2.87  &  2.80  & 0.1(1)   & -0.07 & 0.28$\pm$0.14 &&  3.68 & 0.03(2) &  0.81  & 1.17$\pm$0.11 \\
Sr II & 38  & 2.87  &  3.00  & 0.15(3)  &  0.13 & 0.48$\pm$0.15 &&  3.75 & 0.07(2) &  0.88  & 1.24$\pm$0.13 \\
Y I   & 39  & 2.21  &  1.78  & 0.15(3)  & -0.44 &-0.09$\pm$0.18 &&  2.85 & 0.17(3) &  0.64  & 1.00$\pm$0.19 \\
Y II  & 39  & 2.21  &  2.77  & 0.15(5)  &  0.56 & 0.91$\pm$0.13 &&  3.00 & 0.15(7) &  0.79  & 1.15$\pm$0.13 \\
Zr I  & 40  & 2.58  &  2.55  & 0.10(2)  & -0.03 & 0.32$\pm$0.12 &&  3.50 & 0.00(2) &  0.92  & 1.28$\pm$0.10 \\
Zr II & 40  & 2.58  &    -   &          &   -   &  -            &&  3.83 & 0.11(3) &  1.25  & 1.61$\pm$0.12 \\
Nb I  & 41  & 1.46  &  1.75  & 0.13(9)  &  0.29 & 0.64$\pm$0.16 &&  2.39 & 0.06(4) &  0.93  & 1.29$\pm$0.16 \\
Ba II & 56  & 2.18  &  3.60: &          &  1.42 & 1.77$\pm$0.11 &&  3.70:&         &  1.52  & 1.88$\pm$0.10 \\
La II & 57  & 1.10  &  2.28  & 0.13(7)  &  1.18 & 1.53$\pm$0.12 &&  2.30 & 0.15(6) &  1.20  & 1.56$\pm$0.13 \\
Ce II & 58  & 1.58  &  2.45  & 0.12(7)  &  0.87 & 1.22$\pm$0.11 &&  2.72 & 0.11(8) &  1.14  & 1.50$\pm$0.11 \\
Pr II & 59  & 0.72  &  1.70  & 0.10(4)  &  0.98 & 1.33$\pm$0.11 &&  1.76 & 0.13(4) &  1.04  & 1.40$\pm$0.10 \\
Nd II & 60  & 1.42  &  2.50  & 0.14(7)  &  1.08 & 1.43$\pm$0.11 &&  2.46 & 0.12(14)&  1.04  & 1.40$\pm$0.09 \\
Sm II & 62  & 0.96  &  1.55  & 0.15(2)  &  0.59 & 0.94$\pm$0.16 &&  1.73 & 0.05(3) &  0.77  & 1.13$\pm$0.12 \\
Eu II & 63  & 0.52  &  0.93  & 0.03(2)  &  0.41 & 0.76$\pm$0.13 &&  0.70 & 0.03(4) &  0.18  & 0.54$\pm$0.13 \\
\hline
\end{tabular}

$^{a}$ Asplund et al. (2009) \\
\end{table*}
}

\addtocounter{table}{-1}
{\footnotesize
\begin{table*}
\caption{Continued.}
\begin{tabular}{lllccccccccc}
\hline\\
\multicolumn{3}{c}{}& \multicolumn{4}{c}{HD~50082} && \multicolumn{4}{c}{HD~60197} \\
\cline{4-7}\cline{9-12}\\
\\
&    Z  &    log$_{\odot}{\epsilon}^a$ & log${\epsilon}$&$\sigma_{s}$ (N)& [X/H]  & [X/Fe]$\pm \sigma_{[X/Fe]}$ & &log${\epsilon}$&$\sigma_{s}$(N)& [X/H]  & [X/Fe]$\pm \sigma_{[X/Fe]}$  \\
 &       &                         &                  &        &        \\
\hline

C     &  6  & 8.43 & 8.20 & 0.05(11)&  -0.23 &0.09$\pm$0.11  && 8.00  & -(1)     & -0.43  & 0.17$\pm$0.13  \\
$^{12}$C/$^{13}$C & &&  13 &         &       &               &&   9  &          &        &              \\
N     &  7  & 7.83 & 7.70 & 0.08(49)&  -0.13 & 0.19$\pm$0.21  && 8.39  & 0.13(47) &  0.56  & 1.16$\pm$0.21  \\
O     &  8  & 8.69 & 8.35 & -(2)    &  -0.34 & -0.02$\pm$0.17  && 8.10  & -(2)     & -0.59  & 0.01$\pm$0.17  \\
Na I  & 11  & 6.24 & 6.15 & 0.07(2) & -0.09 & 0.23$\pm$0.13  && 6.18  & 0.04(2)  & -0.06  & 0.54$\pm$0.12  \\
Mg I  & 12  & 7.60 & 7.53 & 0.11(2) & -0.07 & 0.25$\pm$0.27  && -     & -        &        & -              \\
Fe    & 26  & 7.50 & 7.18 & 0.10(36)& -0.32 &                && 6.90  & 0.15(30) & -0.60  &                \\
Rb I  & 37  & 2.52 & 2.1: &         & -0.42 & -0.10$\pm$0.15 && 2.1   & 0.00     & -0.42  & 0.18$\pm$0.13   \\
Sr I  & 38  & 2.87 & 3.40 & 0.03(2) &  0.53 & 0.85$\pm$0.11  && 2.90  & (1)      &  0.03  & 0.63$\pm$0.14  \\
Sr II & 38  & 2.87 & 3.40 & 0.05(1) &  0.53 & 0.85$\pm$0.13  && 3.13  & 0.08(4)  &  0.26  & 0.86$\pm$0.13  \\
Y I   & 39  & 2.21 & 2.75 & 0.07(2) &  0.54 & 0.86$\pm$0.17  && 1.88  & 0.15(4)  & -0.33  & 0.27$\pm$0.18  \\
Y II  & 39  & 2.21 & 2.88 & 0.18(7) &  0.67 & 0.99$\pm$0.13  && 2.55  & 0.12(6)  &  0.34  & 0.94$\pm$0.12  \\
Zr I  & 40  & 2.58 & 3.30 & 0.00(2) &  0.72 & 1.04$\pm$0.10  && 2.85  & 0.00(2)  &  0.27  & 0.87$\pm$0.10  \\
Zr II & 40  & 2.58 & 3.60 & 0.05(2) &  1.02 & 1.34$\pm$0.11  && 3.30  & 0.10(1)  &  0.72  & 1.32$\pm$0.14  \\ 
Nb I  & 41  & 1.46 & 2.28 & 0.12(3) &  0.82 & 1.14$\pm$0.17  && 1.75  & 0.14(9)  &  0.29  & 0.89$\pm$0.16  \\
Ba II & 56  & 2.18 & 3.7: &         &  1.52 & 1.84$\pm$0.10  && 3.3:  &          &  1.12  & 1.72$\pm$0.10  \\
La II & 57  & 1.10 & 2.20 & 0.20(5) &  1.10 & 1.42$\pm$0.14  && 1.55  & 0.10(5)  &  0.45  & 1.05$\pm$0.12  \\
Ce II & 58  & 1.58 & 2.61 & 0.14(11)&  1.03 & 1.35$\pm$0.11  && 1.86  & 0.14(9)  &  0.28  & 0.88$\pm$0.10  \\
Pr II & 59  & 0.72 & 1.73 & 0.16(4) &  1.01 & 1.33$\pm$0.13  && 1.23  & 0.02(4)  &  0.51  & 1.11$\pm$0.10  \\
Nd II & 60  & 1.42 & 2.48 & 0.14(14)&  1.06 & 1.38$\pm$0.11  && 1.99  & 0.08(8)  &  0.57  & 1.17$\pm$0.12  \\
Sm II & 62  & 0.96 & 1.85 & 0.14(2) &  0.89 & 1.21$\pm$0.16  && 1.33  & 0.03(2)  &  0.37  & 0.97$\pm$0.12  \\
Eu II & 63  & 0.52 & 0.65 & 0.03(4) &  0.13 & 0.45$\pm$0.13  && 0.50  & 0.07(2)  & -0.02  & 0.58$\pm$0.14  \\
\hline
\\
\multicolumn{3}{c}{}& \multicolumn{4}{c}{HD~88562} && \multicolumn{4}{c}{HD~100503} \\
\cline{4-7}\cline{9-12}\\
\\
 &    Z  &    log$_{\odot}{\epsilon}^a$ & log${\epsilon}$&$\sigma_{s}$(N)& [X/H]  & [X/Fe]$\pm \sigma_{[X/Fe]}$ & &log${\epsilon}$&$\sigma_{s}$(N)& [X/H]  & [X/Fe]$\pm \sigma_{[X/Fe]}$  \\
 &       &                         &                  &        &        \\
\hline

C     &  6  & 8.43  & 8.00  & -(1)    & -0.43 &  0.10$\pm$0.13  &&  8.00 & -(1)    & -0.43  & 0.29$\pm$0.13  \\ 
$^{12}$C/$^{13}$C & &&  13 &         &       &               &&   10  &          &        &              \\
N     &  7  & 7.83  & 7.50  & 0.13(45)& -0.33 &  0.20$\pm$0.22  &&  8.32 & 0.13(48)&  0.49  & 1.21$\pm$0.21  \\ 
O     &  8  & 8.69  & 8.05  &  -(2)   & -0.64 & -0.11$\pm$0.18  &&  8.03 &  -(2)   & -0.66  & 0.06$\pm$0.18  \\ 
Na I  & 11  & 6.24  & 5.80  & 0.07(2) & -0.44 &  0.09$\pm$0.13  &&  6.10 &0.14(2)  & -0.14  & 0.58$\pm$0.15  \\ 
Mg I  & 12  & 7.60  & 7.00: & -(1)    & -0.60 & -0.07$\pm$0.26  &&  7.20:&         & -0.40  & 0.32$\pm$0.27  \\ 
Fe    & 26  & 7.50  & 6.97  & 0.15(30)& -0.53 &                 &&  6.78 &0.15(35) & -0.72  &      -            \\
Rb I  & 37  & 2.52  & 1.85  & 0.03(2) & -0.67 & -0.14$\pm$0.13  &&  2.00 &-(1)  & -0.52  & 0.20 $\pm$0.13    \\ 
Sr I  & 38  & 2.87  & 2.75  & 0.08(3) & -0.12 &  0.41$\pm$0.11  &&  2.95 &-(1)     &  0.08  & 0.80$\pm$0.14  \\ 
Sr II & 38  & 2.87  & 2.90  & 0.05(3) &  0.05 &  0.56$\pm$0.12  &&  3.10 &0.14(2)  &  0.23  & 0.95$\pm$0.14  \\ 
Y I   & 39  & 2.21  & 1.75  & 0.14(4) & -0.46 &  0.07$\pm$0.17  &&  2.16 &0.12(4)  & -0.05  & 0.67$\pm$0.16  \\ 
Y II  & 39  & 2.21  & 2.61  & 0.10(7) &  0.40 &  0.93$\pm$0.12  &&  2.80 &0.10(3)  &  0.59  & 1.31$\pm$0.13  \\ 
Zr I  & 40  & 2.58  & 2.48  & 0.10(2) & -0.10 &  0.43$\pm$0.12  &&  2.86 &0.10(2)  &  0.18  & 1.00$\pm$0.12  \\ 
Zr II & 40  & 2.58  & 3.42  & 0.04(3) &  0.84 &  1.37$\pm$0.11  &&  3.3:  &         &  0.72  & 1.44$\pm$0.14  \\ 
Nb I  & 41  & 1.46  & 1.46  & 0.10(9) &  0.00 &  0.53$\pm$0.16  &&  2.07 &0.14(7)  &  0.61  & 1.33$\pm$0.16  \\ 
Ba II & 56  & 2.18  & 3.3:  &         &  1.12 &  1.65$\pm$0.10  &&  2.85:&     -   &  0.67  & 1.39$\pm$0.10  \\ 
La II & 57  & 1.10  & 1.72  & 0.10(9) &  0.62 &  1.15$\pm$0.12  &&  2.17 & 0.09(7) &  1.07  & 1.79$\pm$0.12  \\ 
Ce II & 58  & 1.58  & 2.07  & 0.10(10)&  0.49 &  1.02$\pm$0.11  &&  2.49 & 0.12(5) &  0.91  & 1.63$\pm$0.12  \\ 
Pr II & 59  & 0.72  & 1.24  & 0.08(4) &  0.52 &  1.05$\pm$0.11  &&  1.78 & 0.20(4) &  1.06  & 1.78$\pm$0.14  \\ 
Nd II & 60  & 1.42  & 1.90  & 0.14(11)&  0.48 &  1.01$\pm$0.11  &&  2.58 & 0.04(6) &  1.16  & 1.88$\pm$0.10  \\ 
Sm II & 62  & 0.96  & 1.25  & 0.08(2) &  0.29 &  0.82$\pm$0.13  &&  1.91 & 0.13(2) &  0.95  & 1.67$\pm$0.15  \\ 
Eu II & 63  & 0.52  & 0.70  & 0.10(2) &  0.18 &  0.71$\pm$0.15  &&  1.02 & 0.03(2) &  0.50  & 1.22$\pm$0.13  \\ 
\hline
\end{tabular}

$^{a}$ Asplund et al. (2009) \\
\end{table*}
}

\addtocounter{table}{-1}
{\footnotesize
\begin{table*}
\caption{Continued.}
\begin{tabular}{lllccccccccc}
\hline
\\
\multicolumn{3}{c}{}& \multicolumn{4}{c}{HD~116869} && \multicolumn{4}{c}{HD~120620} \\
\cline{4-7}\cline{9-12}\\
 &    Z  &    log$_{\odot}{\epsilon}^a$ & log${\epsilon}$&$\sigma_{s}$ (N)& [X/H]  & [X/Fe]$\pm \sigma_{[X/Fe]}$ & &log${\epsilon}$&$\sigma_{s}$(N)& [X/H]  & [X/Fe]$\pm \sigma_{[X/Fe]}$  \\
 &       &                         &                  &        &        \\
\hline
C  &  6  & 8.43 &  8.3  & 0.03(10)  & -0.13  & 0.31$\pm$0.12 && 8.6  &  0.03(7)  &  0.17  & 0.47$\pm$0.12   \\
$^{12}$C/$^{13}$C & &&  9 &         &       &               &&   90  &          &        &              \\
N     &  7  & 7.83 &  7.7  &   0.05(49)& -0.13  & 0.31$\pm$0.21 && 8.0  &  0.05(52) &  0.17  & 0.47$\pm$0.20   \\
O     &  8  & 8.69 &  8.6  &     - (3) & -0.09  & 0.35$\pm$0.16 && 8.9  &      -(1) &  0.21  & 0.51$\pm$0.17   \\
Na I  & 11 & 6.24 &  6.05 &0.07(2)   & -0.19  & 0.25$\pm$0.13 && 6.22 & 0.03(3)   & -0.02  &0.28$\pm$ 0.15   \\
Mg I  & 12  & 7.60 &  7.60 &- (2)    &  0.00  & 0.44$\pm$0.26 && 7.60:&     -     & 0.00  &0.30 $\pm$0.24   \\
Fe    & 26  & 7.50 &  7.06 &0.15(40) & -0.44  &               && 7.2  &  0.16(42) & -0.30  &                 \\
Rb I  & 37  & 2.52 &    -  &    -  &  -     &        -      &&   -  &    -      &  -     &            -      \\
Sr I  & 38  & 2.87 &  3.30 &-(1) &  0.43  & 0.87$\pm$0.10 && 3.75 &  -(2)     &  0.88  & 1.18$\pm$0.10   \\
Sr II & 38  & 2.87 &  3.30 &-(2) &  0.43  & 0.87$\pm$0.12 && 4.00 &0.10(1)    &  1.13  & 1.43$\pm$0.16   \\   
Y I   & 39  & 2.21 &  2.30 &-(1)&  0.09  & 0.53$\pm$0.19 && 3.03 &0.12(4)    &  0.82  & 1.12$\pm$0.17   \\
Y II  & 39  & 2.21 &  2.48 &0.12(6) &  0.27  & 0.71$\pm$0.12 && 3.20 &0.16(7)    &  0.99  & 1.29$\pm$0.13   \\
Zr I  & 40  & 2.58 &  3.15 &0.20(2) &  0.57  & 1.01$\pm$0.17 && 3.55 &  - (1)    &  0.97  & 1.27$\pm$0.14   \\
Zr II & 40  & 2.58 &  3.18 &0.10(3) & 0.60  & 1.04$\pm$0.12 && 4.04 &0.07(3)    &  1.46  & 1.76$\pm$0.11   \\  
Nb I  & 41  & 1.46 &  2.00 &0.15(1) &  0.54  & 0.98$\pm$0.21 && 2.51 &0.13(8)  &  1.05  & 1.35$\pm$0.16   \\
Ba II & 56  & 2.18 &  3.15:&  -     &  0.97  & 1.41$\pm$0.10 && 3.9: &           &  1.72  & 2.02$\pm$0.10   \\
La II & 57  & 1.10 &  1.82 &0.10(9) &  0.72  & 1.16$\pm$0.12 && 2.40 &0.16(6)    &  1.30  & 1.60$\pm$0.13   \\
Ce II & 58  & 1.58 &  2.21 &0.09(10)&  0.63  & 1.07$\pm$0.11 && 2.80 &0.15(8)    &  1.22  & 1.52$\pm$0.11   \\
Pr II & 59  & 0.72 &  1.20 &0.08(4) &  0.48  & 0.92$\pm$0.11 && 1.93 &0.06(4)    &  1.21  & 1.51$\pm$0.10   \\
Nd II & 60  & 1.42 &  2.10 &0.09(12)&  0.68  & 1.12$\pm$0.10 && 2.60 &0.06(14)   &  1.18  & 1.48$\pm$0.10   \\
Sm II & 62  & 0.96 &  1.58 &0.13(4) &  0.62  & 1.06$\pm$0.14 && 1.85 &0.14(4)    &  0.89  & 1.19$\pm$0.14   \\
Eu II & 63  & 0.52 &  0.65 &0.08(2) &  0.13  & 0.57$\pm$0.14 && 0.97 &0.03(4)    &  0.45  & 0.75$\pm$0.13   \\
\hline
\\
\multicolumn{3}{c}{}& \multicolumn{4}{c}{HD~121447} && \multicolumn{4}{c}{HD~123949} \\
\cline{4-7}\cline{9-12}\\
 &    Z  &    log$_{\odot}{\epsilon}^a$ & log${\epsilon}$&$\sigma_{s}$(N)& [X/H]  & [X/Fe]$\pm \sigma_{[X/Fe]}$ & &log${\epsilon}$&$\sigma_{s}$(N)& [X/H]  & [X/Fe]$\pm \sigma_{[X/Fe]}$  \\
 &       &                         &                  &        &        \\
\hline
C     &  6 & 8.43  & 8.07  & -    (1)& -0.36   & 0.54$\pm$0.13  && 8.4   & 0.05(10)&  -0.03 & 0.28$\pm$0.12 \\
$^{12}$C/$^{13}$C & &&  7 &         &       &               &&   19  &          &        &              \\
N     &  7 & 7.83  & 7.80  & 0.13(45)& -0.03   & 0.87$\pm$0.21  && 7.7   & 0.06(50)&  -0.13 & 0.18$\pm$0.21 \\
O     &  8 & 8.69  & 8.09  & -   (2) &  -0.60  & 0.30$\pm$0.17  && 8.45  &  -(1)   &  -0.24 & 0.07$\pm$0.18 \\
Na I  & 11 & 6.24  & 6.23  & 0.03(2) & -0.01   & 0.89$\pm$0.07  && 6.27  & 0.07(3) &   0.03 & 0.34$\pm$0.12 \\
Mg I  & 12 & 7.60  & -     &         & -       & -              && 7.60  & 0.00(2) &   0.00  & 0.31$\pm$0.26 \\
Fe    & 26 & 7.50  & 6.60  & 0.15(28)& -0.90   &                && 7.19  & 0.18(41)&  -0.31 &               \\
Rb I  & 37 & 2.52  & 2.25  & 0.05(2) & -0.27   & 0.63$\pm$0.13  &&   -   &         &        &       -       \\
Sr I  & 38 & 2.87  & 3.18  & 0.03(2) &  0.31   & 1.21$\pm$0.11  && 3.44  & 0.11(3) &   0.57 &  0.88$\pm$0.12\\
Sr II & 38 & 2.87  & 3.5:  &  -(1)   &  0.63   & 1.53$\pm$0.13  && 3.46  & 0.08(3) &   0.59 &  0.90$\pm$0.13\\
Y I   & 39 & 2.21  & 2.38  & 0.09(4) &  0.17   & 1.07$\pm$0.16  && 2.65  & 0.15(4) &   0.44 &  0.75$\pm$0.18\\
Y II  & 39 & 2.21  & 2.66  & 0.11(5) &  0.45   & 1.35$\pm$0.12  && 2.81  & 0.17(5) &   0.60 &  0.91$\pm$0.14\\
Zr I  & 40 & 2.58  & 3.25  & 0.14(2) &  0.67   & 1.57$\pm$0.14  && 3.15  & 0.14(2) &   0.57 &  0.88$\pm$0.14\\
Zr II & 40 & 2.58  &    -  &         &   -     &   -            && 3.73  & 0.10(3) &   1.15 &  1.46$\pm$0.12\\
Nb I  & 41 & 1.46  & 2.48  & 0.10(7) &  1.02   & 1.92$\pm$0.16  && 2.11  & 0.12(9) &   0.65 &  0.96$\pm$0.16\\
Ba II & 56 & 2.18  & 3.15: &    -    &  0.97   & 1.87$\pm$0.11  && 3.8:  &    -    &   1.62 &  1.93$\pm$0.10\\
La II & 57 & 1.10  & 2.59  & 0.15(5) &  1.49   & 2.39$\pm$0.13  && 2.00  & 0.14(7) &   0.90 &  1.21$\pm$0.13\\
Ce II & 58 & 1.58  & 2.90  & 0.17(3) &  1.32   & 2.22$\pm$0.14  && 2.55  & 0.14(8) &   0.97 &  1.28$\pm$0.11\\
Pr II & 59 & 0.72  & 2.18  & 0.08(4) &  1.46   & 2.36$\pm$0.11  && 1.77  & 0.13(4) &   1.05 &  1.36$\pm$0.12\\
Nd II & 60 & 1.42  & 2.55  & 0.12(6) &  1.13   & 2.03$\pm$0.11  && 2.45  & 0.13(12)&   1.03 &  1.34$\pm$0.11\\
Sm II & 62 & 0.96  & 1.85  & 0.17(2) &  0.89   & 1.79$\pm$0.17  && 1.70  & 0.10(3) &   0.74 &  1.05$\pm$0.13\\
Eu II & 63 & 0.52  & 0.75  & 0.03(2) &  0.23   & 1.13$\pm$0.13  && 0.70  & 0.10(3) &   0.18 &  0.49$\pm$0.14\\
\hline
\end{tabular}

$^{a}$ Asplund et al. (2009) \\
\end{table*}
}
\addtocounter{table}{-1}

{\footnotesize
\begin{table*}
\caption{Continued}
\begin{tabular}{lllccccccccc}
\hline
\\
\multicolumn{3}{c}{}& \multicolumn{4}{c}{HD~178717} && \multicolumn{4}{c}{HD~199939} \\
\cline{4-7}\cline{9-12}\\
\\
 &    Z  &    log$_{\odot}{\epsilon}^a$ & log${\epsilon}$&$\sigma_{s}$(N)& [X/H]  & [X/Fe]$\pm \sigma_{[X/Fe]}$ && log${\epsilon}$&$\sigma_{s}$(N)& [X/H]  & [X/Fe]$\pm \sigma_{[X/Fe]}$  \\
 &       &                         &                  &        &        \\
\hline
C     &  6 & 8.43 & 8.00  & -(1)   & -0.43  &  0.09$\pm$0.13 && 8.5  & 0.06(7) & 0.07    & 0.29$\pm$0.12 \\
$^{12}$C/$^{13}$C & &&  14 &         &       &               &&   19&          &        &              \\
N     &  7 & 7.83 & 7.39  &0.11(45)& -0.44  &  0.08$\pm$0.21 && 8.3  & 0.08(49)& 0.47    & 0.69$\pm$0.20 \\
O     &  8 & 8.69 & 8.003 & (1)    & -0.69  & -0.17$\pm$0.18 && 8.75 &  -(1)   & 0.06    & 0.28$\pm$0.18 \\
Na I  & 11 & 6.24 & 5.70  &0.07(2) & -0.54  & -0.02$\pm$0.13 && 6.38 &0.10(2)  & 0.14    & 0.36$\pm$0.14 \\
Mg I  & 12 & 7.60 & 7.40: &        & -0.20  & 0.32$\pm$0.26  && 7.43 &0.06(3)  & -0.17   & 0.05$\pm$0.24 \\
Fe    & 26 & 7.50 & 6.98  &0.15(30)& -0.52  &                && 7.28 &0.11(48) & -0.22   &     -         \\
Rb I  & 37 & 2.52 & 2.00  &0.00(1) & -0.52  &  0.00$\pm$0.13 &&   -  & -       &     -   &               \\
Sr I  & 38 & 2.87 & 2.95  &0.07(2) &  0.08  &  0.60$\pm$0.11 && 3.70 &- (2)    &  0.83   & 1.05$\pm$0.10 \\
Sr II & 38 & 2.87 & 3.10  &0.08(3) &  0.23  &  0.75$\pm$0.13 && 3.90 &0.15(2)  &  1.03   & 1.25$\pm$0.16 \\
Y I   & 39 & 2.21 & 1.73  &0.17(4) & -0.48  &  0.04$\pm$0.18 && 2.99 &0.11(4)  &  0.78   & 1.00$\pm$0.17 \\
Y II  & 39 & 2.21 & 2.48  &0.10(6) &  0.27  &  0.79$\pm$0.12 && 3.37 &0.18(7)  &  1.16   & 1.38$\pm$0.13 \\
Zr I  & 40 & 2.58 & 2.50  &0.20(2) & -0.08  &  0.44$\pm$0.17 && 3.55 &0.07(2)  &  0.97   & 1.19$\pm$0.13 \\
Zr II & 40 & 2.58 & 3.35  &0.07(2) &  0.77  &  1.29$\pm$0.11 && 4.10 &- (3)    &  1.52   & 1.74$\pm$0.10 \\
Nb I  & 41 & 1.46 & 1.62  &0.12(9) &  0.16  &  0.68$\pm$0.16 && 2.57 &0.09(9)  &  1.11   & 1.33$\pm$0.16 \\
Ba II & 56 & 2.18 & 2.70: &        &  0.52  &  1.04$\pm$0.10 && 4.2: &         &  2.02   & 2.24$\pm$0.10 \\
La II & 57 & 1.10 & 1.43  &0.04(7) &  0.33  &  0.85$\pm$0.11 && 2.60 &0.20(4)  &  1.50   & 1.72$\pm$0.15 \\
Ce II & 58 & 1.58 & 1.80  &0.11(8) &  0.22  &  0.74$\pm$0.11 && 3.03 &0.11(6)  &  1.45   & 1.67$\pm$0.11 \\
Pr II & 59 & 0.72 & 1.08  &0.07(4) &  0.36  &  0.88$\pm$0.10 && 2.30 &0.12(4)  &  1.58   & 1.80$\pm$0.12 \\
Nd II & 60 & 1.42 & 1.82  &0.14(7) &  0.40  &  0.92$\pm$0.11 && 2.92 &0.12(14) &  1.50   & 1.72$\pm$0 .10 \\
Sm II & 62 & 0.96 & 1.08  &0.10(2) &  0.12  &  0.64$\pm$0.14 && 2.30 &0.03(2)  &  1.34   & 1.56$\pm$0.12 \\
Eu II & 63 & 0.52 & 0.55  &0.10(2) &  0.03  &  0.55$\pm$0.15 && 1.0  &0.03(2)  &  0.48   & 0.70$\pm$0.13 \\

\hline
\end{tabular}

$^{a}$ Asplund et al. (2009) \\
\end{table*}
}

\end{appendix}
\end{document}